\documentclass[superscriptaddress,aps,prd,nofootinbib,twocolumn,showpacs]{revtex4-1}
\usepackage[english]{babel}
\usepackage[utf8]{inputenc}
\usepackage{float}
\usepackage{amsmath}
\usepackage{amssymb}
\usepackage{graphicx}
\usepackage{color}

\usepackage[normalem]{ulem}

\newcommand{\ev}[1]{\ensuremath{\big{<}#1\big{>}}}

\begin{document}
	
\title{Effect of the crust on neutron star empirical relations}

\author{Márcio Ferreira}
\email{marcio.ferreira@uc.pt}
\affiliation{CFisUC, 
	Department of Physics, University of Coimbra, P-3004 - 516  Coimbra, Portugal}

\author{Constança Providência}
\email{cp@fis.uc.pt}
\affiliation{CFisUC, 
	Department of Physics, University of Coimbra, P-3004 - 516  Coimbra, Portugal}

\date{\today}

\begin{abstract}

We analyze how the crust equation of state affects several neutron star properties 
and how it impacts on possible constraints inferred from astrophysical observations.
Using three distinct crusts, we generate three sets of model-independent equations
of state describing stellar matter from a Taylor expansion around
saturation density. The equations of state are thermodynamically consistent,
causal, and compatible with astrophysical observations.
The relations
between the tidal deformability $\Lambda$  and compactness $C$, Love
number $k_2$ and radius of neutron star with mass $M$ are studied,
and the effect of the crust equation of state on these relations
analyzed. In most of the relations, the impact of  the crust equation
of state is not larger that 2\%.  If, however, a fixed neutron star
mass is considered, the relation  between the tidal deformability and
the radius  depends on the crust.
We have found that the relation $\Lambda_{M_i}
= \alpha R_{M_i}^{\beta}$ becomes almost exact and crust independent
for massive neutron stars. 
 It is shown that it is possible to determine the  tidal deformability
 of an 1.4$M_\odot$ star from the GW179817  effective tidal deformability
 $\tilde\Lambda$ with an accuracy of
 at least $\approx 10\%$. A high correlation
 between $\tilde\Lambda$ and
 the radius of the most massive star of the neutron star binary was
 confirmed, however, it was demonstrated that the  crust has an effect of
 $\approx 14\%$ on this relation.  
We have found that the relation $\Lambda_1/\Lambda_2=q^a$ 
 depends on $M_{\text{chirp}}$ as $a\sim \sqrt{M_{\text{chirp}}}$.

\end{abstract}

\maketitle
\section{Introduction}

Our knowledge on the equation of state (EoS) of nuclear matter is still very limited.
Its properties above the nuclear saturation density remain an open question in nuclear
physics. Neutron stars (NSs) are special astrophysical objects
through which the properties of cold super-dense neutron-rich nuclear
matter can be investigated. Some massive NSs observed during the last
decade established quite stiff constraints on the EoS of
nuclear matter. The pulsar PSR J1614$-$2230 is the one with the smallest
uncertainty on the mass $M=1.908\pm0.016\, M_\odot$ 
\cite{Arzoumanian2017,Fonseca2017,Demorest2010}.
Two known pulsars with a mass above
two solar masses are PSR J0348$+$0432 with $M=2.01 \pm 0.04 M_\odot$
\cite{Antoniadis2013} and the recently detected MSP J0740$+$6620 
with a mass $2.14 {\scriptsize\begin{array}{c}+0.10\\-0.09\end{array}}M_\odot$ 
\cite{Cromartie2019}. The simultaneous measurement of the mass and
radius of a NS with an uncertainty of the order of 5\% is one of the objectives of the already operating
NICER mission \cite{NICER} and of some of the planned x-ray
observatories like the  Athena x-ray telescope \cite{Athena} and the
eXTP mission
\cite{eXTP}. Recently, the mass and radius of the pulsar  PSR
J0030-0451 has been determined by two different teams of NICER \cite{Miller19,Riley19},
however, still with a larger uncertainty.  It is also expected that
the number of know pulsars, and possibly also pulsars in binary
systems,  will increase enormously when the
radio-telescope  SKA \cite{SKA} will be fully operating.

The gravitational waves (GWs) emitted during the coalescence of binary 
NS systems carry important information on the high density 
properties of the EoS. 
The analysis of the compact binary inspiral event GW170817 has settled
an upper bound on the effective tidal deformability of the binary $\tilde{\Lambda}$
\cite{TheLIGOScientific:2017qsa}. 
Using a low-spin prior, which is consistent
with the observed NS population, the value $\tilde{\Lambda}\le 800$ (with 90\% confidence) was determined from the GW170817 event. 
Tighter constraints were found in a follow up reanalysis \cite{Abbott:2018wiz}, 
with $\tilde{\Lambda}=300^{+420}_{-230}$ (using the 90\% highest posterior density interval), under minimal assumptions about the nature of the compact objects.
The two NS radii for the GW170817 event were estimated in \cite{Abbott18}, 
under the hypothesis that both NS are described by the same EoS and have spins within the range observed in Galactic binary NSs, to be $R_1=11.9^{+1.4}_{-1.4}$ km (heavier star) and $R_2=11.9_{-1.4}^{+1.4}$ km (lighter star). These constraints on $R_{1,2}$ were obtained requiring that the EoS supports NS with masses larger than $1.97M_{\odot}$.
Furthermore, the tidal deformability of a $1.4M_{\odot}$ NS
was estimated to be $70<\Lambda_{1.4M_{\odot}}<580$ at the 90\% level \cite{Abbott18}.

The detection of GWs from the GW170817  event was followed by the  electromagnetic
counterparts,  the gamma-ray burst (GRB) GRB170817A \cite{grb}, and the
electromagnetic transient AT2017gfo \cite{kilo}, that set extra
constraints on the lower limit of the tidal deformability
\cite{Radice2017,Radice2018,Bauswein2019,Coughlin2018,Wang2018}. This last
constraint  seems to rule out very soft
EoS: the lower limit of  the tidal deformability of  a 1.37$M_\odot$
star set by the above studies limits the  tidal deformability
 to  $\Lambda_{1.37M_\odot} > 210$ \cite{Bauswein2019}, 300 \cite{Radice2018},
  279 \cite{Coughlin2018}, and 309 \cite{Wang2018}.\\

Without a reliable theory of dense neutron-rich matter, 
we depend on different parameterizations to describe the EoS
of NS matter. One possibility is characterizing the NS matter by
a Taylor expansion around the saturation density of symmetric nuclear matter 
\cite{Margueron2018a,Margueron2018b,Zhang2018,Margueron2019}. 
Motivated by the empirical quadratic isospin-dependent form \cite{PhysRevC.44.1892},
the EoS of homogeneous nuclear matter is normally characterized by successive derivatives, around saturation density 
and  isospin symmetric matter, of both isoscalar and isovector (symmetry energy) parts.
These derivatives are identified as the empirical parameters of nuclear matter at saturation.
Despite the great effort to determine their values, both from nuclear experiments 
and nuclear theories, most of them, mainly the higher order empirical parameters, 
are still unknown (see \cite{Margueron2018a}).
The parametrization of the EoS of nuclear matter via a Taylor expansion around 
saturation density can be thought of having a dual meaning \cite{Zhang2018}.
When used for describing terrestrial nuclear EoS or predictions of nuclear
energy density functional theories, they are Taylor expansions near saturation
density and symmetric nuclear matter, while at high densities they are
just parameterizations to be constrained by astrophysical observations.
The advantage of this kind of parametrization is that they
satisfy, by construction, all the known constraints for the nuclear matter EoS near saturation density. 
Therefore, by exploring the present theoretical/experimental uncertainty on the possible range of the empirical parameters, we will construct a dataset of possible candidates for the EoS of nuclear matter that  
are thermodynamically consistent, causal, and compatible with astrophysical observations.
Distinct EoS parametrizations, such as piecewise-polytropic 
\cite{PhysRevD.79.124032,PhysRevD.80.103003,Steiner:2010fz,Raithel:2016bux}, 
spectral representation \cite{Lindblom:2010bb},
and speed of sound \cite{Tews:2018chv,Annala:2019puf} are also used.
\\

The main objective of the present study is to understand which is the
role of the crust EoS on the information we extract  from
NS observations. 
 In \cite{Read2008}, a set of
parameterized EoS based on piecewise polytropes was built. These EoS
were 
fitted to  several well known NS EoS, including some with
a quark core. For the
low density EoS the SLy4 crust
was considered and the first politropic curve  was extended to lower
densities until the SLy4  EoS was crossed.
 This  approach was followed in other studies
\cite{Carson:2018xri,Carson:2019xxz}.
In \cite{Margueron2018b}  a cubic spline was built to match the
 SLy4 crust and core EoS between a crust  density of the order of
0.1$\rho_0$  and a core density of the order of $\rho_0$. The authors
have verified that changing the lower limit to the double or the upper
limit to half the  density would not affect  the radius of a low
(high)  mass
star in more than 100m (50m). In this
procedure it is always possible to match the crust and core EoS and
generate a valid model.  In the present study, we will consider
three different inner crust EoS, resulting from nuclear models with a
very different symmetry energy dependence on the density. For the
crust-core transition we will consider  a first order phase transition
and will apply a Maxwell construction. 
We are aware that besides the EoS also the crust-core matching
approach may have an effect on the NS properties. In this work, 
we will focus on the dependence of the NS properties on the crust EoS
taking the same matching procedure for all the EoS. A complete study
that  considers both the crust EoS and the matching procedure will be
left for the future.  However, it has been discussed that the ad-hoc
matching of the crust to the core may give rise to thermodynamic
inconsistencies as discussed in \cite{Fortin16}. \\

The paper is organized as follows. 
In Sec. \ref{sec:eos}, we introduce the EoS parametrization and the different crusts EoS used in this work.
We also detail the procedure of generating our sets composed by possible EoS of nuclear matter in $\beta$-equilibrium.
The neutron stars properties of each set are analyzed and compared in Sec. \ref{sec:eos_res}.
The impact of the crust EoS on several universal relations among NS properties are explored in Sec. \ref{sec:lambda_c}, 
while binary NS quantities are studied in Sec. \ref{sec:binary}. 
The inference of NS properties from the GW170817 event  is carried out in Sec.\ref{Sec:estimate}.
Finally, the conclusions are drawn in Sec. \ref{sec:conclusions}.

\section{EoS parametrization}
\label{sec:eos}
We start from the generic functional form for the energy per particle of homogeneous nuclear matter 
\begin{equation}
{\cal E}(x,\delta)=e_{\text{sat}}(x)+e_{\text{sym}}(x)\delta^2
\end{equation}
with 
\begin{align}
e_{\text{sat}}(x)&=E_{\text{sat}}+\frac{1}{2}K_{\text{sat}}x^2+\frac{1}{6}Q_{\text{sat}}x^3+\frac{1}{24}Z_{\text{sat}}x^4\\
e_{\text{sym}}(x)&=E_{\text{sym}}+L_{\text{sym}}x+\frac{1}{2}K_{\text{sym}}x^2+\frac{1}{6}Q_{\text{sym}}x^3,\nonumber\\
&+\frac{1}{24}Z_{\text{sym}}x^4
\end{align}
where $x=(n-n_{0})/(3n_{0})$. 
The baryon density is given by $n=n_n+n_p$ and $\delta=(n_n-n_p)/n$ is the asymmetry,
with $n_n$ and $n_p$ being the neutron and proton densities, respectively.
This approach of Taylor expanding the energy functional up to fourth order around
the saturation density, $n_{0}$, has been applied recently in several
works \cite{Margueron2018a,Margueron2018b,Margueron2019,Ferreira:2019bgy}.

The empirical parameters can be identified as the coefficients of the expansion.  
The isoscalar empirical parameters are defined as successive density derivatives of $e_{\text{sat}}$,
\begin{equation}
 P_{is}^{k}=(3n_{0})^k\left.\frac{\partial^k e_{\text{sat}}}{\partial n^k}\right|_{\{\delta=0,n=n_{0}\}},
\end{equation}
whereas the isovector parameters measure density derivatives of $e_{\text{sym}}$,
\begin{equation}
 P_{iv}^{k}=(3n_{0})^k\left.\frac{\partial^k e_{\text{sym}}}{\partial n^k}\right|_{\{\delta=0,n=n_{0}\}}.
\end{equation}
The corresponding empirical parameters are then 
\begin{equation}
\small
\{P_{is}^{0}=E_{\text{sat}},P_{is}^{2}=K_{\text{sat}}, P_{is}^{3}=Q_{\text{sat}},P_{is}^{4}=Z_{\text{sat}}\} 
\end{equation}
and
\begin{equation}
\small
 \{P_{iv}^{0}=E_{\text{sym}},P_{iv}^{1}=L_{\text{sym}}, P_{iv}^{2}=K_{\text{sym}},
P_{iv}^{3}=Q_{\text{sym}}, P_{iv}^{4}=Z_{\text{sym}}\} 
\end{equation}

The coefficients of low orders are already quite well
  constrained experimentally \cite{Youngblood1999,Margueron2012,Li2013,Lattimer2013,Stone2014,OertelRMP16}, however  $Q_{\text{sat}},\, Z_{\text{sat}}$ and $K_{\text{sym}}, \, Q_{\text{sym}},
  \, Z_{\text{sym}}$ are only poorly known
 \cite{Farine1997,De2015,Mondal2016,Margueron2018b,Malik2018,Zhang2018,Li2019}. The saturation energy $E_{\text{sat}}$ and  saturation density $n_{0}$ being rather well constrained, we fix their values throughout this work: $E_{\text{sat}}=-15.8$ MeV (the current estimated value is  $-15.8\pm0.3$ MeV \cite{Margueron2018a}), and $n_{0}=0.155$ fm$^{-3}$.
 
Each possible EoS is represented by a point in the 8-dimensional space of parameters.
We use random sampling of models through 
a multivariate Gaussian with zero covariance:
\begin{align}
\text{EoS}_i &= (E_{\text{sym}},L_{\text{sym}},K_{\text{sat}},K_{\text{sym}},Q_{\text{sat}},Q_{\text{sym}},Z_{\text{sat}},Z_{\text{sym}})_i\nonumber\\
 &\sim N(\boldsymbol{\mu},\boldsymbol{\Sigma})
\end{align}
where 
$$\boldsymbol{\mu}^T=(\overline{E}_{\text{sym}},\overline{L}_{\text{sym}},\overline{K}_{\text{sat}},\overline{K}_{\text{sym}},\overline{Q}_{\text{sat}},\overline{Q}_{\text{sym}},\overline{Z}_{\text{sat}},\overline{Z}_{\text{sym}})$$ 
\begin{align}
\boldsymbol{\Sigma}=diag(\sigma_{E_{\text{sym}}},&\sigma_{L_{\text{sym}}},\sigma_{K_{\text{sat}}},\sigma_{K_{\text{sym}}},\nonumber\\
&\sigma_{Q_{\text{sat}}},\sigma_{Q_{\text{sym}}},\sigma_{Z_{\text{sat}}},\sigma_{Z_{\text{sym}}}).
\end{align}
In the present approach, as discussed in \cite{Margueron2018b}, no a priori
correlations exist between the different parameters of the
EoS. However, the requirement that every valid EoS must satisfy a set of experimental and
observational constraints induces correlations among the parameters in the final set of EoS. 
The used parameters values and their standard deviations are in Table \ref{tab:parametros}.

\begin{table}[!htb]
	\begin{center}
		\begin{tabular}{lllllllll}
			\hline
			$P_{i}$  & $E_{\text{sym}}$ &  $L_{\text{sym}}$ & $K_{\text{sat}}$ & $K_{\text{sym}}$ & $Q_{\text{sat}}$ & $Q_{\text{sym}}$ & $Z_{\text{sat}}$ & $Z_{\text{sym}}$ \\
			\hline
			\hline
			$\overline{P}_{i}$ & $32$  & $60$ & $230$ & $-100$ & $300$ & $0$& $-500$ & $-500$ \\
			$\sqrt{\sigma_{{P}_{i}}}$  & $2$ & $15$ & $20$ & $100$ & $400$ & $400$& $1000$ & $1000$\\
			\hline
		\end{tabular}
	\end{center}
	\caption{The mean $\overline{P}_{i}$ and standard deviation
		$\sqrt{\sigma_{{P}_{i}}}$ of the multivariate Gaussian, where
		$\sigma_{{P}_{i}}$ is the variance of the parameter $P_{i}$. 
		Our EoSs are sampled using the initial distribution for 
		$P_i$ assuming that there are no correlations
		among the parameters. All the quantities are in units of MeV.
		These values are from \cite{Margueron2018b}.}
	\label{tab:parametros}
\end{table}

We impose the following conditions to get a valid EoS: 
i) the pressure is an increasing function of density (thermodynamic stability); 
ii)  the speed of sound is smaller than the speed of light (causality);
iii) the EoS supports a maximum mass at least as high as $1.97M_{\odot}$
\cite{Arzoumanian2017,Fonseca2017,Demorest2010,Antoniadis2013}
(observational constraint); 
and iv) the symmetry energy $e_{\text{sym}}(n)$ is positive. 
This may be a too restrictive constraint and a more
  realistic would be that the symmetry energy $e_{\text{sym}}(n)$
is positive  for densities below the central density of the maximum
mass star configuration. We consider, however, that the difference
between both sets of EoS  will not  be significant. 
All EoS describe $npe\mu$ matter in $\beta$-equilibrium. 

\subsection{Adding a crust to the generated EoSs}
\begin{figure}[!htb]
	\centering
	\includegraphics[width=0.8\columnwidth]{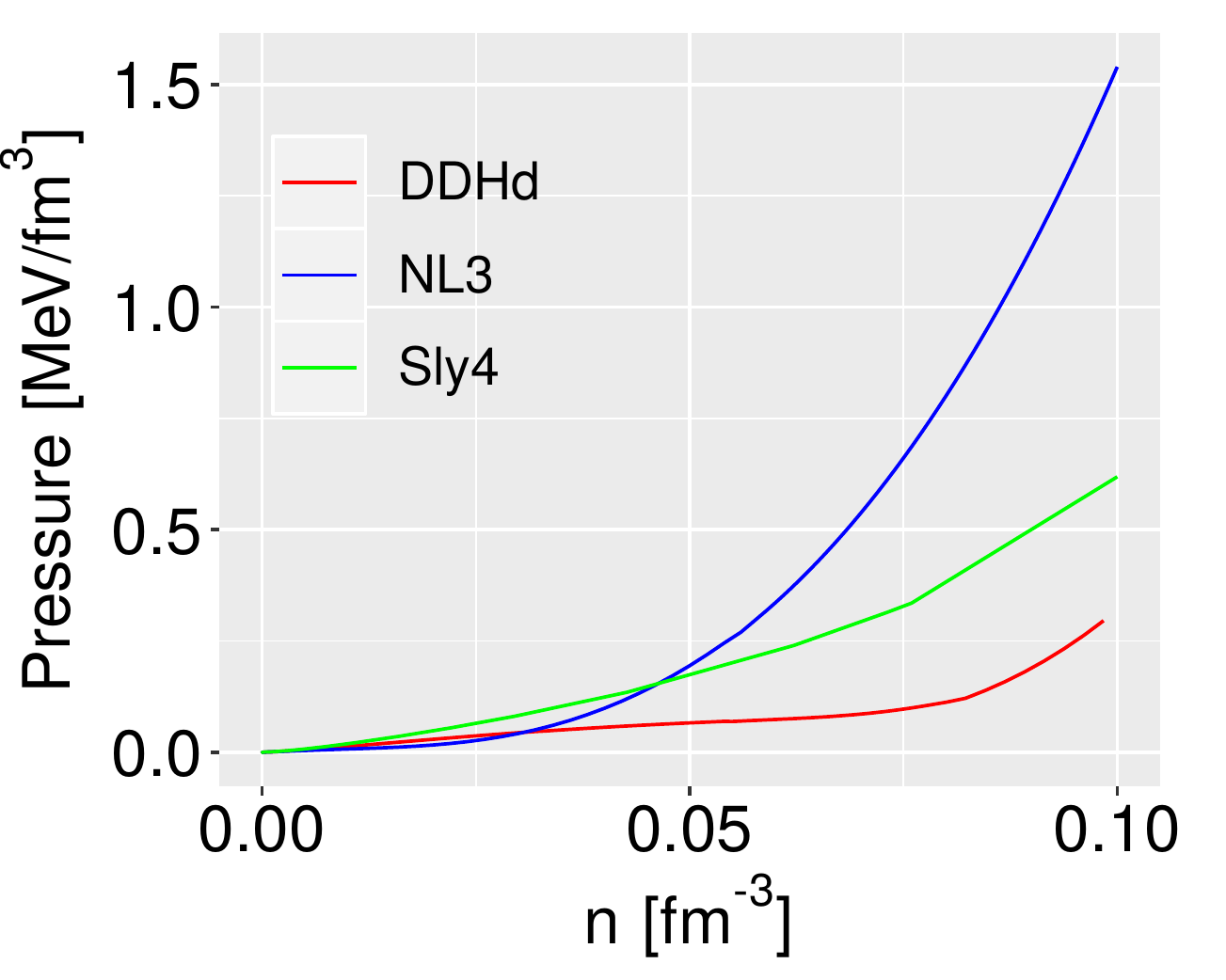}
	\caption{Pressure as a function of density for the DDHd (red), NL3 (blue), and SLy4 (green) crusts.}
\label{fig:crusts_EoS}
\end{figure}

To test the dependence of our results on the crust, we
have built three different sets of NS EoS taking three different crust
EoS: the SLy4 \cite{Douchin2001} and two other EoS obtained from a
Thomas Fermi calculation of the inner crust \cite{Avancini08,Grill12,Grill14} taking as underlying 
models the relativistic mean field (RMF) models NL3 \cite{nl3},
and DDHd \cite{ddhd}, respectively with a stiff and a soft symmetry
energy EoS above saturation density. It has been discussed in \cite{Fortin16,Pais16} that the
inner crust is quite sensitive to the symmetry energy density
dependence. Taking the three inner crust EoS, two of them having
extreme  behaviors of the symmetry energy, will allow us to discuss
how  sensitive are the results to the crust EoS.  In
Fig. \ref{fig:crusts_EoS}, the pressure is plotted as a
function of the baryonic density for these three crusts, and it is
clearly seen their different behavior.

It has been long discussed in the literature whether the
crust-core transition is a first order phase transition or a crossover, see for
instance \cite{Raduta:2010ym,Pais2014} and references therein. Since we  are
considering a two model approach to determine the complete EoS, one for the crust and the other for the core, 
a thermodynamic consistent description of the crust-core transition is to consider the Maxwell
construction for  a first order phase transition. 
We impose that a valid core EoS must cross one of the above crust
EoS in the $P(\mu)$ plane below $n<0.10$ fm$^{-3}$, consistently with the range of core-crust transition densities for a large set of nuclear models \cite{Ducoin2011}.  
The crusts are matched with the generated EoSs by requiring $P_{\text{crust}}(\mu)=P_{\text{core}}(\mu)$, where $\mu$ is the
baryonic chemical potential. All EoS that do not cross  the crust EoS at $n<0.10$ fm$^{-3}$ are discarded.
Our approach is more restrictive than the one proposed in
\cite{Margueron2018b}, where a cubic spline was chosen as a matching procedure
to link the crust and core EoS. 

\section{The EoS set}
\label{sec:eos_res}
In this section, we discuss the properties of the EoS sets used in the present study, 
including the match of the crust EoS to the core EoS.  
The  properties  of the NSs  built from these sets of EoS and their possible dependence on the examined crusts are summarized.

\subsection{EoS dataset}
\begin{figure}[!htb]
	\centering
	\includegraphics[width=1.0\columnwidth]{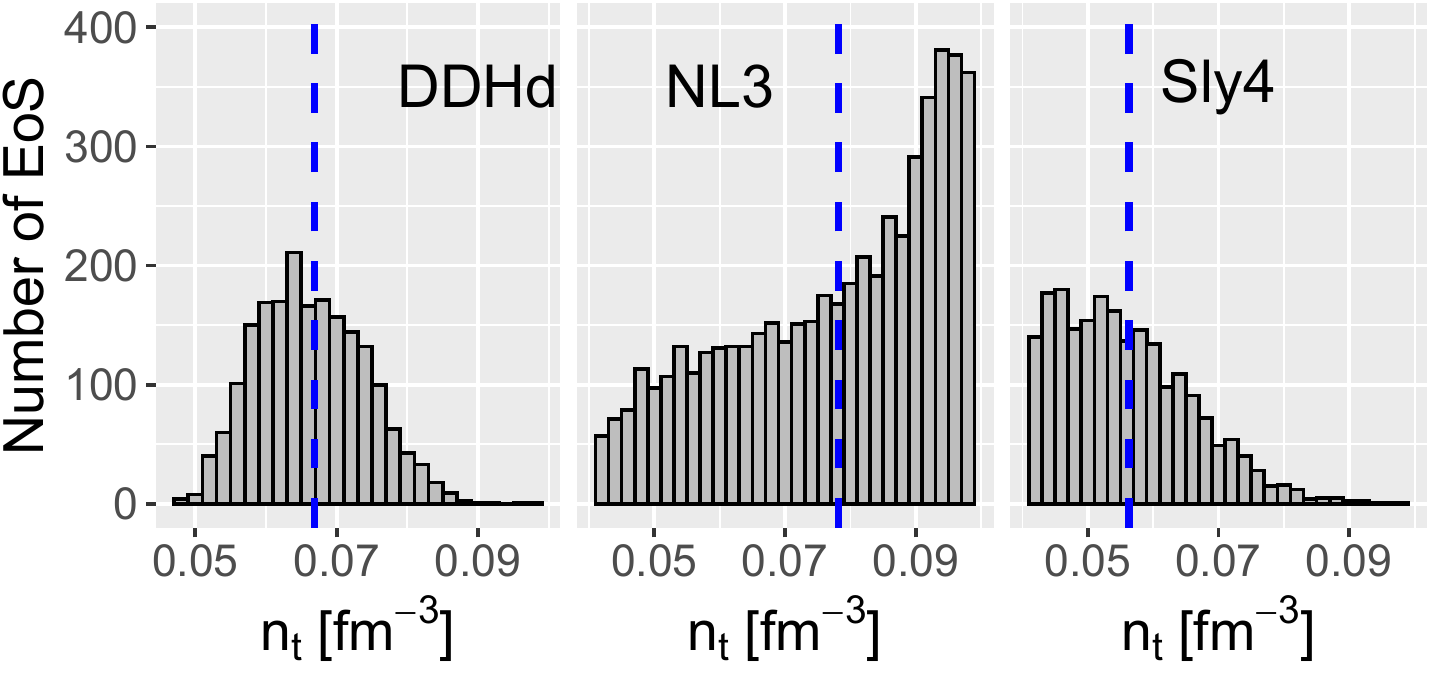}
	\caption{Histograms for the transition density between the
          generated EoSs and the crusts:  DDHd
          (left), NL3 (middle) and SLy4 (right).
           The mean values are represented by blue dashed lines.}
\label{fig:crusts}
\end{figure}

After matching the crust EoS and applying all the conditions at the end of
Sec. \ref{sec:eos}, we ended up with 1956 (DDHd crust EoS),
5167 (NL3  crust EoS), 2158 (SLy4  crust EoS) valid models. 
In Fig. \ref{fig:crusts} the histograms of the crust-core matching
densities are shown for the three crusts considered. 
The requirement that the matching has to occur for densities below 0.1 fm$^{-3}$
has a clear effect only for the EoSs built with the NL3 crust. 
This is due to the fact
that being a quite hard crust EoS it is easier that the crust-core
matching is successful. The SLy4 EoS shows the lowest mean value for the transition density (blue dashed lines),
followed by the DDHd crust, and the NL3 shows the highest
value, around $0.08$ fm$^{-3}$.

The mean values and standard deviations for the empirical parameters of our final sets are shown in Table
\ref{tab:stat}. The mean values of $K_{\text{sat}}$, $E_{\text{sym}}$, and $L_{\text{sym}}$
slightly change when compared to their initial values, indicating that the
conditions applied did not require their values to change considerably.
 As referred in \cite{Margueron2018b}, this is probably due to the fact that these
parameters are already well constrained and most of the conditions
applied probe high densities.   It should be noticed, however, that the
DDHd crust affects quite strongly the $E_{\text{sym}}$
distribution. The effect of the crust on the probability
distributions of the EoS parameters is only reflected on the lower
order isovector parameters. The third and fourth  order parameters
$Q_i$ and $Z_i$ are totally insensitive to the crust. This behavior was
expectable since these parameters control the high density part of the
EoS and the crust slightly affects the high density region of a NS.

\begin{table}[!htb]
  \begin{center}
\begin{tabular}{ccccccc}
  \hline
  & \multicolumn{2}{c}{NL3} & \multicolumn{2}{c}{DDHd}& \multicolumn{2}{c}{SLy4} \\
  $x_i$& $\bar{x}$ & $\sigma_x$ & $\bar{x}$ & $\sigma_x$ & $\bar{x}$ & $\sigma_x$ \\
 \hline
  \hline  
  $E_{\text{sym}}$ & 32.32 & 1.78 & 28.31 & 1.00 & 33.32 & 1.89 \\                 
  $L_{\text{sym}}$ & 59.62 & 10.38 & 69.59 & 13.85 & 51.56 & 11.83 \\              
  $K_{\text{sat}}$ & 230.15 & 20.08 & 229.54 & 21.26 & 233.95 & 18.75 \\   
  $K_{\text{sym}}$ & -61.31 & 70.38 & -88.54 & 80.06 & -43.96 & 63.02 \\  
  $Q_{\text{sat}}$ & 75.76 & 127.64 & 78.54 & 132.28 & 58.62 & 123.33 \\              
  $Q_{\text{sym}}$ & 259.68 & 310.29 & 302.01 & 318.98 & 238.21 & 300.33 \\   
    $Z_{\text{sat}}$ & -199.75 & 145.98 & -201.70 & 147.98 & -181.97 & 143.04 \\
  $Z_{\text{sym}}$ & 348.50 & 676.47 & 344.80 & 719.04 & 371.91 & 698.56 \\        
  \hline
    \end{tabular}
    \caption{The mean $\bar{x}=(1/N)\sum_i x_i$ and standard
      deviation $\sigma_x=\sqrt{1/(N-1)\sum_i(x_i-\bar{x})^2}$ of the
      empirical parameters.  Results obtained
          with the three inner
          crusts discussed in the text, DDHd, NL3 and SLy4, are shown. All the quantities are in units of MeV.}
\label{tab:stat}
  \end{center}
\end{table}

\subsection{Neutron stars properties}

Using our three sets of EoS, we determine the $M(R)$ relations by 
integrating the Tolman-Oppenheimer-Volkoff (TOV) equations  \cite{TOV1,TOV2},
and the dimensionless tidal deformabilities 
 $\Lambda$
\begin{equation}
\label{eq2}
\Lambda = \frac{2}{3}k_2C^{-5},
\end{equation}
where $k_2$ is the quadrupole tidal Love number (determined following Ref.~\cite{Hinderer2008}) and $C=GM/(c^2R)$ is the star's compacteness.

\begin{figure}[!htb]
	\centering
	\includegraphics[width=1.0\columnwidth]{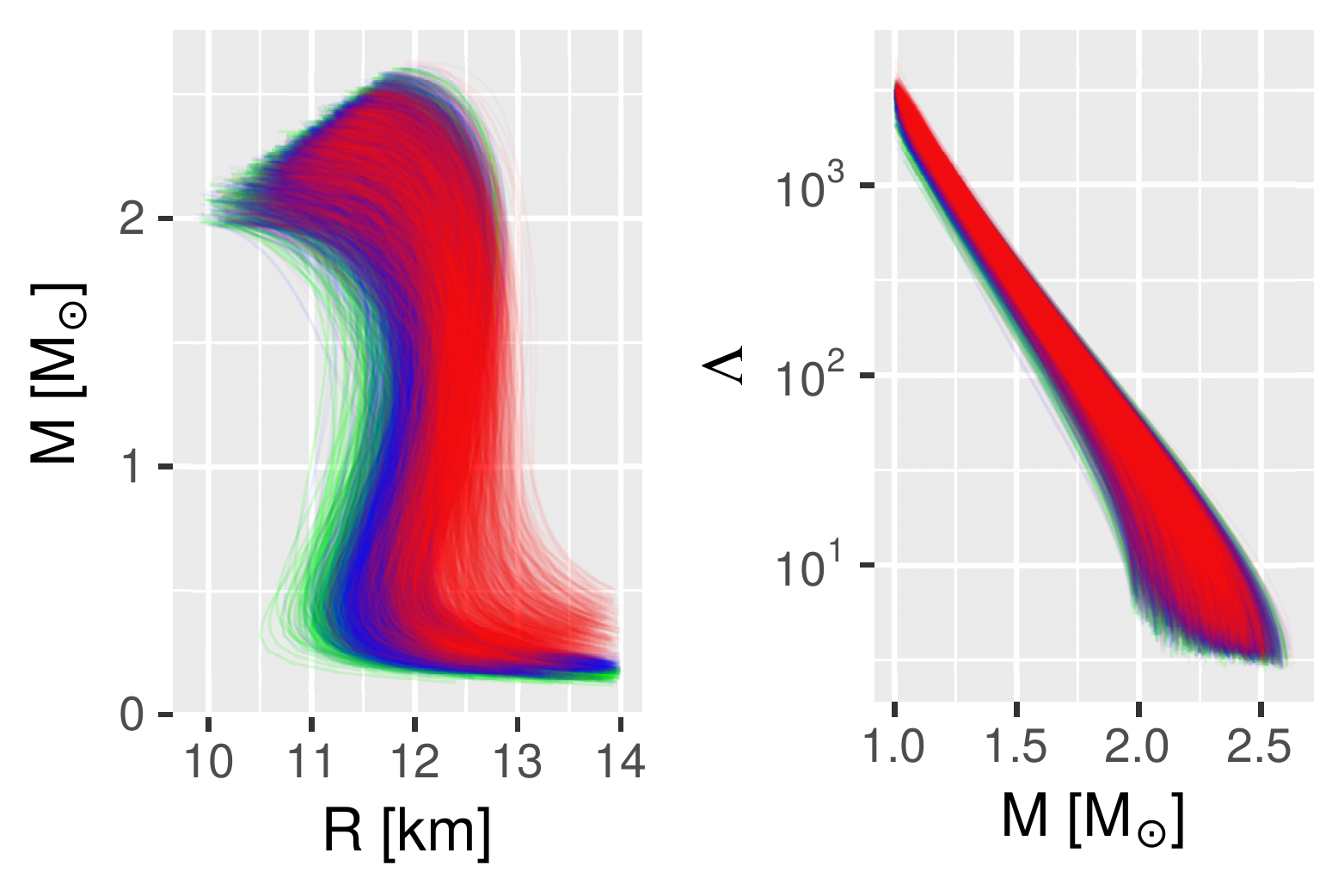}
    \caption{Mass-Radius (left) and $\Lambda$-mass (right) diagrams
      for all valid models with the three different crusts: DDHd
      (red), NL3 (blue), and SLy4 (green).}
\label{lambda}
\end{figure}

Figure \ref{lambda} shows the $M(R)$ and $\Lambda(M)$ diagrams
for all models (the color distinguishes the crust used). 
A summary of the mean values and the standard
deviations for $\Lambda_{M_i}$ and $R_{M_i}$ is in Table \ref{tab:lambda2}.
The NL3 and SLy4 crusts show similar
results for the mean values and standard deviations, however the NL3 crust
shows more extreme values.  
Matching the generated EoSs with the DDHd crust results in
larger and  more scattered radii, mainly for low NSs masses.  

\begin{table}[!ht]
  \begin{center}
\begin{tabular}{c|cc|cc|cc}  
  \hline
  & \multicolumn{2}{c}{NL3} & \multicolumn{2}{c}{DDHd}& \multicolumn{2}{c}{SLy4}\\
 $x_i$& $\bar{x}$ & $\sigma_x$ & $\bar{x}$ & $\sigma_x$  & $\bar{x}$ & $\sigma_x$ \\
 \hline
  \hline
    $\Lambda_{1.0M_{\odot}}$  & 3073.15 & 285.07 & 3323.69 & 347.94 & 2978.05 & 292.00 \\    
    $\Lambda_{1.2M_{\odot}}$  & 1159.03 & 121.49 & 1229.02 & 136.18 & 1133.65 & 122.79 \\       
    $\Lambda_{1.4M_{\odot}}$  & 476.89 & 59.81 & 498.89 & 64.55 & 469.76 & 59.84 \\                
    $\Lambda_{1.6M_{\odot}}$  & 205.28 & 32.50 & 212.68 & 34.78 & 203.12 & 32.32 \\                   
    $\Lambda_{1.7M_{\odot}}$  & 88.90 & 19.02 & 91.53 & 20.21 & 88.18 & 18.9\\                      
    $R_{1.0M_{\odot}}$        & 12.06 & 0.18 & 12.38 & 0.28 & 11.97 & 0.19 \\                            
    $R_{1.2M_{\odot}}$        & 12.17 & 0.19 & 12.43 & 0.25 & 12.10 & 0.20\\                            
    $R_{1.4M_{\odot}}$        & 12.24 & 0.21 & 12.46 & 0.25 & 12.18 & 0.22 \\                            
    $R_{1.6M_{\odot}}$        & 12.26 & 0.25 & 12.43 & 0.28 & 12.21 & 0.25\\                            
    $R_{1.8M_{\odot}}$        &  12.19 & 0.32 & 12.33 & 0.34& 12.15 & 0.32 \\                           
  \hline
\end{tabular}

\caption{Mean and standard deviation  values for $\Lambda_{M_i}$ and $R_{M_i}$
  (km) for the three inner crusts discussed in the text, NL3, DDHd, and SLy4.}
\label{tab:lambda2}
  \end{center}
\end{table}

From the reanalysis of GW170817 data, assuming the same EoS for the two NSs and for a spin range consistent with the one observed in Galactic binary NSs,  
the tidal deformability of a $1.4M_{\odot}$ NS
was estimated to be $70<\Lambda_{1.4M_{\odot}}<580$ at the 90\% level \cite{Abbott18}. 
Almost all EoS in our set are within this interval.    
 Our set of models does not contain models with
 $\Lambda_{1.4M\odot} <200$. However, this is not a drawback since the GRB 170817A \cite{grb}, and the
electromagnetic transient AT2017gfo \cite{kilo} detected immediately
after the GW170817 detection  set an extra
constraint on the lower limit of the tidal deformability
\cite{Radice2017,Radice2019,Bauswein2019,Coughlin2018,Wang2018} ruling out very soft
EoS. In particular, the lower limit of  the tidal deformability of  a 1.37$M_\odot$
star imposes  $\Lambda_{1.37M_\odot} > 210$ \cite{Bauswein2019}, 300 \cite{Radice2018},  279 \cite{Coughlin2018}, and 309 \cite{Wang2018}. 


\section{Universal relations: neutron star properties}
\label{sec:lambda_c}

In the present and following sections,  we investigate several
relations between NS properties using the set of  models
defined in the previous sections. \\

\subsection{Love number $k_2$}

The tidal Love number $k_2$ is restricted to a narrow range of values,
$0.05\lessapprox k_2\lessapprox0.15$, when considering several
hadronic EoS and star masses in the range $1.0<M/M_\odot<1.8$. 
This fact favors the existence of an approximately universal relation 
between the tidal deformability $\Lambda$ and the
compactness $C$ which was first proposed in \cite{Maselli2013}. The
authors of \cite{Yagi2016} have discussed this relation and compared
with other universal relations relating global properties of NS as the
I-Love-Q  and I-Love relations \cite{Yagi2013,Chan2016,Yagi2016} and
have shown that  the 
EoS dependence of the  compactness $C$  relation  with  any of the quantities
I-Love-Q is much larger than  the variation in the I–Love–Q relations.\\

Considering a  set of  three EoS  with quite
different density dependence, it was found that
$C=a_0+a_1\ln\Lambda+a_2(\ln\Lambda)^2$ \cite{Maselli2013}. A fit
using a wider set of EoS was performed in \cite{Yagi2016} and the
maximum deviation obtained was 6.5\%. Still,  in \cite{De:2018uhw} the
tidal deformability was related to the
compactness through $\Lambda= a\, C^{-6}$, where $a=0.0093\pm0.0007$
for stars with a mass in the range  $1.1\le M/M_\odot\le 1.6$.  The
 extra $C^{-1}$  dependence was shown to come from the tidal Love number
$k_2$. 
 The same authors have also verified that $\Lambda$ varies with
$M^{-6}$.  In their study,  the crust was described by the BPS
\cite{bps} and the Negele \& Vautherin \cite{nveos} EoS below a
transition density of the order of one fourth of saturation density. 
In the following,  we investigate the effect of the
crust on the relation between the tidal deformability and compactness
taking our set of EoS.\\

\begin{figure}[!htb]
	\centering
	\includegraphics[width=1.0\columnwidth]{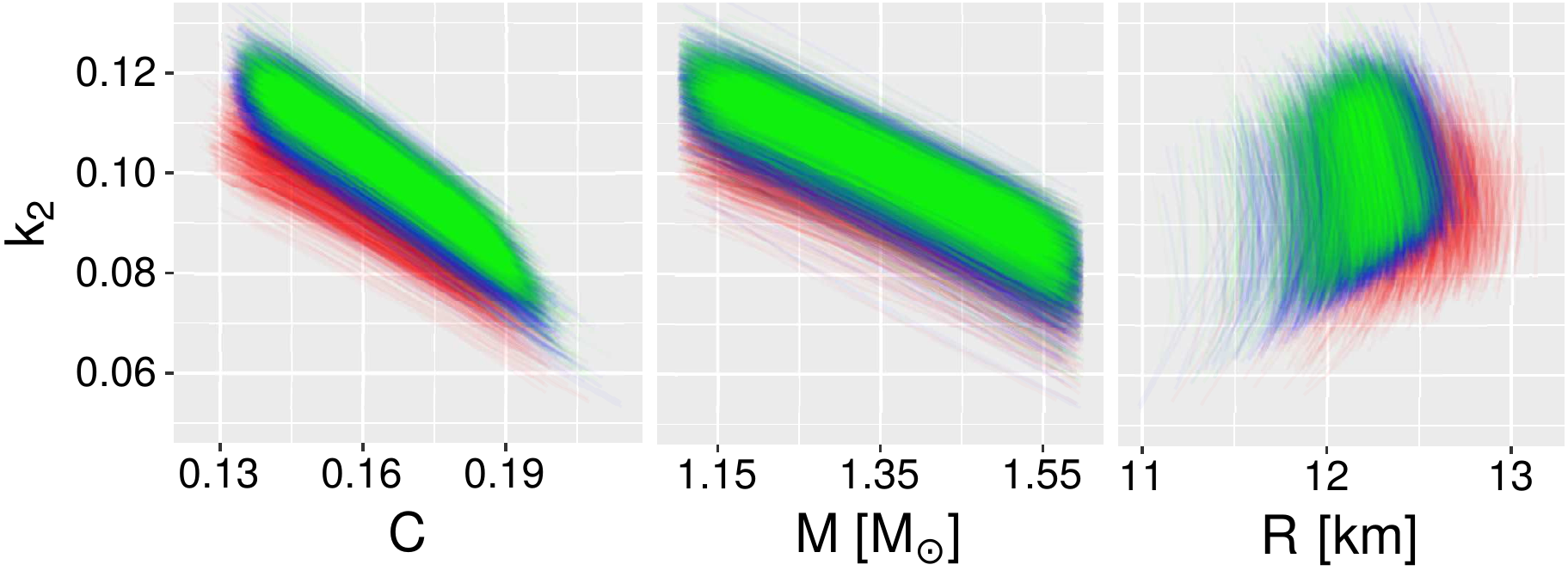}
    \caption{$k_2$ versus $C$ (left), $M$ (middle), and $R$ (right)
      for $1.1\leq M/M_{\odot}\leq1.6$ and  for each crust:
      DDHd (green), SLy4 (blue), and NL3 (red).}%
\label{k2-c}
\end{figure}

As $k_2$ depends on both $C$ and the EoS, a possible relation between
the measurable quantity $\Lambda$ and $k_2$ may give us insights into
the EoS properties. 
In Fig.~\ref{k2-c}, $k_2$ is plotted versus $C$ (left), $M$ (middle), and $R$ (right) for $1.1\leq M/M_{\odot}\leq1.6$, the range of masses covered by the GW170817 event. 
Within the range of masses $1.0< M/M_\odot<1.8$, 
$k_2$ reduces to half, $k_2(1.8M_{\odot})/k_2(1.0M_{\odot})\approx0.5$,
while $C$ doubles, $C(1.8M_{\odot})/C(1.0M_{\odot})\approx2$. 
All crusts give similar mean values and
deviations for both $C$ and $k_2$. 
The largest difference between 
the results obtained for the three crusts lies
on the dispersion of the values.
It is observed a
clear linear relation between  $k_2$ and $C$ or $M$. 
On the other hand, $k_2$ seems to be insensitive to $R$. Let us analyze the
the two first relations. For the compactness $C$, we have performed the linear regression $\ln(k_2)=\beta\ln(C)+\alpha$ for each crust EoS. 
The regression results are
\begin{align}
[\beta&=-1.122, \,\, \alpha=-4.358,\,\, \text{corr}=0.933] \,\, \text{NL3}\\
[\beta&=-1.025, \,\, \alpha=-4.244,\,\, \text{corr}=0.860] \,\, \text{DDHd}\\
[\beta&=-1.137, \,\, \alpha=-4.367,\,\, \text{corr}=0.941] \,\, \text{SLy4},
\end{align}
where corr stands  for correlation.
The correlations are calculated via the Pearson coefficient $ \text{Corr}\,[x,y]=E[(x-\mu _{x})(y-\mu _{y})]/(\sigma _{x}\sigma _{y})$, where $y=\ln(k_2)$ and $x=\ln(C)$, for the above case.
The results indicate that  $k_2\sim C^{-1}$ with a correlation
coefficient that is above $90\%$ for the  NL3 and the  SLy4 crusts. For
DDHd the correlation is slightly weaker. 
The same analysis was repeated for $k_2(M)$ taking the linear regression $\ln(k_2)=\beta\ln(M)+\alpha$. 
We get
\begin{align}
[\beta&=-1.065, \,\, \alpha=-2.006,\,\, \text{corr}=0.892] \,\, \text{NL3}\\
[\beta&=-1.028, \,\, \alpha=-2.061,\,\, \text{corr}=0.843] \,\, \text{DDHd}\\
[\beta&=-1.069, \,\, \alpha=-1.992,\,\, \text{corr}=0.898] \,\, \text{SLy4}.
\end{align}
These results indicate that $k_2\sim M^{-1}$ with a smaller correlation
coefficient than the one obtained above,  but still as high as $88\%$-$90\%$, 
and an exponent deviation from $-1$ of the order of 5\%.\\


\subsection{Tidal deformability $\Lambda$}

\begin{figure}[!htb]
	\centering
	\includegraphics[width=1.0\columnwidth]{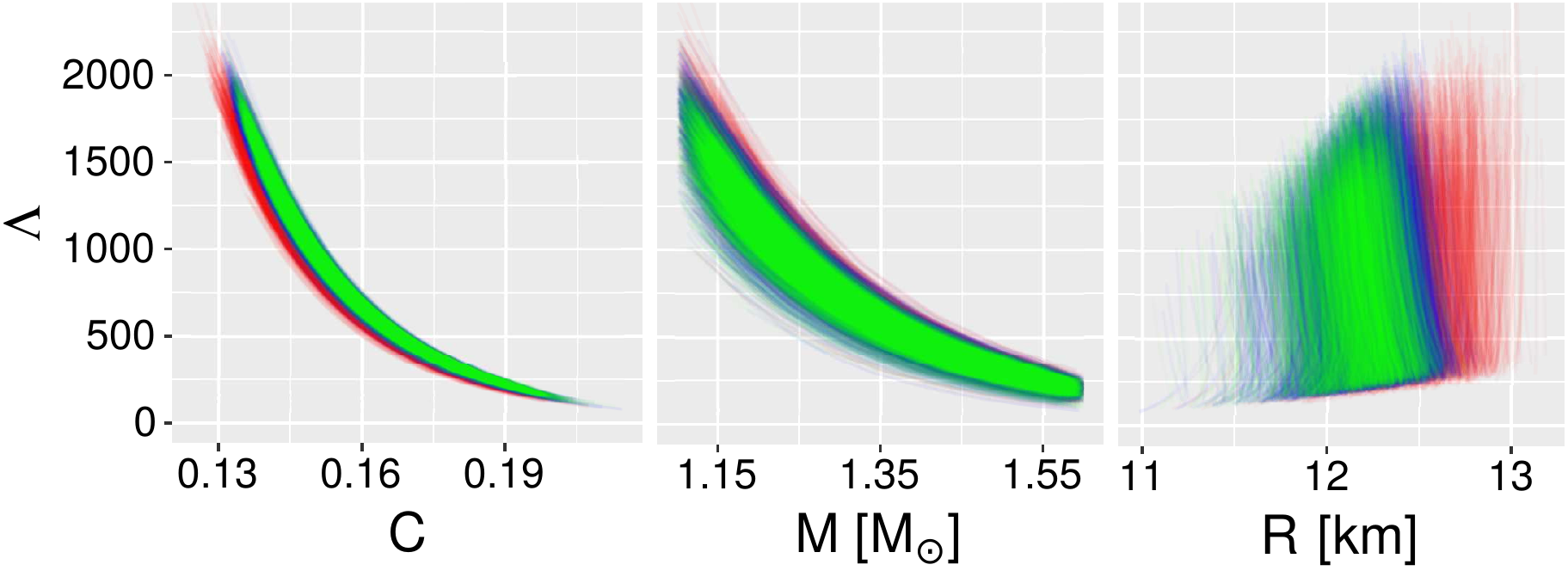}
    \caption{$\Lambda$ versus $C$  (left),   $M$ (right), and $R$
      (right) diagrams for $1.1\leq M/M_{\odot}\leq1.6$ for each
      crust: DDHd (green), SLy4 (blue), and NL3 (red).}%
\label{lambda-c}
\end{figure}

We next analyze the tidal deformability dependences. Figure \ref{lambda-c}
shows the diagrams $\Lambda(C)$  (left), $\Lambda(M)$ (middle) and $\Lambda(R)$
(right), for $1.1\leq M/M_{\odot}\leq1.6$. Similarly to $k_2$, the crusts do not have a strong impact on the 
$\Lambda$ dependences. 
Performing a similar regression analysis as above but now for $\Lambda(C)$, i.e., taking
$\ln(\Lambda)=\beta\ln(C)+\alpha$, we get 
\begin{align}
[\beta&=-6.122, \,\, \alpha=-4.763,\,\, \text{corr}=0.998] \,\, \text{NL3}\\
[\beta&=-6.025, \,\, \alpha=-4.649,\,\, \text{corr}=0.995] \,\, \text{DDHd}\\
[\beta&=-6.137, \,\, \alpha=-4.772,\,\, \text{corr}=0.998] \,\, \text{SLy4}.
\end{align}
The above results indicate that $\Lambda\sim C^{-6}$ with a high quality
fit, similarly to \cite{De:2018uhw}, and a small dependence on the crust. 
The deviation of the exponent  from $-6$ is below 2\%.
This is expected because $\Lambda(C)\sim k_2(C)C^{-5}$ and
we saw that $k_2\sim C^{-1}$ with the  exponent equal to -1 within a
deviation below 13\%. The values obtained for $\alpha$ are in agreement with
$\alpha=-4.6777^{0.078}_{-0.073}$ found in \cite{De:2018uhw}. \\

Let us now analyze the $\Lambda(M)$ dependence through the linear regression $\ln(\Lambda)=\beta\ln(M)+\alpha$. The results are
\begin{align}
[\beta&=-5.962, \,\, \alpha=8.113,\,\, \text{corr}=0.978] \,\, \text{NL3}\\
[\beta&=-6.055, \,\, \alpha=8.188,\,\, \text{corr}=0.977] \,\, \text{DDHd}\\
[\beta&=-5.922, \,\, \alpha=8.088,\,\, \text{corr}=0.977] \,\, \text{SLy4}
\end{align}
The results indicate that the relation obtained in \cite{De:2018uhw},
$\Lambda\sim M^{-6}$, is verified with an exponent slightly different
from $-6$,
the difference not exceeding 4\%, and a small dependence on the crust.\\                                                                                              

On the right panel of Fig. \ref{k2-c}
and \ref{lambda-c} , we also show that the dependence of $k_2$ and
$\Lambda$ on the star radius.                                                                                                                  
Each EoS appears as a vertical line in both  diagrams, 
because each EoS predicts an almost constant $R$ for $1.1\leq
M/M_{\odot}\leq1.6$. 
This is already seen in the left panel of  Fig. \ref{lambda}.
We can easily determine the statistics for $\Delta R=R_{1.6M_{\odot}}-R_{1.1M_{\odot}}$
The results are $\overline{\Delta R}\pm\sigma=0.021\pm 0.185$ km
(DDHd), $0.135\pm 0.143$ km (NL3), and $0.170\pm 0.140$ km (SLy4).
Our set of EoS predicts a larger variation of the
radius for the NL3 and SLy4 crusts than the study presented in 
\cite{De:2018uhw},  where $\Delta R=0.070$ km and
$\sigma=0.11$ km  were obtained with a set of piece-wise polytrope
EoSs.  
The model with the crust DDHd is the only  one 
compatible with $\Delta
R\approx 0$, while the other two models predict $R_{1.6 M_\odot}>R_{1.4 M_\odot}$ on average.\\

However, we normally know the mass of a NS, $M_i$, and we want to infer the $\Lambda_{M_i}$ from $R_{M_i}$ or the other way around. Fixing the NS mass, we have the relation $\Lambda\sim k_2(R)R^5$, and the dependence $k_2(R)$ should be different from $k_2(C)$.    
Let us analyze the following empirical relation between the tidal deformability and the radius of a  NS with mass $M_i$,
$$
\Lambda_{M_i} =  a R_{M_i}^{\beta}.
$$
In \cite{Annala18}, the relation was found to be valid 
 with $\beta=7.5$ (to a rather good accuracy) for $M_i=1.4M_{\odot}$. 
 The authors have
 explored a sample of 250000 EoS constructed by polytropes segments
 that interpolate between the EoS obtained from an effective chiral
 field theory calculation  at low density
and perturbative QCD results at very high densities. In
 \cite{Malik:2018zcf}, using a set of relativistic and
 non-relativistic mean-field models and unified EoS, a value of $\beta=6.13$ was
 obtained but the exponent $\beta$ was seen to be mass dependent: for
 $M=1.17$ and $1.6\,M_\odot$ the values, respectively,  $\beta=5.84$
 and 6.58  were determined.  In \cite{Fattoyev:2017jql}, $\beta=5.28$
 was found for $1.4M_{\odot}$ using several energy density functionals
 within relativistic mean field (RMF) theory.
This relation shows, therefore, some sensitivity to the star
mass and to the set of EoS used.
The dependence on the set of EoS is probably due to the constraints
that have been imposed to build the sets.

In the following, starting from the set of EoS developed for the present
study,  we explore the same relation and its validity for different
NS masses and analyze the impact of the crust. 
Table \ref{tab:lambda-r}
contains the results of the linear regression
$\ln(\Lambda_{M_i})=\beta\ln(R_{M_i})+\alpha$ performed for each mass value and each crust and the
respective correlation coefficient. As in \cite{Malik:2018zcf}, we
see that the exponent $\beta$ is mass dependent but with a stronger
variability. Furthermore, its value depends on the crust used, being
its impact stronger for lighter NSs masses. For the
canonical NS mass, both NL3 and SLy4 crusts show very close
values, $7.25$ and $7.21$ respectively, similar to the  one obtained
in  \cite{Annala18},  but the DDHd crust predicts a smaller
value, $5.81$. 

For massive NSs, the relation $\Lambda_{M_i}
= a R_{M_i}^{\beta}$ becomes almost exact with a small crust dependence:  
for $M\geq1.8M_{\odot}$ NSs a correlation of
$0.99$ for NL3 and  SLy4 and $0.97$ for DDHd was obtained. 
This almost  universal relation shows that the radii of massive neutron stars 
could be precisely determined from their tidal deformability values.
To quantify the uncertainty on $R_{M_i}$, one can determine the 
Residual Standard Error (RSE) of the linear regression $R_{M_i}(\Lambda_{M_i})$
that quantifies the dispersion of $R_{M_i}$ around the regression line.
We get a $\text{RSE}(R_{M_i})$ of $0.08$ km and $0.05$ km for 
$M_i=1.0M_{\odot}$ and $M_i=1.9M_{\odot}$, respectively, using the NL3 crust.

\begin{table}[!htb]
\centering
\begin{tabular}{c|ccc|ccc|ccc} 
  \hline
  &  \multicolumn{3}{c}{$\alpha$} & \multicolumn{3}{c}{$\beta$} & \multicolumn{3}{c}{Correlation} \\ 
    \hline
 $M_i$ & NL3 & SLy4 & DDHd & NL3 & SLy4 & DDHd & NL3 & SLy4 & DDHd \\ 
  \hline
  $1.0$  & -6.14 & -6.90 & -2.02 & 5.69 & 6.00 & 4.02 & 0.89 & 0.93 & 0.84 \\    
  $1.2$  & -9.23 & -9.35 & -4.98 & 6.51 & 6.57 & 4.80 & 0.93 & 0.95 & 0.84 \\    
  $1.4$  & -12.19 & -11.97 & -8.89 & 7.33 & 7.25 & 5.98 & 0.95 & 0.96 & 0.88 \\  
  $1.6$  & -14.94 & -14.65 & -13.08 & 8.08 & 7.97 & 7.31 & 0.97 & 0.98 & 0.93 \\ 
  $1.8$  & -17.56 & -17.40 & -16.74 & 8.81 & 8.75 & 8.45 & 0.99 & 0.99 & 0.97 \\ 
  $1.9$  & -18.90 & -18.82 & -18.38 & 9.19 & 9.16 & 8.95 & 0.99 & 0.99 & 0.98 \\ 
   \hline
\end{tabular}
\caption{Results for the linear regression $\ln(\Lambda_{M_i})=\beta\ln(R_{M_i})+\alpha$ and respective
  correlation coefficient for the three crust EoS. $M_i$ are in units of $M_{\odot}$.}
\label{tab:lambda-r}
\end{table}

We conclude that not only the relation
$\Lambda\sim R^{\beta}$ depends on the NS mass but also on the crust. 
The correlation becomes stronger as the NS mass increases
and for $1.6M_{\odot}$ the relation $\Lambda\sim R^{\beta}$ is
almost an exact power law.  In a sense, this is telling us that $k_2$
only behaves as a power law for massive NSs. Clearly, the $k_2$ has a
nontrivial dependence on $R$ that changes with the NS mass being
considered. 
In fact, for a fixed
mass, $k_2$ is by no means a power law in $R$. 
This becomes the case, only in an approximate way, for $M\ge 1.6M_{\odot}$. 
The regression analysis results using the relation $\ln(k_2)=\beta\ln(R_{M})+\alpha$ are show in Table \ref{tab:k2-r}.
It is striking that for $M\geq 1.8 M_{\odot}$ the correlation is $0.99$ for the crusts NL3 and SLy4.
In \cite{Malik:2018zcf}, the authors have also shown that with a set of 33 RMF 
and Skyrme EoS there was a reasonable correlation between $k_2$ and $R$ for a $1.4M_{\odot}$ star. 
With our set, we get a weaker correlation for a $1.4M_{\odot}$ star but we 
show that the larger the NS mass the stronger the correlation.

\begin{table}[!htb]
\centering
\begin{tabular}{c|ccc|ccc|ccc} 
  \hline
  &  \multicolumn{3}{c}{$\alpha$} & \multicolumn{3}{c}{$\beta$} & \multicolumn{3}{c}{Correlation} \\ 
    \hline
 $M_i$ & NL3 & SLy4 & DDHd & NL3 & SLy4 & DDHd & NL3 & SLy4 & DDHd \\ 
  \hline
    $1.0$  & -3.60 & -4.39 & 0.31 & 0.60 & 0.93 & -0.98 & 0.22 & 0.39 & -0.37 \\         
    $1.2$  & -5.72 & -5.94 & -1.64 & 1.41 & 1.51 & -0.24 & 0.50 & 0.60 & -0.08 \\        
    $1.4$  & -7.87 & -7.76 & -4.69 & 2.21 & 2.17 & 0.91 & 0.72 & 0.76 & 0.28 \\          
    $1.6$  & -10.00 & -9.81 & -8.18 & 2.99 & 2.92 & 2.23 & 0.86 & 0.87 & 0.62 \\         
    $1.8$  & -12.11 & -12.00 & -11.30 & 3.74 & 3.71 & 3.39 & 0.94 & 0.94 & 0.85 \\        
    $1.9$  & -13.22 & -13.18 & -12.71 & 4.14 & 4.13 & 3.91 & 0.96 & 0.96 & 0.91 \\        
   \hline
\end{tabular}
\caption{Parameters of the linear regression, $\ln(k_2)=\beta\ln(R_{M})+\alpha$, and respective
  correlation coefficient. $M_i$ are in units of $M_{\odot}$.}
\label{tab:k2-r}
\end{table}

To conclude this section, we summarize the main results: a) we confirm
the results presented in \cite{De:2018uhw} concerning the relations
$\Lambda(C)$ and $\Lambda(M)$, in particular, that
$\Lambda\sim C^{-6}$  and $\Lambda\sim M^{-6}$  is obtained  with a high quality fit, a deviation  of
the exponent from $-6$, respectively of $\approx 2\%$ and $\approx
4\%$, and  a small dependence on the crust; b)
for a fixed NS mass $M_i$,   the relation
$\Lambda\sim R^{\beta}$ depends on the crust and  on the NS
mass.  The correlation becomes stronger as the NS mass increases
 and for $M>1.6M_{\odot}$ the relation $\Lambda\sim R^{\beta}$ is
 almost an exact power law with $\beta \sim 8-9$, and crust indepedent.


\section{Binary neutron stars}
\label{sec:binary}
In this section, we study the impact of the crust on
several relations between binary quantities and the properties of
the individual NS in the binary.\\

The leading tidal parameter of the gravitational-wave signal of a
NS merger depends on
the effective tidal  deformability of the binary system
\begin{equation}
\tilde{\Lambda}=\frac{16}{13}\frac{(12q+1)\Lambda_1+(12+q)q^4\Lambda_2}{(1+q)^5},
\label{eq:lambda}
\end{equation}
where  $q=M_2/M_1<1$ is the binary mass ratio and 
$\Lambda_{1}\, (M_1)$ and $\Lambda_{2}\, (M2)$ represent the tidal
deformability (mass) of the primary and the secondary NS in the binary, respectively.
The GW170817 event provides an upper bound of
$\tilde{\Lambda}=300^{+420}_{-230}$ (using a 90\% highest posterior
density interval and the waveform model PhenomPNRT, although other models may
predict a larger upper bound) for the low spin-prior
\cite{Abbott:2018wiz}. The binary mass ratio, for the GW170817 event,
is bounded as $0.73\leq q \leq1$ \cite{Abbott:2018wiz}. \\

The chirp mass of the binary system is another quantity that is 
measured with a good accuracy during the
gravitational wave detection. It is given by 
\begin{equation}
M_{\text{chirp}}=\frac{(M_1M_2)^{3/5}}{(M_1+M_2)^{1/5}}=M_1\frac{q^{3/5}}{(1+q)^{1/5}}.
\label{eq:chirp}
\end{equation}
For the GW170817 event,  it was measured to be 
$M_{\text{chirp}}=1.188^{+0.004}_{-0.002}M_{\odot}$ \cite{TheLIGOScientific:2017qsa}, and more recently updated to $1.186^{+0.001}_{-0.001}M_{\odot}$ \cite{Abbott:2018wiz}.\\  
  

\subsection{Relation between $\tilde{\Lambda}$ and $\Lambda_{1,2}$}

\begin{figure}[!htb]
	\centering
	\includegraphics[width=1.0\columnwidth]{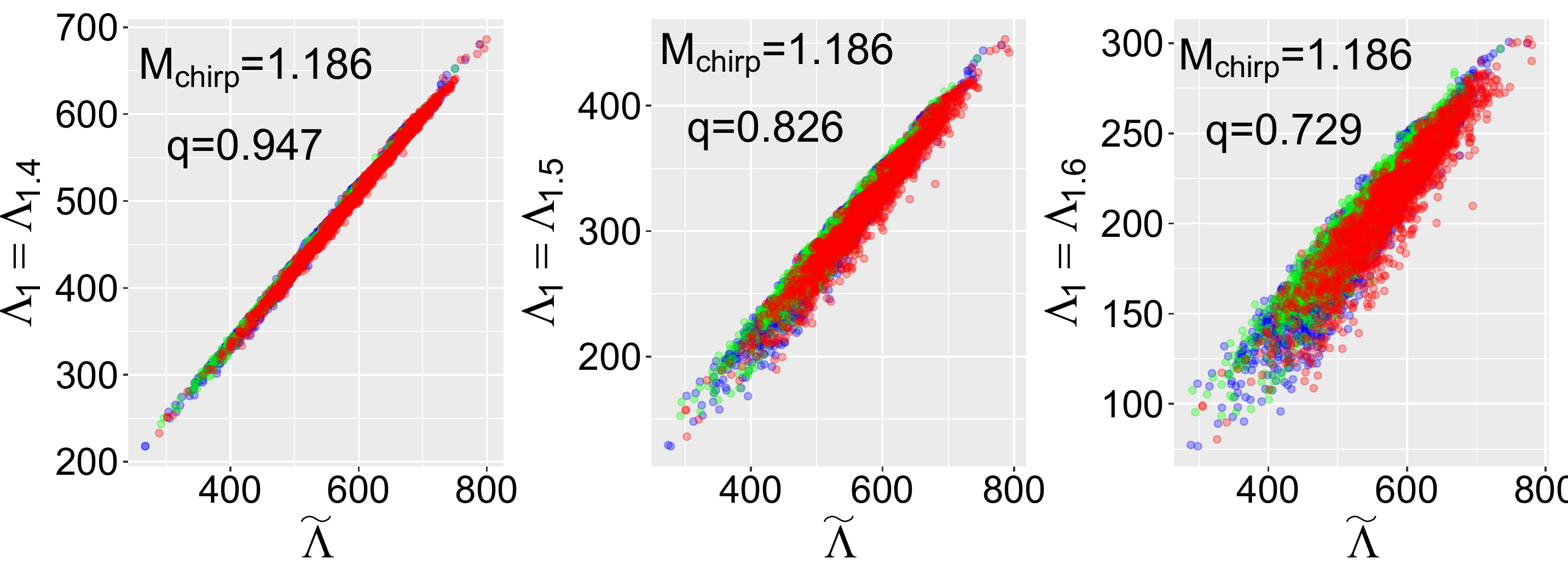}\\
	\includegraphics[width=1.0\columnwidth]{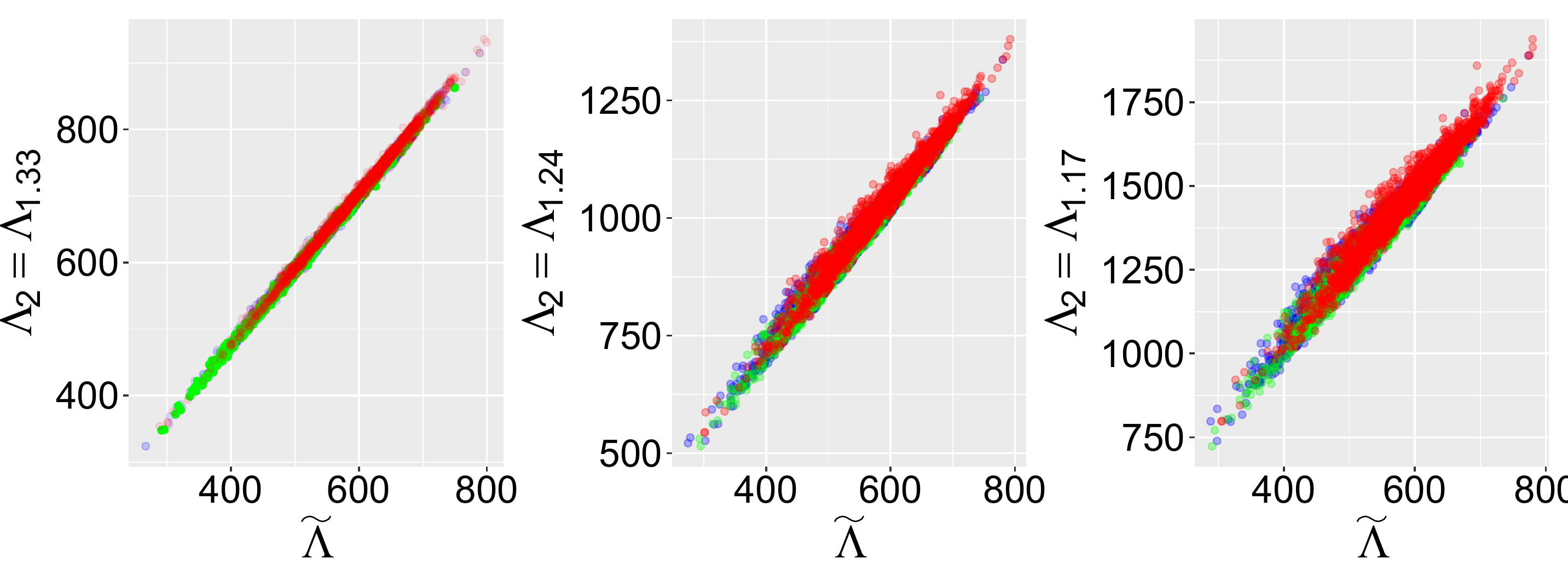}
	\caption{The effective binary tidal deformability $\tilde{\Lambda}$ versus
	the tidal deformability of each component stars $\Lambda_{1,2}$ for several values of the  binary mass ratio, $q=0.947$ (left), $q=0.826$ (middle), and $q=0.729$ (right), and for a fixed chirp mass of $M_{\text{chirp}}=1.186M_{\odot}$.
          The different crusts EoS are shown: DDHd (red), NL3 (green), and SLy4 (blue). }
\label{tildeL}
\end{figure}

If the chirp mass expression (Eq.~(\ref{eq:chirp})) is  rewritten as $M_1(M_{\text{chirp}},q)$,
the binary mass ratio $q$ determines both
$M_1$ and $M_2$ for a fixed $M_{\text{chirp}}$. Knowing the binary NS
masses, their tidal deformabilities $\Lambda_{1,2}$ and the effective
tidal deformability of the binary $\tilde{\Lambda}$ can be  determined.
In Fig. \ref{tildeL}, taking the GW170817 chirp mass, $M_{\text{chirp}}=1.186M_{\odot}$,
we show the tidal deformability of each star $\Lambda_{1,2}$ as a function of the binary tidal deformability $\tilde{\Lambda}$ for three representative values of the binary mass ratio $q$: $0.947, \, 0.826, \, 0.729$.
Since we have shown that $\Lambda_{1,2}\sim M_{1,2}^{-6}$ in Sec. \ref{sec:lambda_c},
we get  $\Lambda_2\sim q^{-6}\Lambda_1$.  
As expected, when the binary mass ratio $q$ decreases, it is the
lighter NS tidal deformability $\Lambda_2$ that dominates
$\tilde{\Lambda}$ and, therefore, the correlation is stronger for the star $M_2$.
This is clearly seen when comparing the top and bottom right panels of Fig. \ref{tildeL}.
For almost  symmetrical binary systems,  as the (1.40 $M_\odot$, 1.33
$M_\odot$) system,  $q\approx 1$, we have $M_1\approx M_2$ and $\Lambda_2 \approx\Lambda_1\approx\Lambda$, which leads to  $\tilde{\Lambda}\propto \Lambda$ (see left panels of Fig. \ref{tildeL}).

Let us consider the binary mass ratio $q=0.947$ (left panels in Fig. \ref{tildeL}),
and analyze what information can be extracted from $\tilde{\Lambda}$
on $\Lambda_{1}=\Lambda_{1.4M_{\odot}}$ (the heavier NS component of the binary). 
Doing the linear regression $\Lambda_{1.4M_{\odot}}=\beta\tilde{\Lambda}+\alpha$, we obtain 
\begin{align}
[\beta&=0.884,  \,\, \alpha=-18.807, \,\, \text{corr}=0.998]  \,\, \text{NL3}\label{eq:L1}\\
[\beta&=0.881,  \,\, \alpha=-18.857, \,\, \text{corr}=0.998]  \,\, \text{DDHd}\\
[\beta&=0.882, \,\, \alpha=-16.847, \,\, \text{corr}=0.999]  \,\, \text{SLy4}\label{eq:L2}
\end{align}
The regression results are similar for all crusts and show very strong
correlations, and, consequently,   $\Lambda_{1.4M_{\odot}}$
can be accurately extracted from $\tilde{\Lambda}$. In
\cite{Malik2018}, the authors have obtained
 $\beta=0.859$ also with a very large correlation
coefficient,  using a set of relativistic and non-relativistic
mean-field models. This seems to indicate that $\Lambda_{1.4M_{\odot}}(\tilde{\Lambda})$ relation is quite insensitive to the EoS parametrization used.


\subsection{Relation between $\tilde{\Lambda}$ and $R_{1,2}$}
Let us now study the relation between the effective tidal deformability  $\tilde{\Lambda}$ and $R_{1,2}$.
In \cite{Raithel2018,Raithel:2019uzi}, the authors have looked at this
problem  taking six nuclear EoS and obtained a strong
correlation between $\tilde{\Lambda}$ and $R_1$, which showed to be quite
independent of
the individual component masses. From this correlation they could
conclude that a upper bound $\tilde\Lambda=800$ would exclude radii
above $\sim$13 km at the 90\% confidence level.  In the following we
examine this same problem with our set of EoS and check how strongly
is it affected by the crust.

\begin{figure}[!htb]
	\centering 
	\includegraphics[width=1.0\columnwidth]{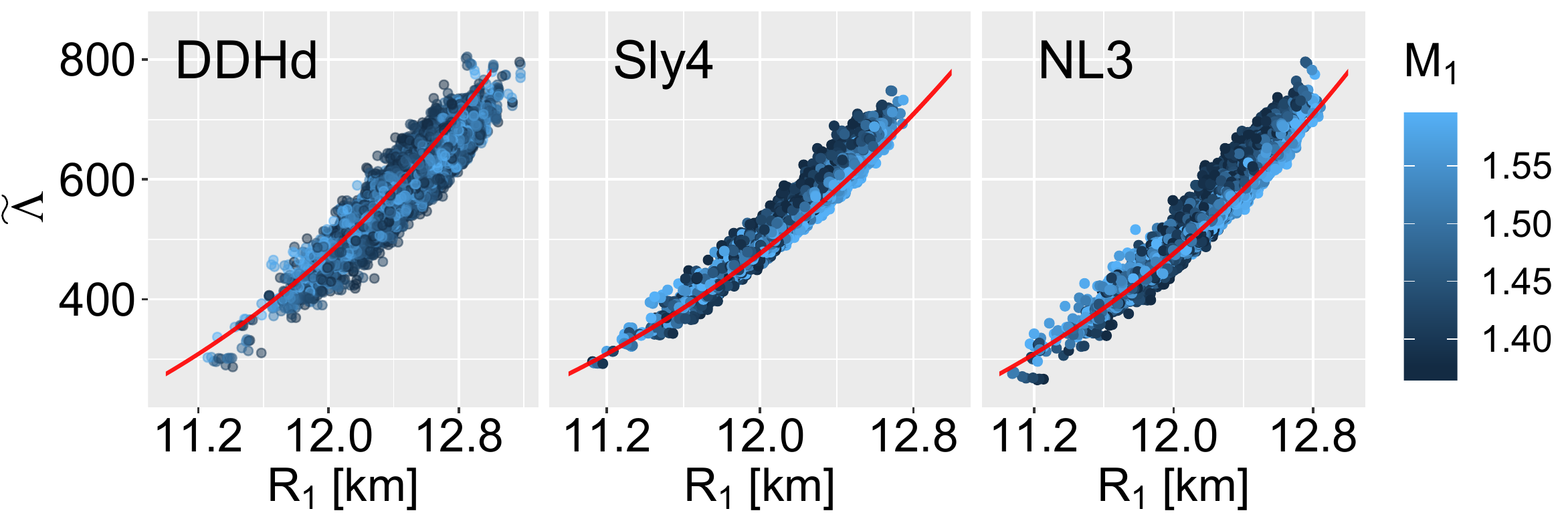}	     
	\caption{ $\tilde{\Lambda}$ as a function of $R_1$ by fixing the 
		$M_{\text{chirp}}=1.186M_{\odot}$. The red curves in three panels are the fit proposed
		in \cite{Raithel2018,Raithel:2019uzi}.}
	\label{L-r}
\end{figure}

In Fig. \ref{L-r}, we plot the $\tilde{\Lambda}$ as a function of the radius of the primary NS in the binary,
$R_1$, for $M_{\text{chirp}}=1.186M_{\odot}$ and $0.73<q<1.0$. All three crusts are shown:
DDHd (left), NL3 (middle) and SLy4 (right). 
Our set of EoS only describes $\tilde{\Lambda}<800$ values, however the
plot confirms that there is a strong correlation between
$\tilde{\Lambda}$ and $R_1$ that is not destroyed by
the individual component masses. A plot of the
$\tilde{\Lambda}$ versus the radius of the secondary NS in the binary,
$R_2$, would give the same results. The red curve in these panels represents
the fit proposed in \cite{Raithel2018,Raithel:2019uzi}. 
Although the  
correlation between the effective tidal deformability and $R_1$ is 
strong, we conclude that it is crust dependent and the fitting curve
depends on the EoS set, SLy4 giving the strongest correlation.

Performing the linear regression $\ln(\tilde{\Lambda})=\beta\ln(R_1)+\alpha$, we get
\begin{align}
[\beta&=6.554, \,\, \alpha=-10.108, \,\, \text{corr}=0.966]  \,\, \text{NL3}\\
[\beta&=5.729,  \,\, \alpha=-8.083, \,\, \text{corr}=0.926]  \,\, \text{DDHd}\\
[\beta&=6.553,  \,\, \alpha=-10.093, \,\, \text{corr}=0.973]  \,\, \text{SLy4}
\end{align}

The crust has an impact of the order $\approx14\%$ on the exponent
that describes the dependence of $\tilde{\Lambda}$ on $R_1$, when the only constraint set on the
star mass is that  $0.73<M_2/M_1<1.0$ and $1.1<M/M_\odot<1.6$. 


\subsection{Relation between $\tilde{\Lambda}$ and $M_{\text{chirp}}$}

\begin{figure}[!htb]
	\centering
	\includegraphics[width=0.49\columnwidth]{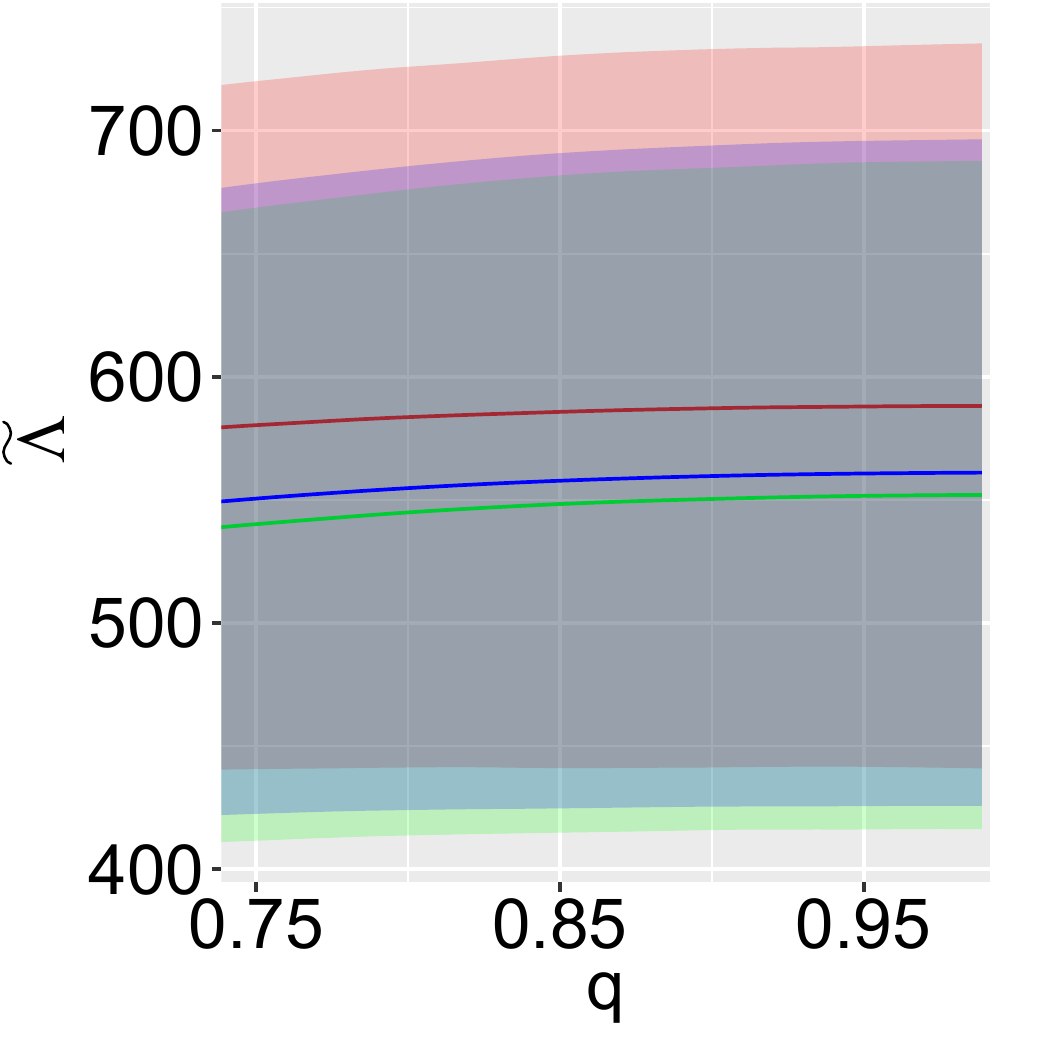}
	\includegraphics[width=0.49\columnwidth]{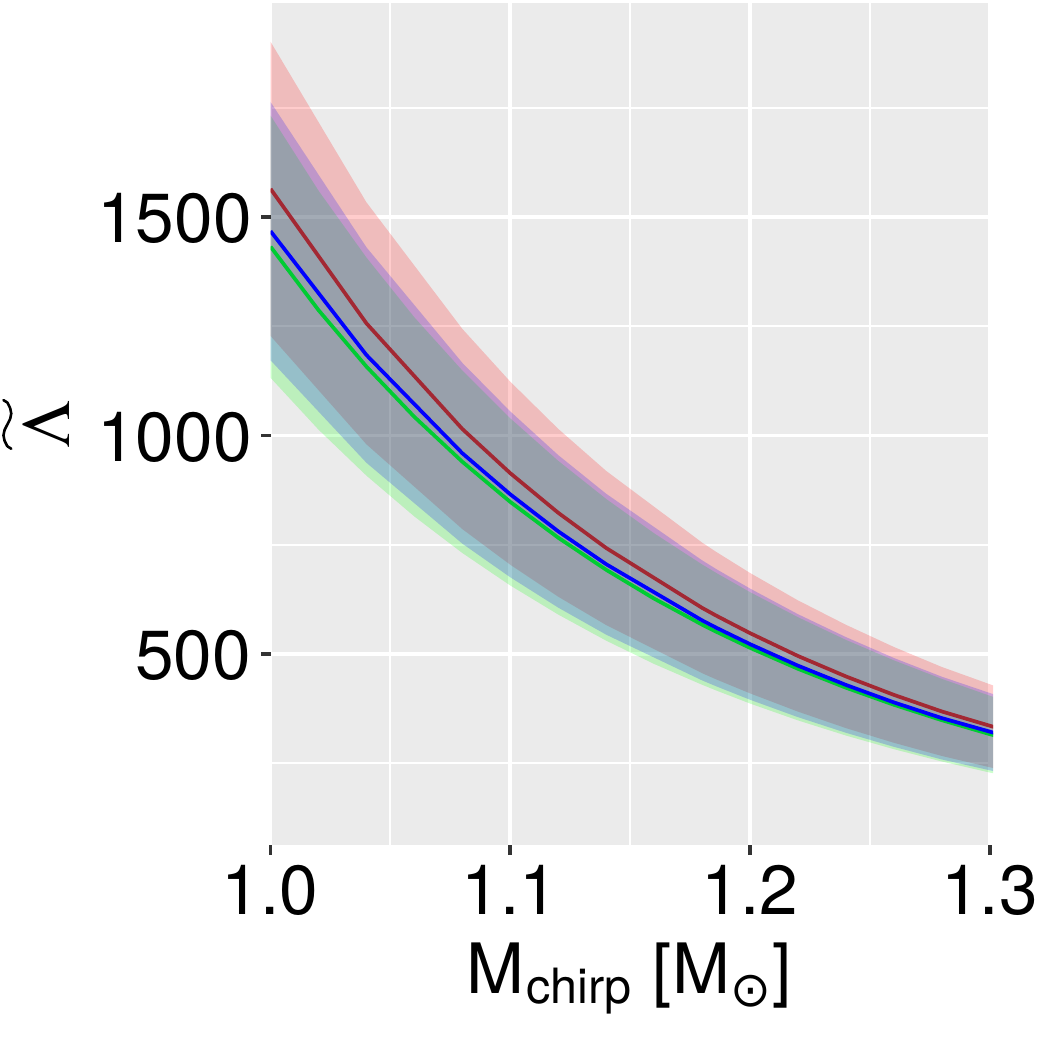}
	\caption{ $\tilde{\Lambda}$ density as a function of $q$ by fixing the 
$M_{\text{chirp}}=1.186M_{\odot}$ (left) and as a function of $M_{\text{chirp}}$ by fixing $q=0.90$ (right).
	The solid lines represent the mean values and the shaded
        region encloses $\pm2\sigma$ for DDHd (blue), SLy4
        (red), and NL3 (black) crusts.}
\label{chirp-q}
\end{figure}

We next analyze the impact of $M_{\text{chirp}}$ on $\tilde{\Lambda}$.  
From Eq. (\ref{eq:lambda}), the effective tidal deformability of the binary, $\tilde{\Lambda}$,
is written as a function of $q$ and $M_{\text{chirp}}$, $\tilde{\Lambda} (M_{\text{chirp}},q)$.
In Fig. \ref{chirp-q}, $\tilde{\Lambda}$ is plotted as a function of $q$ for a fixed  
$M_{\text{chirp}}=1.186M_{\odot}$ (left) and as a function of $M_{\text{chirp}}$ for a fixed
$q=0.90$ (right). 
We display both the average values (solid lines) and
dispersion (shaded region corresponds to $\pm2\sigma$). Each set with
a different crust EoS is represented by a different color.
The mean value of $\tilde{\Lambda}$ is   quite insensitive to the
value of $q$, as previously discussed in \cite{Carson:2019xxz}, with the DDHd pushing the whole distribution to higher values of $\tilde{\Lambda}$.
However, as also shown in \cite{Carson:2019xxz}, the
$\tilde{\Lambda}(M_{\text{chirp}})$ dependence for fixed $q$ is
strongly dependent on the binary chirp mass $M_{\text{chirp}}$, as shown
on the  right panel of Fig. \ref{chirp-q} for $q=0.9$. We further
conclude that the  impact of the crust on these results  is small. 

To quantify the differences imposed by each crust, we perform the linear regression $\ln(\tilde{\Lambda})=\beta\ln(M_{\text{chirp}})+\alpha$. 
The results are
\begin{align}
[\beta&=-5.824,  \,\, \alpha=7.312, \,\, \text{corr}=-1.000]  \,\, \text{NL3}\\
[\beta&=-5.910,  \,\, \alpha=7.375, \,\, \text{corr}=-1.000]  \,\, \text{DDHd}\\
[\beta&=-5.781,  \,\, \alpha=7.288, \,\, \text{corr}=-1.000]  \,\, \text{SLy4,} 
\end{align}
which indicate that $\tilde\Lambda$ and $M_{\text{chirp}}$ are highly
correlated. Moreover,   we get a
similar dependence as the one discussed in  \cite{De:2018uhw}, where  $\tilde{\Lambda}\sim
M_{\text{chirp}}^{-6}$ was determined,  with a exponent about 3\%
smaller. The above results are only slightly
dependent on the crust.

\subsection{Relation $\Lambda_1/\Lambda_2=q^a$ and dependence on $M_{\text{chirp}}$}

In \cite{De:2018uhw}, a reanalysis of the GW170817 event was made, assuming the relation $\Lambda_1/\Lambda_2=q^6$, based on the assumption that the two stars of the binary have the same EoS. 
The above relation was a consequence of two empirical relations $R_1\approx R_2$ (for $1.1\le M/M_\odot\le 1.6$) and $\Lambda\sim C^6$ within the EoS piecewise-polytrope methodology \cite{De:2018uhw}. 
In Sec. \ref{sec:lambda_c}, from the regression analysis of $\Lambda(M)$ (see Fig. \ref{lambda-c}), we got $\Lambda \sim C^{a}$ with $a=\{-6.122,-6.025,-6.137\}$,
for the NL3, DDHd, and SLy4 crusts, respectively, with an almost
perfect correlation ($\text{corr}\approx 1$). 
Let us recall that, as shown in  Sec. \ref{sec:lambda_c},  
only the model  with the crust DDHd gives results 
  compatible with $\Delta
R=R_{1.6 M_\odot}-R_{1.1 M_\odot}\approx 0$.

\begin{figure}[!htb]
	\centering
	\includegraphics[width=0.49\columnwidth]{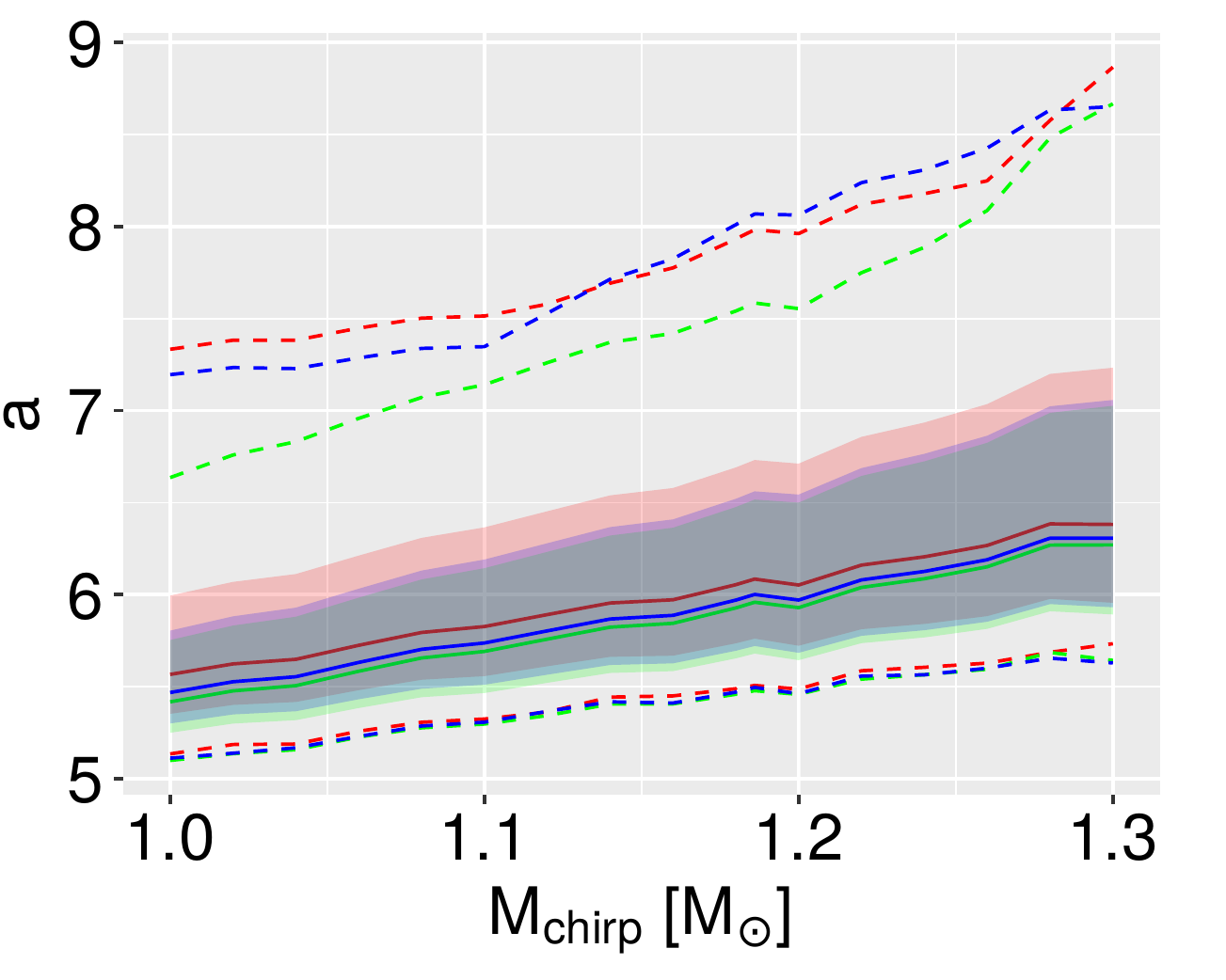}
	\includegraphics[width=0.49\columnwidth]{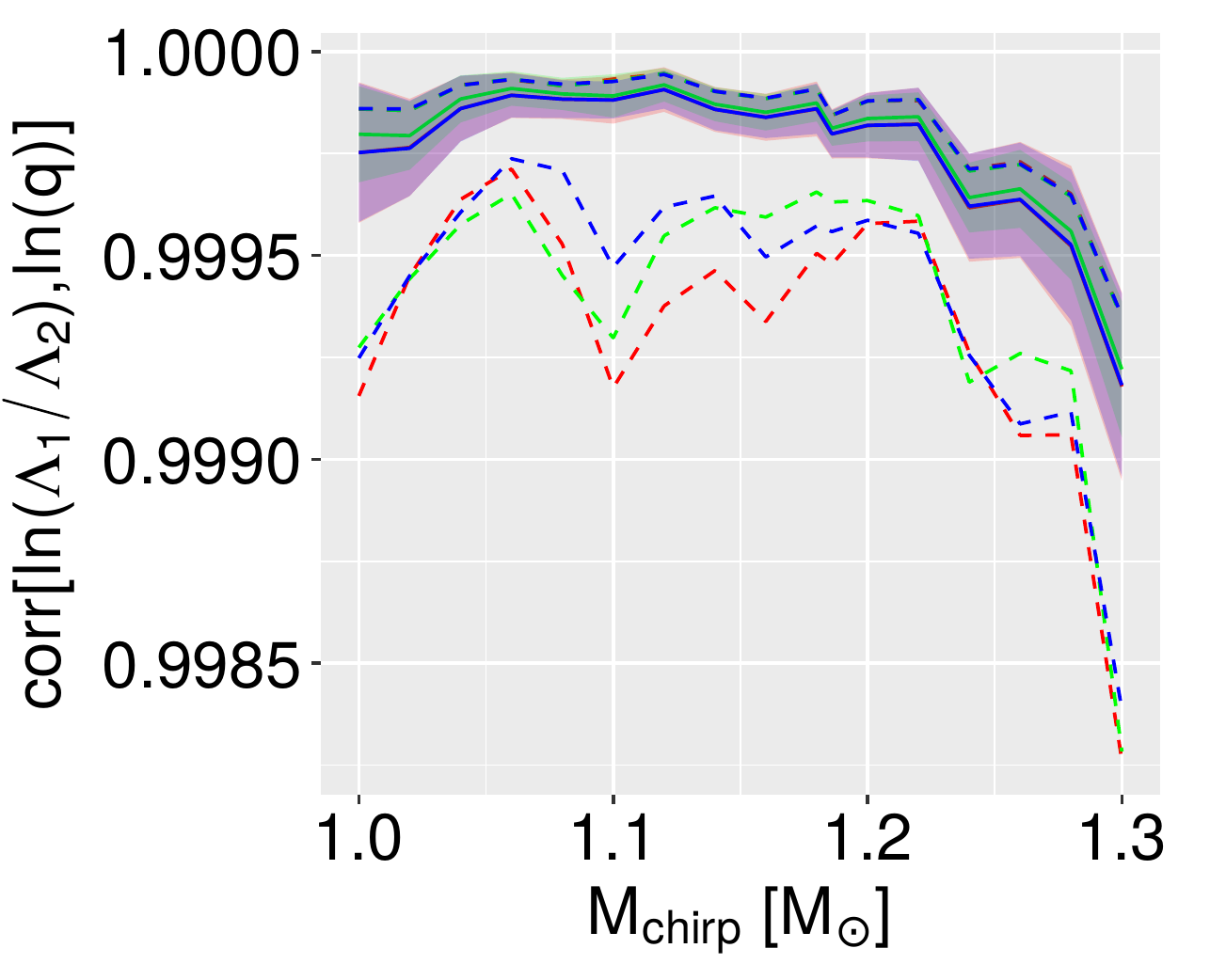}
	\caption{Dependence of $a$ (left) and $\text{Corr}[\ln(\Lambda_1/\Lambda_2),\ln(q)]$ (right) on $M_{\text{chirp}}$ (in units of $M_{\odot}$). It is shown the mean value (solid line), $\pm\sigma$ region (shaded region), and minimum/maximum values (dashed lines). 
		The results for all crust are shown: NL3 (green), SLy4 (red), and DDHd (blue).}
	\label{fig:a_Mchirp}
\end{figure}

To test the relation $\Lambda_1/\Lambda_2=q^a$, we determine the linear regression $\ln(\Lambda_1/\Lambda_2)\sim a\ln(q)$ for a given $M_{\text{chirp}}$. 
In Fig. \ref{fig:a_Mchirp} left panel, we show the
mean value  of $a$,  the $\pm\sigma$ deviation, and the
minimum/maximum values as a function of
$M_{\text{chirp}}$. 
The correlation value (right panel) shows that the relation
$\ln(\Lambda_1/\Lambda_2)\sim a\ln(q)$ perfectly captures the
dependence between $\Lambda_1/\Lambda_2$ and $q$. The exponent $a$ is
an increasing function of $M_{\text{chirp}}$. However, there is a
considerable spread, and the standard deviation is always larger than
0.5 and gets larger with increasing
$M_{\text{chirp}}$. Some EoS show $a$ values that deviate more than
38\% from the mean value (dashed lines). 
We have obtained $5.25<a<6.91$  for $1.0\leq
M_{\text{chirp}}/M_{\odot}\leq 1.3$,  and including the uncertainty
that is attributed to the crust. Considering only the SLy4 crust, 
this interval would reduce to  $5.25<a<6.68$. 
In \cite{De:2018uhw}, in a similar study the authors have obtained
$5.76<a<7.48$. Although not very different, our results correspond 
to  larger  values  of the
ratio $ \Lambda_1/\Lambda_2$,  up to  50\%  (15\%)  larger at the
lower (upper) limit.

We have further studied the dependence of the exponent $a$,  the solid
lines in  Fig. \ref{fig:a_Mchirp} left panel, on the chirp
mass $M_{\text{chirp}}$.
Performing the  linear regression
$\ln(\bar{a})=\beta\ln(M_{\text{chirp}})+\alpha$, where $\bar{a}$
represents the mean value  of the exponent $a$, we have obtained
\begin{align}
[\beta&=0.551,  \,\, \alpha=1.696, \,\, \text{corr}=0.996]  \,\, \text{NL3}\\
[\beta&=0.528,  \,\, \alpha=1.714, \,\, \text{corr}=0.926]  \,\, \text{DDHd}\\
[\beta&=0.564,  \,\, \alpha=1.687, \,\, \text{corr}=0.996]  \,\, \text{SLy4}.
\end{align}
The regression values for $\beta$ of $\approx 0.5$ show that $a\sim \sqrt{M_{\text{chirp}}}$,
to a very good approximation. Besides, the correlation  obtained with the
SLy4 and NL3 crusts  is very  strong, close to 1.


\section{Estimation of NS properties from $\tilde{\Lambda}$}
\label{Sec:estimate}

In the present section, the full information from  regression analysis will be used in predicting NS properties from GW observables.\\
 
Since $\tilde{\Lambda}$ and $M_{\text{chirp}}$ are both extractable from 
gravitational wave detections, it is convenient
to analyze what information $\tilde{\Lambda}$ contains about a given NS.
In the following, we focus on the GW170817 event, i.e.,
we fix $M_{\text{chirp}}=1.186M_{\odot}$, and analyze what we can infer about 
the tidal deformability and the radius of a NS.

\subsection{Tidal deformability of a $1.4M_{\odot}$ NS}

\begin{figure}[!htb]
	\centering
	\includegraphics[width=1.0\columnwidth]{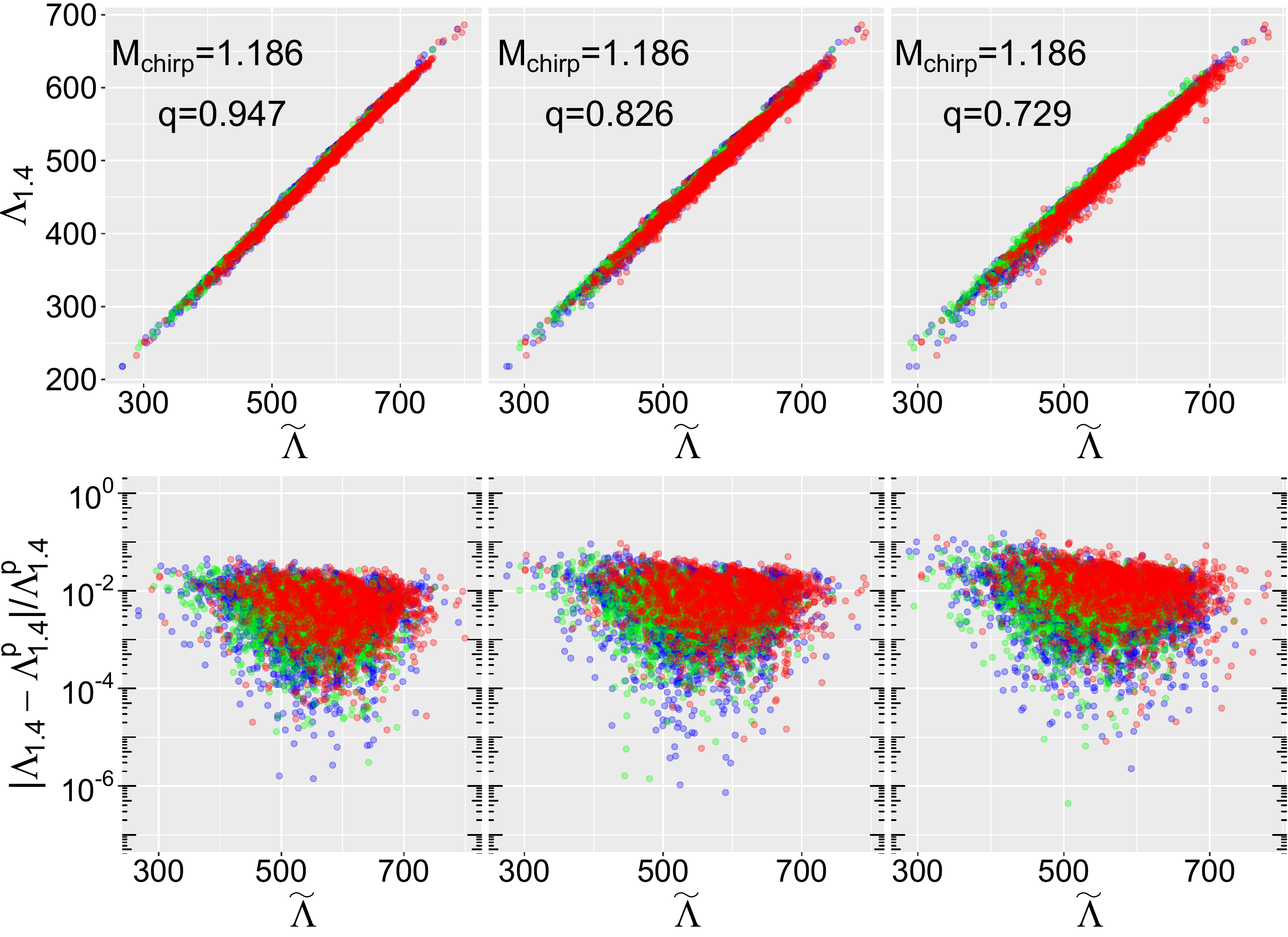}
	\caption{$\Lambda_{1.4M_{\odot}}$ vs. $\tilde{\Lambda}$ (top
          panels) for different binary mass ratio values: $0.948$
          (left), $0.826$ (center), and $0.729$ (right), with fixed
          $M_{\text{chirp}}=1.186M_{\odot}$. We also show the
          respective scaled residuals
          $(\Lambda_{1.4M_{\odot}}-\Lambda_{1.4M_{\odot}}^{p})/\Lambda_{1.4M_{\odot}}^{p}$
          (bottom panels), where $\Lambda_{1.4M_{\odot}}^{p}$ are the
          linear regressions model predictions  (see text). The
          different crust EoS are shown: DDHd (red), NL3 (green), and
          SLy4 (blue).}
	\label{fig:ll}
\end{figure}

We first analyze the tidal deformability of a $1.4M_{\odot}$ NS.
Figure \ref{fig:ll} shows $\Lambda_{1.4M_{\odot}}$ vs. $\tilde{\Lambda}$
(top panels) and respective respective scaled residual (bottom panels) for $q=0.947$ (left), $0.826$ (center), and $0.729$ (right).

Performing the same linear regression as done for $q=0.947$ in
Eqs. (\ref{eq:L1})-(\ref{eq:L2}), for
$\Lambda_{1.4M_{\odot}}=\beta\tilde{\Lambda}+\alpha$, we get for $q=0.823$ (middle plot of Fig. \ref{fig:ll}), 
\begin{align}
[\beta&=0.902,  \,\, \alpha=-25.278, \,\, \text{corr}=0.997]  \,\, \text{NL3}\label{eq:LL1}\\
[\beta&=0.895,  \,\, \alpha=-24.780, \,\, \text{corr}=0.995]  \,\, \text{DDHd}\\
[\beta&=0.901,  \,\, \alpha=-22.797, \,\, \text{corr}=0.998]  \,\, \text{SLy4}\label{eq:LL2},
\end{align}
and for $q=0.729$ (right plot of Fig. \ref{fig:ll}),
\begin{align}
[\beta&=0.938,  \,\, \alpha=-37.374, \,\, \text{corr}=0.994]  \,\, \text{NL3} \label{eq:LLL1}\\
[\beta&=0.923,  \,\, \alpha=-35.053, \,\, \text{corr}=0.989]  \,\, \text{DDHd}\\
[\beta&=0.936,  \,\, \alpha=-33.375, \,\, \text{corr}=0.995]  \,\, \text{SLy4}\label{eq:LLL2}.
\end{align}
Regardless of the binary mass ratio $q$, 
there is always a strong linear relation between $\tilde{\Lambda}$ and $\Lambda_{1.4M_{\odot}}$.
The regression analysis shows that both NL3 and SLy4 crusts have similar results while
the the DDHd predicts a smaller slope value $\beta$. Increasing the
 asymmetry of the binary systems $q<1$ (i.e., $M_1> M_2$), the slope
increases  and there is a decrease on the correlation
coefficient. Thus, the higher is the binary asymmetry $q$ the less
information $\tilde{\Lambda}$ carries about a $1.4M_{\odot}$ NS. 

The scaled residuals of the linear regressions,
$(\Lambda_{1.4M_{\odot}}-\Lambda_{1.4M_{\odot}}^{p})/\Lambda_{1.4M_{\odot}}^{p}$,
where the $\Lambda_{1.4M_{\odot}}^{p}$ are the linear regression predictions,
are shown in the bottom panels of Fig. \ref{fig:ll}.
The regression $\Lambda_{1.4M_{\odot}}^p=\beta\tilde{\Lambda}+\alpha$  has an overall uncertainty below $5\%$ for $q=0.947$, while it is around $10\%$ for lower $q$ values.
If we consider the extreme lower bound of $q=0.73$ from
\cite{Abbott18}, we can infer the value of $\Lambda_{1.4M_{\odot}}$
with an $10\%$ accuracy. The precision gets better as one increases
the mass ratio value $q$. This shows that 
even if a $1.4 M_\odot$ NS is not part of the binary, it is still possible to determine its tidal deformability,
$\Lambda_{1.4 M_\odot}$, with an accuracy of at least $\sim 10\%$. For
the case of the GW170817, the closer the ratio $q$ to one the larger is
the accuracy. 
This is the case because $q=1$ corresponds to two
  stars with a mass $1.37\,M_\odot$ very close to  $1.4\,M_\odot$. On
  the other hand,
  the smaller $q$ the larger the mass difference between both stars,
  and this blurs to some extent the information that can be drawn from $\tilde\Lambda$.  

\subsection{Constraining $\Lambda_{M_i}$ from  $\tilde{\Lambda}$}

In this section we study the correlation between $\tilde{\Lambda}$  and
$\Lambda_{M_i}$,  i.e. we calculate
$\text{Corr}[\Lambda_{M_i},\tilde{\Lambda}]$, for different NS masses
as a function of the binary mass ratio $q$, in order to  answer the question
what can we learn from the $\tilde{\Lambda}$ about
NSs that have a mass different from $1.4M_{\odot}$.

\begin{figure}[!htb]
	\centering
	\includegraphics[width=0.9\columnwidth]{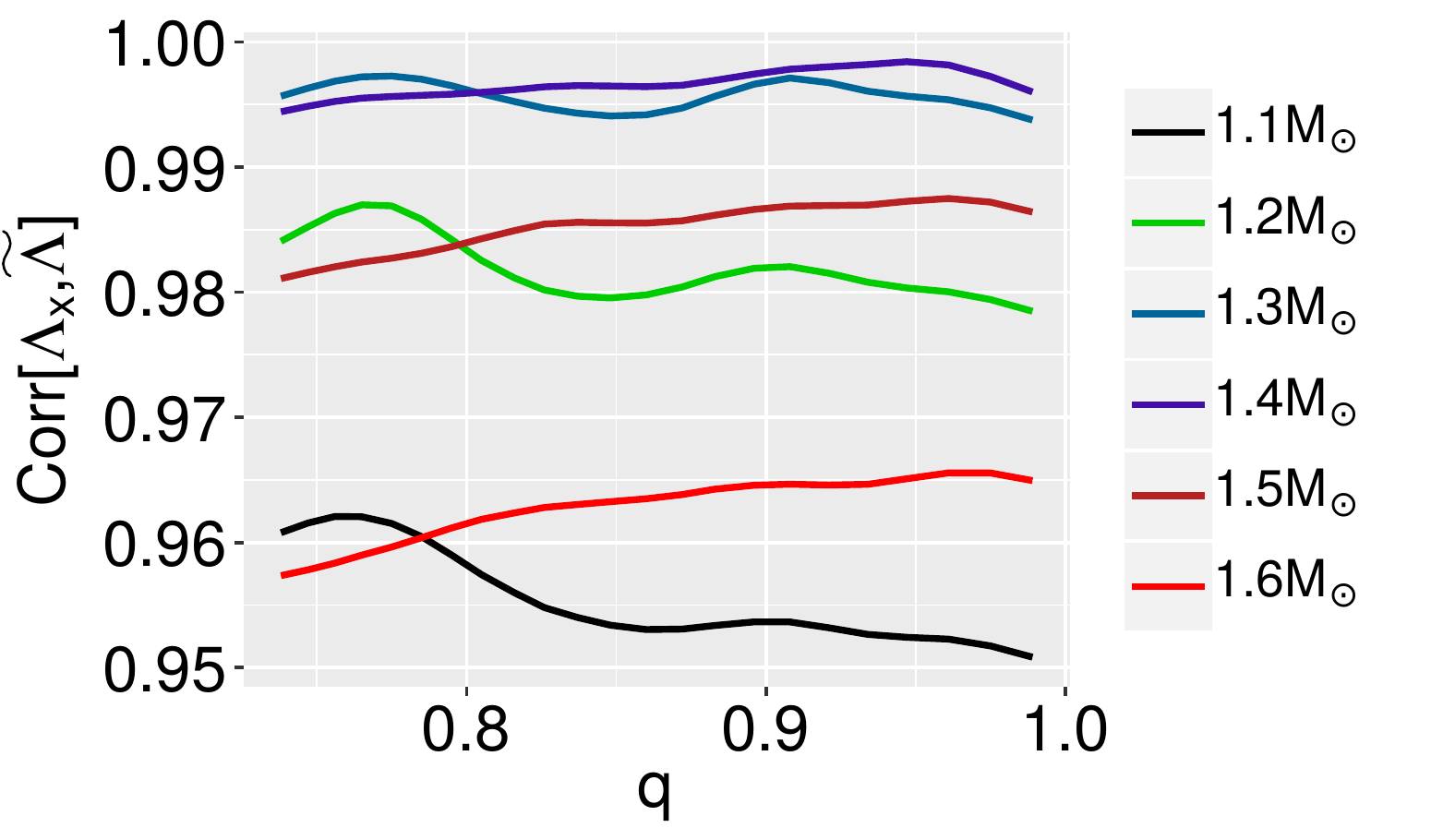}
	\caption{$\text{Corr}[\Lambda_{M_i},\tilde{\Lambda}]$ as a function of $q$ for $M_i/M_{\odot}=1.1,1.2,1.3,1.4,1.5$, and $1.6$ using the NL3 crust.}
	\label{fig:utl1}
\end{figure}

The results for $M_{\text{chirp}}=1.186M_{\odot}$ are in Fig. \ref{fig:utl1} for the NL3 crust (the results are similar for the other two crusts). 
A strong correlation is found between $\tilde{\Lambda}$ and any $\Lambda_{M_i}$ with 
$1.1<M_i/M_{\odot}<1.6$. However, depending on the $q$ value,  the strongest constraint linearly
extractable from $\tilde{\Lambda}$ is
either for $\Lambda_{1.3M_{\odot}}$ or  for
$\Lambda_{1.4M_{\odot}}$. $\tilde{\Lambda}$ contains more information
on NS with these masses because  they are the  intermediate masses for
the two extremes defined by $q$.

This analysis may be used to constrain $\Lambda_{M_i}$ from
observational bounds on $\tilde{\Lambda}$. 
Our sample does not describe  
large $\tilde{\Lambda}$ values:
for $q=0.947$ it predicts
$\ev{\tilde{\Lambda}}\pm2\sigma=560.67\pm135.11$. 
 However, the strong correlation
 $\text{Corr}[\Lambda_{M_i},\tilde{\Lambda}]$ indicates a universal
 behavior that we can explore to extract constraints on
 $\Lambda_{M_i}$ from any observational bound on $\tilde{\Lambda}$. 

In the following, we use first the LIGO/Virgo upper bound
  $\tilde{\Lambda}=720$  obtained from the waveform model
  PhenomPNRT\cite{Abbott:2018wiz} to set an upper bound on
  $\Lambda_{M_i}$ as a function of $q$. Taking next the constraint
  $\tilde{\Lambda}\ge 300$
\cite{Radice2019} obtained from the  electromagnetic  (EM)
counterpart of the GW170817 event,  the  AT2017gfo/GRB  170817A, we
will also derive a lower bound. Although  $\tilde{\Lambda}$ does not
depend much on $q$,  as  shown before  (see Fig. \ref{chirp-q}),
$\Lambda_{M_i}$   is $q$ dependent.

\begin{figure}[!htb]
	\centering
	\includegraphics[width=1.0\columnwidth]{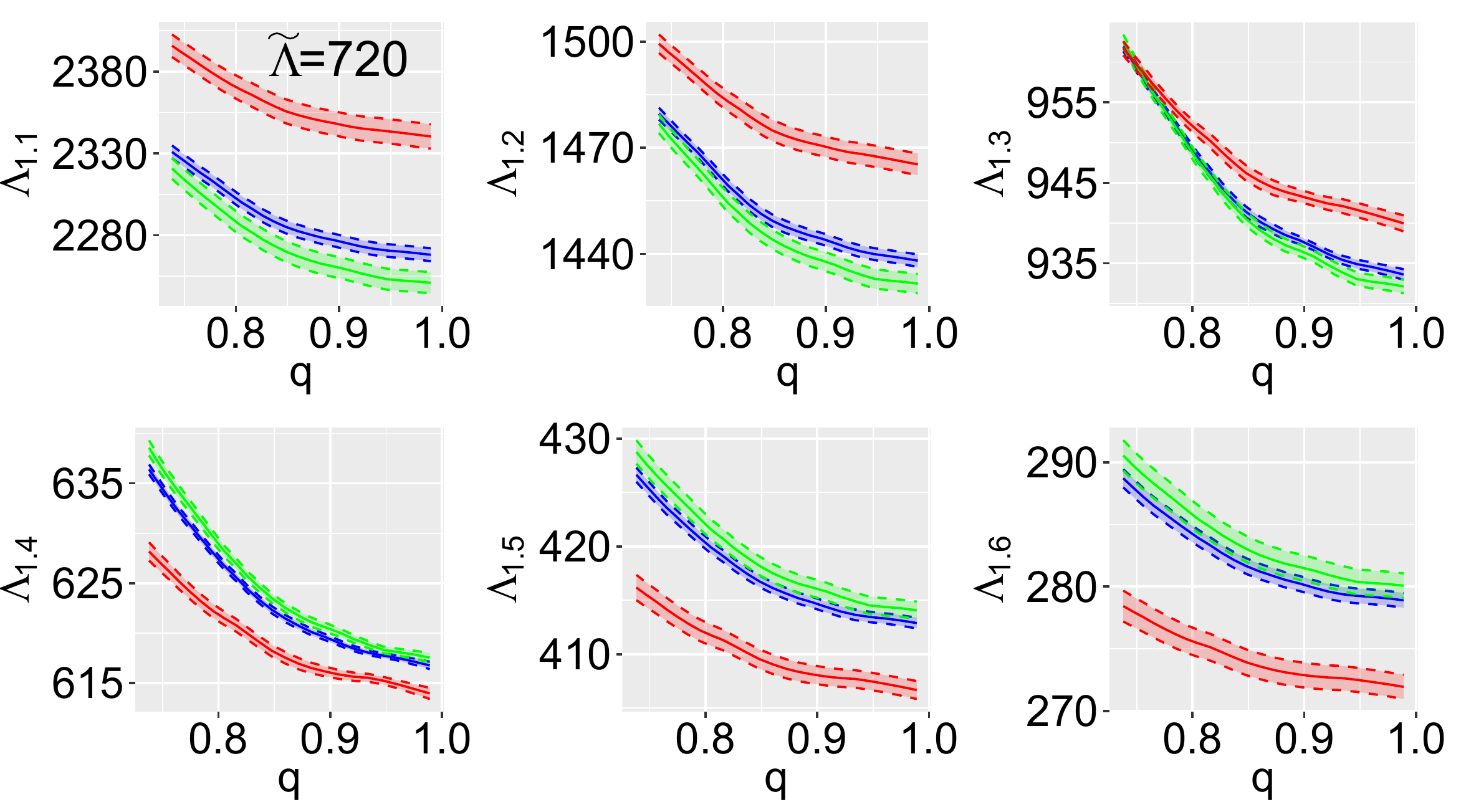}
	\caption{Prediction for $\Lambda_{M_i}(q,\tilde{\Lambda}=720)$ for $M_i=1.1,1.2,1.3,1.4,1.5$, and $1.6M_{\odot}$. The solid lines indicates the prediction value while the shaded region represents the 95\% confidence interval region. 
	The results for the three crusts are shown: NL3 (blue), SLy4 (red), and DDHd (green).}
	\label{fig:utl2}
\end{figure}

Figure \ref{fig:utl2} shows the upper bound on $\Lambda_{M}(q)$ 
for each crust EoS from the $\tilde{\Lambda}=720$ constraint.
The predictions $\Lambda_{M}(q)$ for both NL3 and SLy4 crusts are similar. 
For $M>1.4M_{\odot}$, the DDHd predicts higher values than the
other two crusts while the opposite happens for higher $M$ values.
The width of the confidence intervals reflects the  strength of
$\text{Corr}[\Lambda_{M_i},\tilde{\Lambda}]$ 
 displayed in Fig.~\ref{fig:utl1}.
$\Lambda_{M}$ decreases when  $q$ increases, due
to the
$\beta$ dependence   on $q$. We found, from the regression analysis
on $\Lambda_{M_i}=\beta\tilde{\Lambda}+\alpha$ (see Eqs. (\ref{eq:L1})-(\ref{eq:L2}), (\ref{eq:LL1})-(\ref{eq:LL2}), and (\ref{eq:LLL1})-(\ref{eq:LLL2}) for the case of $M_i=1.4M_{\odot}$) that      
$\beta$ decreases as $q$ grows. This is an indirect effect of the fact
that $\tilde{\Lambda}$ increases slightly when $q$ increases, see Fig.
\ref{chirp-q}.

The values of $\Lambda_{1.4M_{\odot}}(q,\tilde{\Lambda}=720)$ for the 
extreme values of $q$ are shown in Table \ref{tab:constraints_Lambdas} (Appendix \ref{appendix:a}).
The predictions lie between  $636$  at $q=0.74$ and $617$ at $q=0.99$ for the NL3 crust (SLy4 gives slightly higher values), while
it is $628$ at $q=0.74$ and $614$ at $q=0.99$ for the DDHd
crust. We, therefore, get as an upper bound  
 $\Lambda_{1.4M_{\odot}}<640$ at 95\% confidence 
  interval, already taking into account the  crust  effect, that
  brings  an uncertainty not larger than 1\%, and the $q$ dependence
  which  brings an uncertainty below 3\%.

Table \ref{tab:constraints_Lambdas} also contains a possible lower bound  on $\Lambda_{M_i}(q,\tilde{\Lambda})$, for
$q=0.74$ and $0.99$, by using the constraint $\tilde{\Lambda}=300$
\cite{Radice2019} obtained from the  electromagnetic  (EM)
counterpart of the GW170817 event.  In
this case, the lower bound on  $\Lambda_{1.4M_{\odot}}$ would be  $\Lambda_{1.4M_{\odot}}<240$ at 95\% confidence
  interval  including the crust  and $q$  dependence  uncertainties.

Taking the above  upper and lower bound constraints, we  have obtained $240<
\Lambda_{1.4M_{\odot}}< 640$. This prediction is compatible with
  results from
  other studies.
Assuming a common EoS for the two NSs in the binary, the
LIGO/Virgo collaborations constrained $70<\Lambda_{1.4M_{\odot}}<580$
at the 90\% level \cite{Abbott18}. This constraint was
obtained without requiring that the EoS should support NSs up to at
least $1.97\, M_{\odot}$. A higher upper bound on
$\Lambda_{1.4M_{\odot}}$ is expected by requiring that the EoS should comply with
massive stars due to the positive correlation between $\Lambda_{M_i}$
and the maximum NS mass. 

\subsection{Constraining $R_{M_i}$ from  $\tilde{\Lambda}$}

\begin{figure}[!htb]
	\centering 
	\includegraphics[width=0.8\columnwidth]{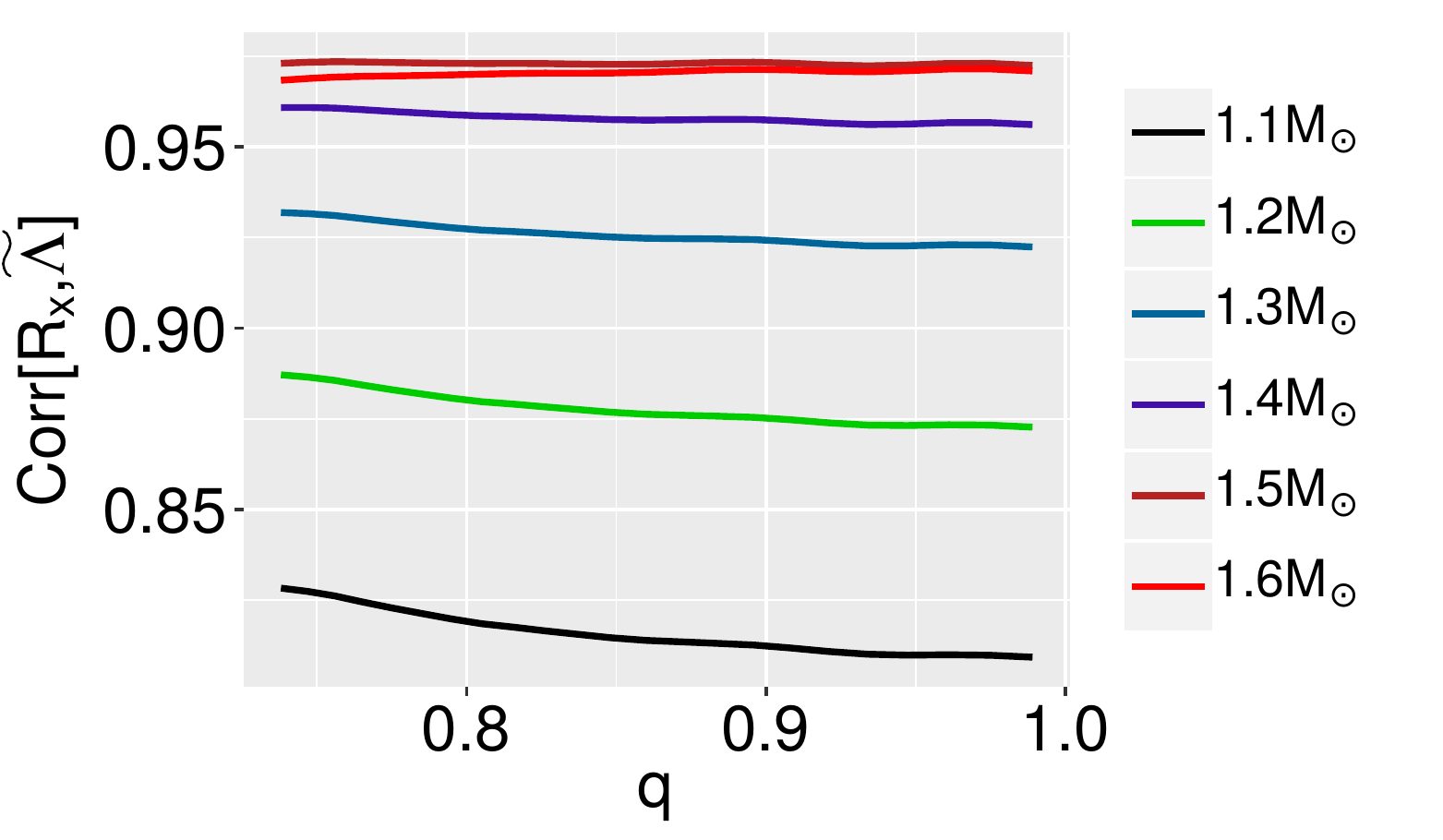}	     	
	\caption{$\text{Corr}[R_{M_i},\tilde{\Lambda}]$ as a function of $q$ for $M_i=1.1,1.2,1.3,1.4,1.5$, and $1.6M_{\odot}$ using the NL3 crust.}
\label{fig:R_corr1}
\end{figure}

We perform a similar analysis of $\text{Corr}[\Lambda_{M_i},\tilde{\Lambda}]$,
but this time on $\text{Corr}[R_{M_i},\tilde{\Lambda}]$. In this way, we
are able to constrain the NS radius for any NS mass.  The results are
given in Fig. \ref{fig:R_corr1}, showing that the correlations
$\text{Corr}[R_{M_i},\tilde{\Lambda}]$ are weaker than
$\text{Corr}[\Lambda_{M_i},\tilde{\Lambda}]$ (see
Fig. \ref{fig:utl1}). This indicates that $\Lambda_{M_i}$ has a
stronger linear dependence on  $\tilde{\Lambda}$ than on $R_{M_i}$.
The $1.4M_{\odot}$, $1.5M_{\odot}$, and $1.6M_{\odot}$ NS are the ones that show the strongest correlations, above $0.95$.

\begin{figure}[!htb]
	\centering
	\includegraphics[width=1.0\columnwidth]{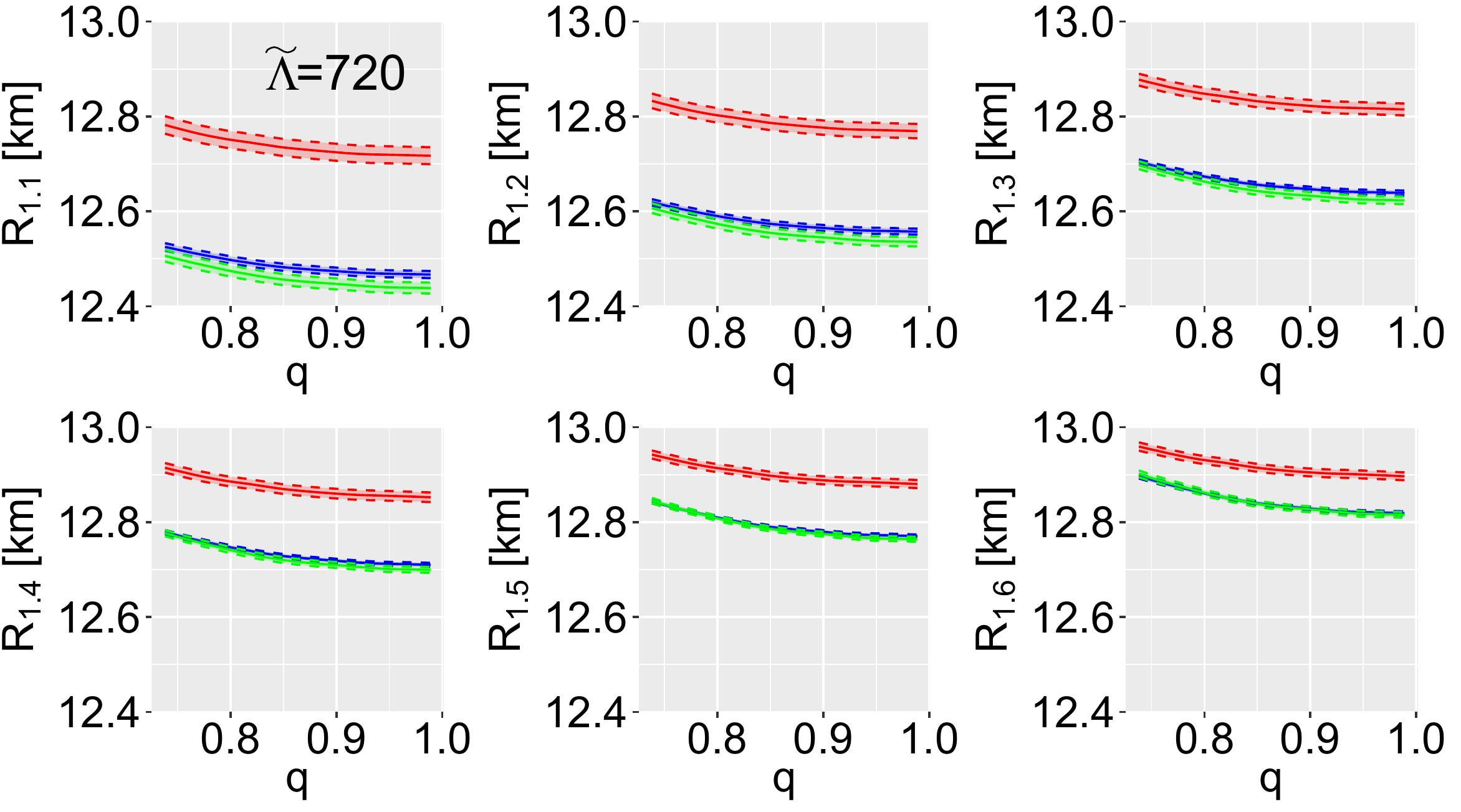}
	\caption{Prediction for $R_{M_i}(q,\tilde{\Lambda}=720)$ for $M_i=1.1,1.2,1.3,1.4,1.5$, and $1.6M_{\odot}$. The solid lines indicates the prediction value while the shaded region represents the 95\% confidence interval region. 
	The results for the three crusts are shown: NL3 (blue), SLy4
        (red), and DDHd (green).}
	\label{fig:R_corr2}
\end{figure}

In Fig. \ref{fig:R_corr2}, we show the upper bound on $R_{M_i}$ for
$M_i=1.1,\,1.2,\,  1.3,\,  1.4,\, 1.5$, and $1.6M_{\odot}$ using the LIGO/Virgo
upper bound $\tilde{\Lambda}=720$ \cite{Abbott:2018wiz}.  The  
predicted value for the radius decreases when $q$ increases.
Results obtained for
$R_M(q,\tilde{\Lambda}=720)$ with the NL3 and SLy4 crust are
compatible with each other. However, 
DDHd crust predicts larger  radii: a difference
that  can be as large as $\sim 300$m for
$M=1.1M_\odot$,  but that reduces to $\sim 50$m for
$M=1.6M_\odot$. 

The  upper bound predictions obtained for the 1.4$M_\odot$
star radius, $R^{\text{upper}}_{1.4M_{\odot}}(q,\tilde{\Lambda}=720)$, are given in
Table \ref{tab:constraints_Rs} (Appendix \ref{appendix:a}) for the extreme values of $q$.
They  lie between $12.78$ km at $q=0.74$ and $12.71$ km at $q=0.99$
for the NL3 crust and within 0.01 km for the SLy4 crusts. The $q$
dependence introduces an uncertainty of the order of 0.5\%.
For the DDHd crust, $R^{\text{upper}}_{1.4M_{\odot}}$ is  at least 100  m larger.
All the values obtained are 
compatible with $R=11.9\pm 1.4$ km (at the 90\%
credible level) of the LIGO/Virgo collaborations \citep{Abbott18}, and
also with maximum value $R_{1.4M_{\odot}}=13.6$ km reported in \cite{Annala18}, where a generic family of EoS interpolating between chiral effective field theory results (low densities) and perturbative QCD (high densities) was used. 
Our results are also consistent with a mean value of $R_{1.4M_{\odot}}=12.39$ km and a $2\sigma$ confidence of $12.00<R_{1.4M_{\odot}}/\text{km}<13.45$ in \cite{Most18},
where a piecewise polytrope parametrization of the EoS, which took
into account nuclear matter calculations of the outer crust  and close
to  the
saturation density, as well as perturbative QCD at very high densities.

  Considering the lower
bound defined by $\tilde{\Lambda}=300$,  we obtain
$R^{\text{lower}}_{1.4M_{\odot}}> 11.39$  km.
 In this limit the three crusts predict
slightly different radii, with SLy4 predicting the smallest values and
DDHd the largest ones.

\section{Conclusions}
\label{sec:conclusions}
We have analyzed the impact of the crust EoS on several
empirical relations. We have used three sets of EoS in our analyzes, each set with a
different inner crust. The core EoS was determined from a Taylor expansion
of the energy functional around
the saturation density, $n_{0}$, until fourth order, and different models were generated 
through random sampling of the empirical parameters via a multivariate Gaussian with zero
covariance.  Thermodynamic stability, causality, a maximum star mass of
1.97 $M_\odot$ and a positive symmetry energy were
the conditions imposed to validate the generated EoS.
 For the crust EoS, we have considered the crusts determined within the models SLy4, NL3
and DDHd, characterized by a quite different density dependence of the
symmetry energy at sub-saturation densities.  
The matching of the crust
was imposed to occur below 0.1 fm$^{-3}$ and it was required that the crust
and the core pressures are the same at the transition
baryonic chemical potential, i.e., a first order phase transition was
considered.  This matching procedure gives  an EoS that is
thermodynamically consistent.   
While other matching procedures are frequently used, and although we left for future work
a complete study that considers both the crust EoS and the matching procedure, 
preliminary results show that the matching procedure only has a very weak impact. 
   The three sets of EoS satisfy the
constraints obtained from the analysis of the GW170817 event \cite{Abbott:2018wiz}.

First, the relations between the tidal deformability
$\Lambda$ and compactness $C$, Love number $k_2$ and radius $R_{M_i}$
were studied. 
It was shown that $k\propto C^{\beta}$ with $\beta\approx -1$ for
$1.1\leq M/M_{\odot}\leq1.6$, where the crust shows an impact of the
order of $11\%$. The same behavior is found for $k_2(M)$, i.e.,
$k_2\propto1/M$, but with a much smaller impact of the crust,
$4\%$. As a consequence, it was established a correlation of almost
one between the compactness $C$ 
and  $\Lambda$ via the dependence $\Lambda\propto C^\beta$ with $\beta=-6.122$, $-6.025$, and $-6.137$ for the NL3, DDHd, and SLy4 crusts, respectively. The impact of the crust is only of $2\%$. 
A similar dependency is seen for $\Lambda(M)$ but with a lower
correlation value. We have, therefore, confirmed with our EoS sets the results presented in \cite{De:2018uhw} concerning the relations
$\Lambda\sim C^{-6}$  and $\Lambda\sim M^{-6}$, and shown that the
impact of the crust is small. It was also shown that, for a fixed NS mass $M_i$,   the relation
$\Lambda\sim R^{\beta}$ depends on the crust and  on the NS
mass.  The correlation becomes stronger as the NS mass increases, and
it is
 almost exact and crust independent for massive NS: a correlation of
 $0.99$ for both NL3 and SLy4 crusts and $0.97$ for DDHd was obtained for
 $M\geq1.8M_{\odot}$ with $\beta \sim 8-9$.

The effective tidal deformability  $\tilde\Lambda$ is directly
determined from the  gravitational-wave signal of a neutron star
merger, and, therefore it is important to identify possible
correlations between this quantity and NS properties.
We have found a perfect power relation between $M_{\text{chirp}}$ and $\tilde{\Lambda}$.
The dependence $M_{\text{chirp}}\propto \tilde{\Lambda}^\beta$ has
$\beta=-5.824$, $-5.910$, and $-5.781$ for, the NL3, DDHd, and SLy4 crusts,
respectively. The crust impact is around $2\%$. We have also analyzed
how $\Lambda_{M_i}$ can be predicted from the effective binary tidal
deformability $\tilde{\Lambda}$ for $M_{\text{chirp}}=1.186M_{\odot}$,
the value obtained for the GW170817 event. We first studied the
$M_i=1.4M_{\odot}$ case, showing an almost perfect correlation and
crust independence between $\Lambda_{1.4M_{\odot}}$ and
$\tilde{\Lambda}$. The relation
$\Lambda_{1.4M_{\odot}}=\beta\tilde{\Lambda}+\alpha$ shows an
overall uncertainty below $5\%$ for $q=0.947$ and $10\%$ for lower $q$
values. This result is interesting because even if a 1.4$M_\odot$ NS
is not part of the binary, it is still possible to determine its tidal
deformability, $\Lambda_{1.4 M\odot}$ , with an accuracy of at least
$\approx 10\%$.

In \cite{Raithel2018,Raithel:2019uzi},  the authors have obtained a high
correlation between  $\tilde{\Lambda}$ and the radius of the $M_1$
star, $R_1$.  We have confirmed the validity of the empirical
relation $\tilde{\Lambda}\propto R_1^{\beta}$ for $0.73<M_1/M_2<1.0$,
however, we could also verify that the crust shows an impact of about
$14\%$. The values $\beta=6.554$, $5.729$, and $6.553$  have been
determined for the NL3, DDHd, and SLy4 crusts, respectively.  

We have explored the possibility of constraining the upper limits of
both $\Lambda_{M_i}$ and $R_{M_i}$ from the LIGO/Virgo upper bound
$\tilde{\Lambda}=720$ \cite{Abbott:2018wiz}. For a $1.4M_{\odot}$
NS, the constraints obtained from
$\Lambda_{1.4M_{\odot}}(q,\tilde{\Lambda}=720)$ are in between $636$
at $q=0.74$ and $617$ at $q=0.99$ for the NL3 crust; similar values were found 
 for the SLY4 crust and slightly smaller values for the DDHd crust.
Imposing $\tilde{\Lambda}=720$, we have obtained
$12.70<R^{\text{upper}}_{1.4M_{\odot}}<12.78$ km as an upper bound
for the NL3 and SLy4 crusts, while DDHd shows values $\approx 1\%$ larger. 
If, besides, we consider $300$ as a lower bound on $\tilde{\Lambda}$ for
the GW170817 event, as determined from the
analysis of the electromagnetic counterpart in \cite{Radice2019}, a
lower limit of $11.40<R^{\text{lower}}_{1.4M_{\odot}}<11.48$ km
was established for the NL3 and SLy4 crusts, and  $\approx 1\%$ larger for the DDHd crust.

We confirm the very strong correlation between
$\tilde\Lambda$ and $M_{\text{chirp}}$ obtained in \cite{De:2018uhw}.
For our EoS sets, the dependence $\tilde{\Lambda}\sim
M_{\text{chirp}}^{\beta}$ was obtained, with $-5.92<\beta<-5.78$ very
close to $-6$.
In  \cite{De:2018uhw}, it was proposed that  $\Lambda_1/\Lambda_2=q^a$,
with $a=6$
since $R_1\approx R_2$. Allowing the chirp mass to vary in a
reasonable range taking into account the event GW170817, we have
obtained $5.25<a<6.91$ which give rise to   larger  values  of the
ratio $ \Lambda_1/\Lambda_2$,  up to  50\%  (15\%)  larger at the
lower (upper) limit than when compared to the results obtained in
\cite{De:2018uhw}.
Besides, we have shown that this correlation  depends on the chirp mass $M_{\text{chirp}}$  and that it
can be described  to a very good
approximation as $\Lambda_1/\Lambda_2=q^a$, with $a\sim
\sqrt{M_{\text{chirp}}}$, for $1.0\le M_{\text{chirp}}/M_{\odot}\le1.3$.  \\ 

{\bf Acknowledgments}:
This work was partly supported by 
 Fundação  para  a Ciência e Tecnologia,  Portugal,
under the projects UID/FIS/04564/2016 and POCI-01-0145-FEDER-029912
with financial support from POCI, in its FEDER component,
and by the FCT/MCTES budget through national funds  (OE),  and by PHAROS COST Action
CA16214. \\

\appendix

\section{Predictions for $\Lambda_{M_i}(q,\tilde{\Lambda})$ and $R_{M_i}(q,\tilde{\Lambda})$}
\label{appendix:a}
Herein, we show the predictions for $\Lambda_{M_i}(q,\tilde{\Lambda})$ and $R_{M_i}(q,\tilde{\Lambda})$, using 
the correlation analysis $\text{Corr}[\Lambda_{M_i},\tilde{\Lambda}]$ and $\text{Corr}[R_{M_i},\tilde{\Lambda}]$.
Tables \ref{tab:constraints_Lambdas} and \ref{tab:constraints_Rs} contain the $\Lambda_{M_i}$ and $R_{M_i}$ bounds, 
respectively, using  $\tilde{\Lambda}=300$ and $720$ as constraints.

\begin{table*}[ht]
\begin{tabular}{cccccccc}
 \multicolumn{8}{l}{Bound: $\tilde{\Lambda}=720$}\\
  \hline
 Crust&$q$ & $\Lambda_{1.1M_{\odot}}$ & $\Lambda_{1.2M_{\odot}}$ & $\Lambda_{1.3M_{\odot}}$  & $\Lambda_{1.4M_{\odot}}$  & $\Lambda_{1.5M_{\odot}}$  & $\Lambda_{1.6M_{\odot}}$  \\ 
  \hline
NL3  & $0.74$ & $2331 \pm 4$ & $1480 \pm 2$ & $962 \pm 1$ & $636 \pm 0$ & $427 \pm 1$ & $289 \pm 1$ \\ 
NL3  & $0.99$ & $2268 \pm 4$ & $1438 \pm 2$ & $934 \pm 1$ & $617 \pm 0$ & $413 \pm 1$ & $279 \pm 1$ \\ 
  \hline
SLy4  & $0.74$ & $2321 \pm 6$ & $1477 \pm 3$ & $963 \pm 1$ & $639 \pm 1$ & $429 \pm 1$ & $291 \pm 1$ \\ 
SLy4  & $0.99$ & $2251 \pm 6$ & $1432 \pm 3$ & $932 \pm 1$ & $618 \pm 0$ & $414 \pm 1$ & $280 \pm 1$ \\ 
  \hline
DDHd  & $0.74$ & $2396 \pm 7$ & $1500 \pm 3$ & $962 \pm 1$ & $628 \pm 1$ & $416 \pm 1$ & $278 \pm 1$ \\ 
DDHd  & $0.99$ & $2340 \pm 7$ & $1465 \pm 3$ & $940 \pm 1$ & $614 \pm 1$ & $407 \pm 1$ & $272 \pm 1$ \\ 
   \hline\\
    \multicolumn{8}{l}{Bound: $\tilde{\Lambda}=300$}\\
  \hline
 Crust&$q$ & $\Lambda_{1.1M_{\odot}}$ & $\Lambda_{1.2M_{\odot}}$ & $\Lambda_{1.3M_{\odot}}$  & $\Lambda_{1.4M_{\odot}}$  & $\Lambda_{1.5M_{\odot}}$  & $\Lambda_{1.6M_{\odot}}$  \\ 
  \hline
  NL3 & $0.74$& $1175\pm 6$ & $691 \pm 2$ & $410 \pm 1$ & $244 \pm 1$ & $144 \pm 1$ & $83 \pm 1$ \\ 
  NL3 & $0.99$ & $1192\pm 6$ & $701 \pm 2$ & $415 \pm 1$ & $247 \pm 1$ & $146 \pm 1$ & $84 \pm 1$ \\ 
   \hline
  SLy4 & $0.74$ & $1143 \pm 8$ & $681 \pm 3$ & $410 \pm 1$ & $247 \pm 1$ & $149 \pm 1$ & $88 \pm 2$ \\ 
  SLy4 & $0.99$ & $1156 \pm 9$ & $687 \pm 4$ & $412 \pm 1$ & $248 \pm 1$ & $149 \pm 1$ & $88 \pm 1$ \\ 
   \hline
  DDHd & $0.74$ & $1187 \pm 13$ & $692 \pm 5$ & $409 \pm 2$ & $242 \pm 2$ & $142 \pm 2$ & $82 \pm 2$ \\ 
  DDHd & $0.99$ & $1228 \pm 15$ & $712 \pm 6$ & $419 \pm 2$ & $247 \pm 1$ & $145 \pm 2$ & $83 \pm 2$ \\ 
   \hline
   \end{tabular}
\caption{Predictions for $\Lambda_{M_i}(q,\tilde{\Lambda})$ for
  $q=0.74$ and $0.99$ using $\tilde{\Lambda}=720$ and $300$. The uncertainties define the  95\% confidence interval region.}
\label{tab:constraints_Lambdas}
\end{table*}

\begin{table*}[ht]
\centering
\begin{tabular}{cccccccc}
 \multicolumn{8}{l}{Radius Upper Bound: $\tilde{\Lambda}=720$}\\
  \hline
 Crust&$q$ & $R_{1.1M_{\odot}}$ & $R_{1.2M_{\odot}}$ & $R_{1.3M_{\odot}}$  & $R_{1.4M_{\odot}}$  & $R_{1.5M_{\odot}}$  & $R_{1.6M_{\odot}}$  \\ 
  \hline
  NL3 & $0.74$ & $12.53 \pm 0.01$ & $12.62 \pm 0.01$ & $12.70 \pm 0.01$ & $12.78 \pm 0.00$ & $12.84 \pm 0.00$ & $12.90 \pm 0.00$ \\ 
  NL3 & $0.99$ & $12.47 \pm 0.01$ & $12.56 \pm 0.01$ & $12.64 \pm 0.01$ & $12.71 \pm 0.00$ & $12.77 \pm 0.00$ & $12.82 \pm 0.00$ \\ 
  \hline
  SLy4 & $0.74$ & $12.51 \pm 0.01$ & $12.61 \pm 0.01$ & $12.70 \pm 0.01$ & $12.78 \pm 0.01$ & $12.85 \pm 0.01$ & $12.90 \pm 0.01$ \\ 
  SLy4 & $0.99$ & $12.44 \pm 0.01$ & $12.54 \pm 0.01$ & $12.62 \pm 0.01$ & $12.70 \pm 0.01$ & $12.76 \pm 0.01$ & $12.82 \pm 0.01$ \\ 
    \hline
  DDHd & $0.74$ & $12.78 \pm 0.02$ & $12.83 \pm 0.02$ & $12.88 \pm 0.01$ & $12.91 \pm 0.01$ & $12.94 \pm 0.01$ & $12.96 \pm 0.01$ \\  
  DDHd & $0.99$ & $12.72 \pm 0.02$ & $12.77 \pm 0.02$ & $12.81 \pm 0.01$ & $12.85 \pm 0.01$ & $12.88 \pm 0.01$ & $12.90 \pm 0.01$ \\  
   \hline\\
    \multicolumn{8}{l}{Radius Lower Bound: $\tilde{\Lambda}=300$}\\
  \hline
 Crust&$q$ & $R_{1.1M_{\odot}}$ & $R_{1.2M_{\odot}}$ & $R_{1.3M_{\odot}}$  & $R_{1.4M_{\odot}}$  & $R_{1.5M_{\odot}}$  & $R_{1.6M_{\odot}}$  \\  
  \hline
  NL3 & $0.74$ & $11.53 \pm 0.01$ & $11.52 \pm 0.01$ & $11.50 \pm 0.01$ & $11.46 \pm 0.01$ & $11.40 \pm 0.01$ & $11.32 \pm 0.01$ \\ 
  NL3 & $0.99$ & $11.55 \pm 0.01$ & $11.54 \pm 0.01$ & $11.52 \pm 0.01$ & $11.48 \pm 0.01$ & $11.42 \pm 0.01$ & $11.33 \pm 0.01$ \\ 
   \hline
  SLy4 & $0.74$ & $11.42\pm 0.02$ & $11.43 \pm 0.01$ & $11.43 \pm 0.01$ & $11.40 \pm 0.01$ & $11.36 \pm 0.01$ & $11.29 \pm 0.01$ \\ 
  SLy4 & $0.99$ & $11.44\pm 0.02$ & $11.45 \pm 0.01$ & $11.44 \pm 0.01$ & $11.41 \pm 0.01$ & $11.37 \pm 0.01$ & $11.30 \pm 0.01$ \\ 
   \hline
  DDHd & $0.74$ & $11.67 \pm 0.03$ & $11.64 \pm 0.03$ & $11.60 \pm 0.02$ & $11.54 \pm 0.02$ & $11.47 \pm 0.02$ & $11.38 \pm 0.02$ \\ 
  DDHd & $0.99$ & $11.74 \pm 0.04$ & $11.70 \pm 0.03$ & $11.65 \pm 0.02$ & $11.59 \pm 0.02$ & $11.51 \pm 0.02$ & $11.41 \pm 0.02$ \\ 
   \hline
\end{tabular}
\caption{Predictions for $R_{M_i}(q,\tilde{\Lambda})$ (in km) for
  $q=0.74$ and $0.99$ using $\tilde{\Lambda}=720$ and
  $300$. The uncertainties define the  95\% confidence interval
  region.}
\label{tab:constraints_Rs}
\end{table*}


\begin{thebibliography}{76}%
	\makeatletter
	\providecommand \@ifxundefined [1]{%
		\@ifx{#1\undefined}
	}%
	\providecommand \@ifnum [1]{%
		\ifnum #1\expandafter \@firstoftwo
		\else \expandafter \@secondoftwo
		\fi
	}%
	\providecommand \@ifx [1]{%
		\ifx #1\expandafter \@firstoftwo
		\else \expandafter \@secondoftwo
		\fi
	}%
	\providecommand \natexlab [1]{#1}%
	\providecommand \enquote  [1]{``#1''}%
	\providecommand \bibnamefont  [1]{#1}%
	\providecommand \bibfnamefont [1]{#1}%
	\providecommand \citenamefont [1]{#1}%
	\providecommand \href@noop [0]{\@secondoftwo}%
	\providecommand \href [0]{\begingroup \@sanitize@url \@href}%
	\providecommand \@href[1]{\@@startlink{#1}\@@href}%
	\providecommand \@@href[1]{\endgroup#1\@@endlink}%
	\providecommand \@sanitize@url [0]{\catcode `\\12\catcode `\$12\catcode
		`\&12\catcode `\#12\catcode `\^12\catcode `\_12\catcode `\%12\relax}%
	\providecommand \@@startlink[1]{}%
	\providecommand \@@endlink[0]{}%
	\providecommand \url  [0]{\begingroup\@sanitize@url \@url }%
	\providecommand \@url [1]{\endgroup\@href {#1}{\urlprefix }}%
	\providecommand \urlprefix  [0]{URL }%
	\providecommand \Eprint [0]{\href }%
	\providecommand \doibase [0]{http://dx.doi.org/}%
	\providecommand \selectlanguage [0]{\@gobble}%
	\providecommand \bibinfo  [0]{\@secondoftwo}%
	\providecommand \bibfield  [0]{\@secondoftwo}%
	\providecommand \translation [1]{[#1]}%
	\providecommand \BibitemOpen [0]{}%
	\providecommand \bibitemStop [0]{}%
	\providecommand \bibitemNoStop [0]{.\EOS\space}%
	\providecommand \EOS [0]{\spacefactor3000\relax}%
	\providecommand \BibitemShut  [1]{\csname bibitem#1\endcsname}%
	\let\auto@bib@innerbib\@empty
	\bibitem [{\citenamefont {Arzoumanian}\ \emph {et~al.}(2018)\citenamefont
		{Arzoumanian} \emph {et~al.}}]{Arzoumanian2017}%
	\BibitemOpen
	\bibfield  {author} {\bibinfo {author} {\bibfnamefont {Z.}~\bibnamefont
			{Arzoumanian}} \emph {et~al.} (\bibinfo {collaboration} {NANOGrav}),\ }\href
	{\doibase 10.3847/1538-4365/aab5b0} {\bibfield  {journal} {\bibinfo
			{journal} {Astrophys. J. Suppl.}\ }\textbf {\bibinfo {volume} {235}},\
		\bibinfo {pages} {37} (\bibinfo {year} {2018})},\ \Eprint
	{http://arxiv.org/abs/1801.01837} {arXiv:1801.01837 [astro-ph.HE]}
	\BibitemShut {NoStop}%
	\bibitem [{\citenamefont {Fonseca}\ \emph {et~al.}(2016)\citenamefont {Fonseca}
		\emph {et~al.}}]{Fonseca2017}%
	\BibitemOpen
	\bibfield  {author} {\bibinfo {author} {\bibfnamefont {E.}~\bibnamefont
			{Fonseca}} \emph {et~al.},\ }\href {\doibase 10.3847/0004-637X/832/2/167}
	{\bibfield  {journal} {\bibinfo  {journal} {Astrophys. J.}\ }\textbf
		{\bibinfo {volume} {832}},\ \bibinfo {pages} {167} (\bibinfo {year}
		{2016})},\ \Eprint {http://arxiv.org/abs/1603.00545} {arXiv:1603.00545
		[astro-ph.HE]} \BibitemShut {NoStop}%
	\bibitem [{\citenamefont {Demorest}\ \emph {et~al.}(2010)\citenamefont
		{Demorest}, \citenamefont {Pennucci}, \citenamefont {Ransom}, \citenamefont
		{Roberts},\ and\ \citenamefont {Hessels}}]{Demorest2010}%
	\BibitemOpen
	\bibfield  {author} {\bibinfo {author} {\bibfnamefont {P.}~\bibnamefont
			{Demorest}}, \bibinfo {author} {\bibfnamefont {T.}~\bibnamefont {Pennucci}},
		\bibinfo {author} {\bibfnamefont {S.}~\bibnamefont {Ransom}}, \bibinfo
		{author} {\bibfnamefont {M.}~\bibnamefont {Roberts}}, \ and\ \bibinfo
		{author} {\bibfnamefont {J.}~\bibnamefont {Hessels}},\ }\href {\doibase
		10.1038/nature09466} {\bibfield  {journal} {\bibinfo  {journal} {Nature}\
		}\textbf {\bibinfo {volume} {467}},\ \bibinfo {pages} {1081} (\bibinfo {year}
		{2010})}\BibitemShut {NoStop}%
	\bibitem [{\citenamefont {{Antoniadis}}\ \emph {et~al.}(2013)\citenamefont
		{{Antoniadis}}, \citenamefont {{Freire}}, \citenamefont {{Wex}},
		\citenamefont {{Tauris}}, \citenamefont {{Lynch}}, \citenamefont {{van
				Kerkwijk}}, \citenamefont {{Kramer}}, \citenamefont {{Bassa}}, \citenamefont
		{{Dhillon}}, \citenamefont {{Driebe}}, \citenamefont {{Hessels}},
		\citenamefont {{Kaspi}}, \citenamefont {{Kondratiev}}, \citenamefont
		{{Langer}}, \citenamefont {{Marsh}}, \citenamefont {{McLaughlin}},
		\citenamefont {{Pennucci}}, \citenamefont {{Ransom}}, \citenamefont
		{{Stairs}}, \citenamefont {{van Leeuwen}}, \citenamefont {{Verbiest}},\ and\
		\citenamefont {{Whelan}}}]{Antoniadis2013}%
	\BibitemOpen
	\bibfield  {author} {\bibinfo {author} {\bibfnamefont {J.}~\bibnamefont
			{{Antoniadis}}}, \bibinfo {author} {\bibfnamefont {P.~C.~C.}\ \bibnamefont
			{{Freire}}}, \bibinfo {author} {\bibfnamefont {N.}~\bibnamefont {{Wex}}},
		\bibinfo {author} {\bibfnamefont {T.~M.}\ \bibnamefont {{Tauris}}}, \bibinfo
		{author} {\bibfnamefont {R.~S.}\ \bibnamefont {{Lynch}}}, \bibinfo {author}
		{\bibfnamefont {M.~H.}\ \bibnamefont {{van Kerkwijk}}}, \bibinfo {author}
		{\bibfnamefont {M.}~\bibnamefont {{Kramer}}}, \bibinfo {author}
		{\bibfnamefont {C.}~\bibnamefont {{Bassa}}}, \bibinfo {author} {\bibfnamefont
			{V.~S.}\ \bibnamefont {{Dhillon}}}, \bibinfo {author} {\bibfnamefont
			{T.}~\bibnamefont {{Driebe}}}, \bibinfo {author} {\bibfnamefont {J.~W.~T.}\
			\bibnamefont {{Hessels}}}, \bibinfo {author} {\bibfnamefont {V.~M.}\
			\bibnamefont {{Kaspi}}}, \bibinfo {author} {\bibfnamefont {V.~I.}\
			\bibnamefont {{Kondratiev}}}, \bibinfo {author} {\bibfnamefont
			{N.}~\bibnamefont {{Langer}}}, \bibinfo {author} {\bibfnamefont {T.~R.}\
			\bibnamefont {{Marsh}}}, \bibinfo {author} {\bibfnamefont {M.~A.}\
			\bibnamefont {{McLaughlin}}}, \bibinfo {author} {\bibfnamefont {T.~T.}\
			\bibnamefont {{Pennucci}}}, \bibinfo {author} {\bibfnamefont {S.~M.}\
			\bibnamefont {{Ransom}}}, \bibinfo {author} {\bibfnamefont {I.~H.}\
			\bibnamefont {{Stairs}}}, \bibinfo {author} {\bibfnamefont {J.}~\bibnamefont
			{{van Leeuwen}}}, \bibinfo {author} {\bibfnamefont {J.~P.~W.}\ \bibnamefont
			{{Verbiest}}}, \ and\ \bibinfo {author} {\bibfnamefont {D.~G.}\ \bibnamefont
			{{Whelan}}},\ }\href {\doibase 10.1126/science.1233232} {\bibfield  {journal}
		{\bibinfo  {journal} {Science}\ }\textbf {\bibinfo {volume} {340}},\ \bibinfo
		{pages} {448} (\bibinfo {year} {2013})}\BibitemShut {NoStop}%
	\bibitem [{\citenamefont {Cromartie}\ \emph {et~al.}(2019)\citenamefont
		{Cromartie} \emph {et~al.}}]{Cromartie2019}%
	\BibitemOpen
	\bibfield  {author} {\bibinfo {author} {\bibfnamefont {H.~T.}\ \bibnamefont
			{Cromartie}} \emph {et~al.},\ }\href {\doibase 10.1038/s41550-019-0880-2} {\
		(\bibinfo {year} {2019}),\ 10.1038/s41550-019-0880-2},\ \Eprint
	{http://arxiv.org/abs/1904.06759} {arXiv:1904.06759 [astro-ph.HE]}
	\BibitemShut {NoStop}%
	\bibitem [{\citenamefont {{Arzoumanian}}\ \emph {et~al.}(2014)\citenamefont
		{{Arzoumanian}}, \citenamefont {{Gendreau}}, \citenamefont {{Baker}},
		\citenamefont {{Cazeau}}, \citenamefont {{Hestnes}}, \citenamefont
		{{Kellogg}}, \citenamefont {{Kenyon}}, \citenamefont {{Kozon}}, \citenamefont
		{{Liu}}, \citenamefont {{Manthripragada}}, \citenamefont {{Markwardt}},
		\citenamefont {{Mitchell}}, \citenamefont {{Mitchell}}, \citenamefont
		{{Monroe}}, \citenamefont {{Okajima}}, \citenamefont {{Pollard}},
		\citenamefont {{Powers}}, \citenamefont {{Savadkin}}, \citenamefont
		{{Winternitz}}, \citenamefont {{Chen}}, \citenamefont {{Wright}},
		\citenamefont {{Foster}}, \citenamefont {{Prigozhin}}, \citenamefont
		{{Remillard}},\ and\ \citenamefont {{Doty}}}]{NICER}%
	\BibitemOpen
	\bibfield  {author} {\bibinfo {author} {\bibfnamefont {Z.}~\bibnamefont
			{{Arzoumanian}}}, \bibinfo {author} {\bibfnamefont {K.~C.}\ \bibnamefont
			{{Gendreau}}}, \bibinfo {author} {\bibfnamefont {C.~L.}\ \bibnamefont
			{{Baker}}}, \bibinfo {author} {\bibfnamefont {T.}~\bibnamefont {{Cazeau}}},
		\bibinfo {author} {\bibfnamefont {P.}~\bibnamefont {{Hestnes}}}, \bibinfo
		{author} {\bibfnamefont {J.~W.}\ \bibnamefont {{Kellogg}}}, \bibinfo {author}
		{\bibfnamefont {S.~J.}\ \bibnamefont {{Kenyon}}}, \bibinfo {author}
		{\bibfnamefont {R.~P.}\ \bibnamefont {{Kozon}}}, \bibinfo {author}
		{\bibfnamefont {K.~C.}\ \bibnamefont {{Liu}}}, \bibinfo {author}
		{\bibfnamefont {S.~S.}\ \bibnamefont {{Manthripragada}}}, \bibinfo {author}
		{\bibfnamefont {C.~B.}\ \bibnamefont {{Markwardt}}}, \bibinfo {author}
		{\bibfnamefont {A.~L.}\ \bibnamefont {{Mitchell}}}, \bibinfo {author}
		{\bibfnamefont {J.~W.}\ \bibnamefont {{Mitchell}}}, \bibinfo {author}
		{\bibfnamefont {C.~A.}\ \bibnamefont {{Monroe}}}, \bibinfo {author}
		{\bibfnamefont {T.}~\bibnamefont {{Okajima}}}, \bibinfo {author}
		{\bibfnamefont {S.~E.}\ \bibnamefont {{Pollard}}}, \bibinfo {author}
		{\bibfnamefont {D.~F.}\ \bibnamefont {{Powers}}}, \bibinfo {author}
		{\bibfnamefont {B.~J.}\ \bibnamefont {{Savadkin}}}, \bibinfo {author}
		{\bibfnamefont {L.~B.}\ \bibnamefont {{Winternitz}}}, \bibinfo {author}
		{\bibfnamefont {P.~T.}\ \bibnamefont {{Chen}}}, \bibinfo {author}
		{\bibfnamefont {M.~R.}\ \bibnamefont {{Wright}}}, \bibinfo {author}
		{\bibfnamefont {R.}~\bibnamefont {{Foster}}}, \bibinfo {author}
		{\bibfnamefont {G.}~\bibnamefont {{Prigozhin}}}, \bibinfo {author}
		{\bibfnamefont {R.}~\bibnamefont {{Remillard}}}, \ and\ \bibinfo {author}
		{\bibfnamefont {J.}~\bibnamefont {{Doty}}},\ }\enquote {\bibinfo {title}
		{{The neutron star interior composition explorer (NICER): mission
				definition}},}\ in\ \href {\doibase 10.1117/12.2056811} {\emph {\bibinfo
			{booktitle} {Proceedings of the SPIE, Volume 9144, id. 914420 9 pp.
				(2014).}}},\ \bibinfo {series} {Society of Photo-Optical Instrumentation
		Engineers (SPIE) Conference Series}, Vol.\ \bibinfo {volume} {9144}\
	(\bibinfo {year} {2014})\ p.\ \bibinfo {pages} {914420}\BibitemShut {NoStop}%
	\bibitem [{\citenamefont {{Motch}}\ \emph {et~al.}(2013)\citenamefont
		{{Motch}}, \citenamefont {{Wilms}}, \citenamefont {{Barret}}, \citenamefont
		{{Becker}}, \citenamefont {{Bogdanov}}, \citenamefont {{Boirin}},
		\citenamefont {{Corbel}}, \citenamefont {{Cackett}}, \citenamefont
		{{Campana}}, \citenamefont {{de Martino}}, \citenamefont {{Haberl}},
		\citenamefont {{in't Zand}}, \citenamefont {{M{\'e}ndez}}, \citenamefont
		{{Mignani}}, \citenamefont {{Miller}}, \citenamefont {{Orio}}, \citenamefont
		{{Psaltis}}, \citenamefont {{Rea}}, \citenamefont {{Rodriguez}},
		\citenamefont {{Rozanska}}, \citenamefont {{Schwope}}, \citenamefont
		{{Steiner}}, \citenamefont {{Webb}}, \citenamefont {{Zampieri}},\ and\
		\citenamefont {{Zane}}}]{Athena}%
	\BibitemOpen
	\bibfield  {author} {\bibinfo {author} {\bibfnamefont {C.}~\bibnamefont
			{{Motch}}}, \bibinfo {author} {\bibfnamefont {J.}~\bibnamefont {{Wilms}}},
		\bibinfo {author} {\bibfnamefont {D.}~\bibnamefont {{Barret}}}, \bibinfo
		{author} {\bibfnamefont {W.}~\bibnamefont {{Becker}}}, \bibinfo {author}
		{\bibfnamefont {S.}~\bibnamefont {{Bogdanov}}}, \bibinfo {author}
		{\bibfnamefont {L.}~\bibnamefont {{Boirin}}}, \bibinfo {author}
		{\bibfnamefont {S.}~\bibnamefont {{Corbel}}}, \bibinfo {author}
		{\bibfnamefont {E.}~\bibnamefont {{Cackett}}}, \bibinfo {author}
		{\bibfnamefont {S.}~\bibnamefont {{Campana}}}, \bibinfo {author}
		{\bibfnamefont {D.}~\bibnamefont {{de Martino}}}, \bibinfo {author}
		{\bibfnamefont {F.}~\bibnamefont {{Haberl}}}, \bibinfo {author}
		{\bibfnamefont {J.}~\bibnamefont {{in't Zand}}}, \bibinfo {author}
		{\bibfnamefont {M.}~\bibnamefont {{M{\'e}ndez}}}, \bibinfo {author}
		{\bibfnamefont {R.}~\bibnamefont {{Mignani}}}, \bibinfo {author}
		{\bibfnamefont {J.}~\bibnamefont {{Miller}}}, \bibinfo {author}
		{\bibfnamefont {M.}~\bibnamefont {{Orio}}}, \bibinfo {author} {\bibfnamefont
			{D.}~\bibnamefont {{Psaltis}}}, \bibinfo {author} {\bibfnamefont
			{N.}~\bibnamefont {{Rea}}}, \bibinfo {author} {\bibfnamefont
			{J.}~\bibnamefont {{Rodriguez}}}, \bibinfo {author} {\bibfnamefont
			{A.}~\bibnamefont {{Rozanska}}}, \bibinfo {author} {\bibfnamefont
			{A.}~\bibnamefont {{Schwope}}}, \bibinfo {author} {\bibfnamefont
			{A.}~\bibnamefont {{Steiner}}}, \bibinfo {author} {\bibfnamefont
			{N.}~\bibnamefont {{Webb}}}, \bibinfo {author} {\bibfnamefont
			{L.}~\bibnamefont {{Zampieri}}}, \ and\ \bibinfo {author} {\bibfnamefont
			{S.}~\bibnamefont {{Zane}}},\ }\href@noop {} {\bibfield  {journal} {\bibinfo
			{journal} {arXiv e-prints}\ ,\ \bibinfo {eid} {arXiv:1306.2334}} (\bibinfo
		{year} {2013})},\ \Eprint {http://arxiv.org/abs/1306.2334} {arXiv:1306.2334
		[astro-ph.HE]} \BibitemShut {NoStop}%
	\bibitem [{\citenamefont {{Watts}}\ \emph {et~al.}(2019)\citenamefont
		{{Watts}}, \citenamefont {{Yu}}, \citenamefont {{Poutanen}}, \citenamefont
		{{Zhang}}, \citenamefont {{Bhattacharyya}}, \citenamefont {{Bogdanov}},
		\citenamefont {{Ji}}, \citenamefont {{Patruno}}, \citenamefont {{Riley}},
		\citenamefont {{Bakala}}, \citenamefont {{Baykal}}, \citenamefont
		{{Bernardini}}, \citenamefont {{Bombaci}}, \citenamefont {{Brown}},
		\citenamefont {{Cavecchi}}, \citenamefont {{Chakrabarty}}, \citenamefont
		{{Chenevez}}, \citenamefont {{Degenaar}}, \citenamefont {{Del Santo}},
		\citenamefont {{Di Salvo}}, \citenamefont {{Doroshenko}}, \citenamefont
		{{Falanga}}, \citenamefont {{Ferdman}}, \citenamefont {{Feroci}},
		\citenamefont {{Gambino}}, \citenamefont {{Ge}}, \citenamefont {{Greif}},
		\citenamefont {{Guillot}}, \citenamefont {{Gungor}}, \citenamefont
		{{Hartmann}}, \citenamefont {{Hebeler}}, \citenamefont {{Heger}},
		\citenamefont {{Homan}}, \citenamefont {{Iaria}}, \citenamefont {{Zand}},
		\citenamefont {{Kargaltsev}}, \citenamefont {{Kurkela}}, \citenamefont
		{{Lai}}, \citenamefont {{Li}}, \citenamefont {{Li}}, \citenamefont {{Li}},
		\citenamefont {{Linares}}, \citenamefont {{Lu}}, \citenamefont
		{{Mahmoodifar}}, \citenamefont {{M{\'e}ndez}}, \citenamefont {{Coleman
				Miller}}, \citenamefont {{Morsink}}, \citenamefont {{N{\"a}ttil{\"a}}},
		\citenamefont {{Possenti}}, \citenamefont {{Prescod-Weinstein}},
		\citenamefont {{Qu}}, \citenamefont {{Riggio}}, \citenamefont {{Salmi}},
		\citenamefont {{Sanna}}, \citenamefont {{Santangelo}}, \citenamefont
		{{Schatz}}, \citenamefont {{Schwenk}}, \citenamefont {{Song}}, \citenamefont
		{{{\v{S}}r{\'a}mkov{\'a}}}, \citenamefont {{Stappers}}, \citenamefont
		{{Stiele}}, \citenamefont {{Strohmayer}}, \citenamefont {{Tews}},
		\citenamefont {{Tolos}}, \citenamefont {{T{\"o}r{\"o}k}}, \citenamefont
		{{Tsang}}, \citenamefont {{Urbanec}}, \citenamefont {{Vacchi}}, \citenamefont
		{{Xu}}, \citenamefont {{Xu}}, \citenamefont {{Zane}}, \citenamefont
		{{Zhang}}, \citenamefont {{Zhang}}, \citenamefont {{Zhang}}, \citenamefont
		{{Zheng}},\ and\ \citenamefont {{Zhou}}}]{eXTP}%
	\BibitemOpen
	\bibfield  {author} {\bibinfo {author} {\bibfnamefont {A.~L.}\ \bibnamefont
			{{Watts}}}, \bibinfo {author} {\bibfnamefont {W.}~\bibnamefont {{Yu}}},
		\bibinfo {author} {\bibfnamefont {J.}~\bibnamefont {{Poutanen}}}, \bibinfo
		{author} {\bibfnamefont {S.}~\bibnamefont {{Zhang}}}, \bibinfo {author}
		{\bibfnamefont {S.}~\bibnamefont {{Bhattacharyya}}}, \bibinfo {author}
		{\bibfnamefont {S.}~\bibnamefont {{Bogdanov}}}, \bibinfo {author}
		{\bibfnamefont {L.}~\bibnamefont {{Ji}}}, \bibinfo {author} {\bibfnamefont
			{A.}~\bibnamefont {{Patruno}}}, \bibinfo {author} {\bibfnamefont {T.~E.}\
			\bibnamefont {{Riley}}}, \bibinfo {author} {\bibfnamefont {P.}~\bibnamefont
			{{Bakala}}}, \bibinfo {author} {\bibfnamefont {A.}~\bibnamefont {{Baykal}}},
		\bibinfo {author} {\bibfnamefont {F.}~\bibnamefont {{Bernardini}}}, \bibinfo
		{author} {\bibfnamefont {I.}~\bibnamefont {{Bombaci}}}, \bibinfo {author}
		{\bibfnamefont {E.}~\bibnamefont {{Brown}}}, \bibinfo {author} {\bibfnamefont
			{Y.}~\bibnamefont {{Cavecchi}}}, \bibinfo {author} {\bibfnamefont
			{D.}~\bibnamefont {{Chakrabarty}}}, \bibinfo {author} {\bibfnamefont
			{J.}~\bibnamefont {{Chenevez}}}, \bibinfo {author} {\bibfnamefont
			{N.}~\bibnamefont {{Degenaar}}}, \bibinfo {author} {\bibfnamefont
			{M.}~\bibnamefont {{Del Santo}}}, \bibinfo {author} {\bibfnamefont
			{T.}~\bibnamefont {{Di Salvo}}}, \bibinfo {author} {\bibfnamefont
			{V.}~\bibnamefont {{Doroshenko}}}, \bibinfo {author} {\bibfnamefont
			{M.}~\bibnamefont {{Falanga}}}, \bibinfo {author} {\bibfnamefont {R.~D.}\
			\bibnamefont {{Ferdman}}}, \bibinfo {author} {\bibfnamefont {M.}~\bibnamefont
			{{Feroci}}}, \bibinfo {author} {\bibfnamefont {A.~F.}\ \bibnamefont
			{{Gambino}}}, \bibinfo {author} {\bibfnamefont {M.}~\bibnamefont {{Ge}}},
		\bibinfo {author} {\bibfnamefont {S.~K.}\ \bibnamefont {{Greif}}}, \bibinfo
		{author} {\bibfnamefont {S.}~\bibnamefont {{Guillot}}}, \bibinfo {author}
		{\bibfnamefont {C.}~\bibnamefont {{Gungor}}}, \bibinfo {author}
		{\bibfnamefont {D.~H.}\ \bibnamefont {{Hartmann}}}, \bibinfo {author}
		{\bibfnamefont {K.}~\bibnamefont {{Hebeler}}}, \bibinfo {author}
		{\bibfnamefont {A.}~\bibnamefont {{Heger}}}, \bibinfo {author} {\bibfnamefont
			{J.}~\bibnamefont {{Homan}}}, \bibinfo {author} {\bibfnamefont
			{R.}~\bibnamefont {{Iaria}}}, \bibinfo {author} {\bibfnamefont {J.~i.}\
			\bibnamefont {{Zand}}}, \bibinfo {author} {\bibfnamefont {O.}~\bibnamefont
			{{Kargaltsev}}}, \bibinfo {author} {\bibfnamefont {A.}~\bibnamefont
			{{Kurkela}}}, \bibinfo {author} {\bibfnamefont {X.}~\bibnamefont {{Lai}}},
		\bibinfo {author} {\bibfnamefont {A.}~\bibnamefont {{Li}}}, \bibinfo {author}
		{\bibfnamefont {X.}~\bibnamefont {{Li}}}, \bibinfo {author} {\bibfnamefont
			{Z.}~\bibnamefont {{Li}}}, \bibinfo {author} {\bibfnamefont {M.}~\bibnamefont
			{{Linares}}}, \bibinfo {author} {\bibfnamefont {F.}~\bibnamefont {{Lu}}},
		\bibinfo {author} {\bibfnamefont {S.}~\bibnamefont {{Mahmoodifar}}}, \bibinfo
		{author} {\bibfnamefont {M.}~\bibnamefont {{M{\'e}ndez}}}, \bibinfo {author}
		{\bibfnamefont {M.}~\bibnamefont {{Coleman Miller}}}, \bibinfo {author}
		{\bibfnamefont {S.}~\bibnamefont {{Morsink}}}, \bibinfo {author}
		{\bibfnamefont {J.}~\bibnamefont {{N{\"a}ttil{\"a}}}}, \bibinfo {author}
		{\bibfnamefont {A.}~\bibnamefont {{Possenti}}}, \bibinfo {author}
		{\bibfnamefont {C.}~\bibnamefont {{Prescod-Weinstein}}}, \bibinfo {author}
		{\bibfnamefont {J.}~\bibnamefont {{Qu}}}, \bibinfo {author} {\bibfnamefont
			{A.}~\bibnamefont {{Riggio}}}, \bibinfo {author} {\bibfnamefont
			{T.}~\bibnamefont {{Salmi}}}, \bibinfo {author} {\bibfnamefont
			{A.}~\bibnamefont {{Sanna}}}, \bibinfo {author} {\bibfnamefont
			{A.}~\bibnamefont {{Santangelo}}}, \bibinfo {author} {\bibfnamefont
			{H.}~\bibnamefont {{Schatz}}}, \bibinfo {author} {\bibfnamefont
			{A.}~\bibnamefont {{Schwenk}}}, \bibinfo {author} {\bibfnamefont
			{L.}~\bibnamefont {{Song}}}, \bibinfo {author} {\bibfnamefont
			{E.}~\bibnamefont {{{\v{S}}r{\'a}mkov{\'a}}}}, \bibinfo {author}
		{\bibfnamefont {B.}~\bibnamefont {{Stappers}}}, \bibinfo {author}
		{\bibfnamefont {H.}~\bibnamefont {{Stiele}}}, \bibinfo {author}
		{\bibfnamefont {T.}~\bibnamefont {{Strohmayer}}}, \bibinfo {author}
		{\bibfnamefont {I.}~\bibnamefont {{Tews}}}, \bibinfo {author} {\bibfnamefont
			{L.}~\bibnamefont {{Tolos}}}, \bibinfo {author} {\bibfnamefont
			{G.}~\bibnamefont {{T{\"o}r{\"o}k}}}, \bibinfo {author} {\bibfnamefont
			{D.}~\bibnamefont {{Tsang}}}, \bibinfo {author} {\bibfnamefont
			{M.}~\bibnamefont {{Urbanec}}}, \bibinfo {author} {\bibfnamefont
			{A.}~\bibnamefont {{Vacchi}}}, \bibinfo {author} {\bibfnamefont
			{R.}~\bibnamefont {{Xu}}}, \bibinfo {author} {\bibfnamefont {Y.}~\bibnamefont
			{{Xu}}}, \bibinfo {author} {\bibfnamefont {S.}~\bibnamefont {{Zane}}},
		\bibinfo {author} {\bibfnamefont {G.}~\bibnamefont {{Zhang}}}, \bibinfo
		{author} {\bibfnamefont {S.}~\bibnamefont {{Zhang}}}, \bibinfo {author}
		{\bibfnamefont {W.}~\bibnamefont {{Zhang}}}, \bibinfo {author} {\bibfnamefont
			{S.}~\bibnamefont {{Zheng}}}, \ and\ \bibinfo {author} {\bibfnamefont
			{X.}~\bibnamefont {{Zhou}}},\ }\href {\doibase 10.1007/s11433-017-9188-4}
	{\bibfield  {journal} {\bibinfo  {journal} {Science China Physics, Mechanics,
				and Astronomy}\ }\textbf {\bibinfo {volume} {62}},\ \bibinfo {eid} {29503}
		(\bibinfo {year} {2019})},\ \Eprint {http://arxiv.org/abs/1812.04021}
	{arXiv:1812.04021 [astro-ph.HE]} \BibitemShut {NoStop}%
	\bibitem [{\citenamefont {Miller}\ \emph {et~al.}(2019)\citenamefont {Miller},
		\citenamefont {Lamb}, \citenamefont {Dittmann}, \citenamefont {Bogdanov},
		\citenamefont {Arzoumanian}, \citenamefont {Gendreau}, \citenamefont
		{Guillot}, \citenamefont {Harding}, \citenamefont {Ho}, \citenamefont
		{Lattimer}, \citenamefont {Ludlam}, \citenamefont {Mahmoodifar},
		\citenamefont {Morsink}, \citenamefont {Ray}, \citenamefont {Strohmayer},
		\citenamefont {Wood}, \citenamefont {Enoto}, \citenamefont {Foster},
		\citenamefont {Okajima}, \citenamefont {Prigozhin},\ and\ \citenamefont
		{Soong}}]{Miller19}%
	\BibitemOpen
	\bibfield  {author} {\bibinfo {author} {\bibfnamefont {M.~C.}\ \bibnamefont
			{Miller}}, \bibinfo {author} {\bibfnamefont {F.~K.}\ \bibnamefont {Lamb}},
		\bibinfo {author} {\bibfnamefont {A.~J.}\ \bibnamefont {Dittmann}}, \bibinfo
		{author} {\bibfnamefont {S.}~\bibnamefont {Bogdanov}}, \bibinfo {author}
		{\bibfnamefont {Z.}~\bibnamefont {Arzoumanian}}, \bibinfo {author}
		{\bibfnamefont {K.~C.}\ \bibnamefont {Gendreau}}, \bibinfo {author}
		{\bibfnamefont {S.}~\bibnamefont {Guillot}}, \bibinfo {author} {\bibfnamefont
			{A.~K.}\ \bibnamefont {Harding}}, \bibinfo {author} {\bibfnamefont
			{W.~C.~G.}\ \bibnamefont {Ho}}, \bibinfo {author} {\bibfnamefont {J.~M.}\
			\bibnamefont {Lattimer}}, \bibinfo {author} {\bibfnamefont {R.~M.}\
			\bibnamefont {Ludlam}}, \bibinfo {author} {\bibfnamefont {S.}~\bibnamefont
			{Mahmoodifar}}, \bibinfo {author} {\bibfnamefont {S.~M.}\ \bibnamefont
			{Morsink}}, \bibinfo {author} {\bibfnamefont {P.~S.}\ \bibnamefont {Ray}},
		\bibinfo {author} {\bibfnamefont {T.~E.}\ \bibnamefont {Strohmayer}},
		\bibinfo {author} {\bibfnamefont {K.~S.}\ \bibnamefont {Wood}}, \bibinfo
		{author} {\bibfnamefont {T.}~\bibnamefont {Enoto}}, \bibinfo {author}
		{\bibfnamefont {R.}~\bibnamefont {Foster}}, \bibinfo {author} {\bibfnamefont
			{T.}~\bibnamefont {Okajima}}, \bibinfo {author} {\bibfnamefont
			{G.}~\bibnamefont {Prigozhin}}, \ and\ \bibinfo {author} {\bibfnamefont
			{Y.}~\bibnamefont {Soong}},\ }\href {\doibase 10.3847/2041-8213/ab50c5}
	{\bibfield  {journal} {\bibinfo  {journal} {The Astrophysical Journal}\
		}\textbf {\bibinfo {volume} {887}},\ \bibinfo {pages} {L24} (\bibinfo {year}
		{2019})}\BibitemShut {NoStop}%
	\bibitem [{\citenamefont {Riley}\ \emph {et~al.}(2019)\citenamefont {Riley},
		\citenamefont {Watts}, \citenamefont {Bogdanov}, \citenamefont {Ray},
		\citenamefont {Ludlam}, \citenamefont {Guillot}, \citenamefont {Arzoumanian},
		\citenamefont {Baker}, \citenamefont {Bilous}, \citenamefont {Chakrabarty},
		\citenamefont {Gendreau}, \citenamefont {Harding}, \citenamefont {Ho},
		\citenamefont {Lattimer}, \citenamefont {Morsink},\ and\ \citenamefont
		{Strohmayer}}]{Riley19}%
	\BibitemOpen
	\bibfield  {author} {\bibinfo {author} {\bibfnamefont {T.~E.}\ \bibnamefont
			{Riley}}, \bibinfo {author} {\bibfnamefont {A.~L.}\ \bibnamefont {Watts}},
		\bibinfo {author} {\bibfnamefont {S.}~\bibnamefont {Bogdanov}}, \bibinfo
		{author} {\bibfnamefont {P.~S.}\ \bibnamefont {Ray}}, \bibinfo {author}
		{\bibfnamefont {R.~M.}\ \bibnamefont {Ludlam}}, \bibinfo {author}
		{\bibfnamefont {S.}~\bibnamefont {Guillot}}, \bibinfo {author} {\bibfnamefont
			{Z.}~\bibnamefont {Arzoumanian}}, \bibinfo {author} {\bibfnamefont {C.~L.}\
			\bibnamefont {Baker}}, \bibinfo {author} {\bibfnamefont {A.~V.}\ \bibnamefont
			{Bilous}}, \bibinfo {author} {\bibfnamefont {D.}~\bibnamefont {Chakrabarty}},
		\bibinfo {author} {\bibfnamefont {K.~C.}\ \bibnamefont {Gendreau}}, \bibinfo
		{author} {\bibfnamefont {A.~K.}\ \bibnamefont {Harding}}, \bibinfo {author}
		{\bibfnamefont {W.~C.~G.}\ \bibnamefont {Ho}}, \bibinfo {author}
		{\bibfnamefont {J.~M.}\ \bibnamefont {Lattimer}}, \bibinfo {author}
		{\bibfnamefont {S.~M.}\ \bibnamefont {Morsink}}, \ and\ \bibinfo {author}
		{\bibfnamefont {T.~E.}\ \bibnamefont {Strohmayer}},\ }\href {\doibase
		10.3847/2041-8213/ab481c} {\bibfield  {journal} {\bibinfo  {journal} {The
				Astrophysical Journal}\ }\textbf {\bibinfo {volume} {887}},\ \bibinfo {pages}
		{L21} (\bibinfo {year} {2019})}\BibitemShut {NoStop}%
	\bibitem [{\citenamefont {Watts}\ \emph {et~al.}(2015)\citenamefont {Watts}
		\emph {et~al.}}]{SKA}%
	\BibitemOpen
	\bibfield  {author} {\bibinfo {author} {\bibfnamefont {A.}~\bibnamefont
			{Watts}} \emph {et~al.},\ }\bibfield  {booktitle} {\emph {\bibinfo
			{booktitle} {{Proceedings, Advancing Astrophysics with the Square Kilometre
					Array (AASKA14): Giardini Naxos, Italy, June 9-13, 2014}}},\ }\href {\doibase
		10.22323/1.215.0043} {\bibfield  {journal} {\bibinfo  {journal} {PoS}\
		}\textbf {\bibinfo {volume} {AASKA14}},\ \bibinfo {pages} {043} (\bibinfo
		{year} {2015})},\ \Eprint {http://arxiv.org/abs/1501.00042} {arXiv:1501.00042
		[astro-ph.SR]} \BibitemShut {NoStop}%
	\bibitem [{\citenamefont {Abbott}\ \emph
		{et~al.}(2017{\natexlab{a}})\citenamefont {Abbott} \emph
		{et~al.}}]{TheLIGOScientific:2017qsa}%
	\BibitemOpen
	\bibfield  {author} {\bibinfo {author} {\bibfnamefont {B.~P.}\ \bibnamefont
			{Abbott}} \emph {et~al.} (\bibinfo {collaboration} {LIGO Scientific,
			Virgo}),\ }\href {\doibase 10.1103/PhysRevLett.119.161101} {\bibfield
		{journal} {\bibinfo  {journal} {Phys. Rev. Lett.}\ }\textbf {\bibinfo
			{volume} {119}},\ \bibinfo {pages} {161101} (\bibinfo {year}
		{2017}{\natexlab{a}})},\ \Eprint {http://arxiv.org/abs/1710.05832}
	{arXiv:1710.05832 [gr-qc]} \BibitemShut {NoStop}%
	\bibitem [{\citenamefont {Abbott}\ \emph {et~al.}(2019)\citenamefont {Abbott}
		\emph {et~al.}}]{Abbott:2018wiz}%
	\BibitemOpen
	\bibfield  {author} {\bibinfo {author} {\bibfnamefont {B.~P.}\ \bibnamefont
			{Abbott}} \emph {et~al.} (\bibinfo {collaboration} {LIGO Scientific,
			Virgo}),\ }\href {\doibase 10.1103/PhysRevX.9.011001} {\bibfield  {journal}
		{\bibinfo  {journal} {Phys. Rev.}\ }\textbf {\bibinfo {volume} {X9}},\
		\bibinfo {pages} {011001} (\bibinfo {year} {2019})},\ \Eprint
	{http://arxiv.org/abs/1805.11579} {arXiv:1805.11579 [gr-qc]} \BibitemShut
	{NoStop}%
	\bibitem [{\citenamefont {Abbott}\ \emph {et~al.}(2018)\citenamefont {Abbott}
		\emph {et~al.}}]{Abbott18}%
	\BibitemOpen
	\bibfield  {author} {\bibinfo {author} {\bibfnamefont {B.~P.}\ \bibnamefont
			{Abbott}} \emph {et~al.} (\bibinfo {collaboration} {Virgo, LIGO
			Scientific}),\ }\href {\doibase 10.1103/PhysRevLett.121.161101} {\bibfield
		{journal} {\bibinfo  {journal} {Phys. Rev. Lett.}\ }\textbf {\bibinfo
			{volume} {121}},\ \bibinfo {pages} {161101} (\bibinfo {year} {2018})},\
	\Eprint {http://arxiv.org/abs/1805.11581} {arXiv:1805.11581 [gr-qc]}
	\BibitemShut {NoStop}%
	\bibitem [{\citenamefont {Abbott}\ \emph
		{et~al.}(2017{\natexlab{b}})\citenamefont {Abbott} \emph {et~al.}}]{grb}%
	\BibitemOpen
	\bibfield  {author} {\bibinfo {author} {\bibfnamefont {B.~P.}\ \bibnamefont
			{Abbott}} \emph {et~al.} (\bibinfo {collaboration} {LIGO Scientific, Virgo,
			Fermi-GBM, INTEGRAL}),\ }\href {\doibase 10.3847/2041-8213/aa920c} {\bibfield
		{journal} {\bibinfo  {journal} {Astrophys. J.}\ }\textbf {\bibinfo {volume}
			{848}},\ \bibinfo {pages} {L13} (\bibinfo {year} {2017}{\natexlab{b}})},\
	\Eprint {http://arxiv.org/abs/1710.05834} {arXiv:1710.05834 [astro-ph.HE]}
	\BibitemShut {NoStop}%
	\bibitem [{\citenamefont {Abbott}\ \emph
		{et~al.}(2017{\natexlab{c}})\citenamefont {Abbott} \emph {et~al.}}]{kilo}%
	\BibitemOpen
	\bibfield  {author} {\bibinfo {author} {\bibfnamefont {B.~P.}\ \bibnamefont
			{Abbott}} \emph {et~al.} (\bibinfo {collaboration} {LIGO Scientific, Virgo,
			Fermi GBM, INTEGRAL, IceCube, AstroSat Cadmium Zinc Telluride Imager Team,
			IPN, Insight-Hxmt, ANTARES, Swift, AGILE Team, 1M2H Team, Dark Energy Camera
			GW-EM, DES, DLT40, GRAWITA, Fermi-LAT, ATCA, ASKAP, Las Cumbres Observatory
			Group, OzGrav, DWF (Deeper Wider Faster Program), AST3, CAASTRO, VINROUGE,
			MASTER, J-GEM, GROWTH, JAGWAR, CaltechNRAO, TTU-NRAO, NuSTAR, Pan-STARRS,
			MAXI Team, TZAC Consortium, KU, Nordic Optical Telescope, ePESSTO, GROND,
			Texas Tech University, SALT Group, TOROS, BOOTES, MWA, CALET, IKI-GW
			Follow-up, H.E.S.S., LOFAR, LWA, HAWC, Pierre Auger, ALMA, Euro VLBI Team, Pi
			of Sky, Chandra Team at McGill University, DFN, ATLAS Telescopes, High Time
			Resolution Universe Survey, RIMAS, RATIR, SKA South Africa/MeerKAT}),\ }\href
	{\doibase 10.3847/2041-8213/aa91c9} {\bibfield  {journal} {\bibinfo
			{journal} {Astrophys. J.}\ }\textbf {\bibinfo {volume} {848}},\ \bibinfo
		{pages} {L12} (\bibinfo {year} {2017}{\natexlab{c}})},\ \Eprint
	{http://arxiv.org/abs/1710.05833} {arXiv:1710.05833 [astro-ph.HE]}
	\BibitemShut {NoStop}%
	\bibitem [{\citenamefont {Radice}\ \emph {et~al.}(2017)\citenamefont {Radice},
		\citenamefont {Bernuzzi}, \citenamefont {Del~Pozzo}, \citenamefont
		{Roberts},\ and\ \citenamefont {Ott}}]{Radice2017}%
	\BibitemOpen
	\bibfield  {author} {\bibinfo {author} {\bibfnamefont {D.}~\bibnamefont
			{Radice}}, \bibinfo {author} {\bibfnamefont {S.}~\bibnamefont {Bernuzzi}},
		\bibinfo {author} {\bibfnamefont {W.}~\bibnamefont {Del~Pozzo}}, \bibinfo
		{author} {\bibfnamefont {L.~F.}\ \bibnamefont {Roberts}}, \ and\ \bibinfo
		{author} {\bibfnamefont {C.~D.}\ \bibnamefont {Ott}},\ }\href {\doibase
		10.3847/2041-8213/aa775f} {\bibfield  {journal} {\bibinfo  {journal}
			{Astrophys. J.}\ }\textbf {\bibinfo {volume} {842}},\ \bibinfo {pages} {L10}
		(\bibinfo {year} {2017})},\ \Eprint {http://arxiv.org/abs/1612.06429}
	{arXiv:1612.06429 [astro-ph.HE]} \BibitemShut {NoStop}%
	\bibitem [{\citenamefont {Radice}\ \emph {et~al.}(2018)\citenamefont {Radice},
		\citenamefont {Perego}, \citenamefont {Zappa},\ and\ \citenamefont
		{Bernuzzi}}]{Radice2018}%
	\BibitemOpen
	\bibfield  {author} {\bibinfo {author} {\bibfnamefont {D.}~\bibnamefont
			{Radice}}, \bibinfo {author} {\bibfnamefont {A.}~\bibnamefont {Perego}},
		\bibinfo {author} {\bibfnamefont {F.}~\bibnamefont {Zappa}}, \ and\ \bibinfo
		{author} {\bibfnamefont {S.}~\bibnamefont {Bernuzzi}},\ }\href {\doibase
		10.3847/2041-8213/aaa402} {\bibfield  {journal} {\bibinfo  {journal}
			{Astrophys. J.}\ }\textbf {\bibinfo {volume} {852}},\ \bibinfo {pages} {L29}
		(\bibinfo {year} {2018})},\ \Eprint {http://arxiv.org/abs/1711.03647}
	{arXiv:1711.03647 [astro-ph.HE]} \BibitemShut {NoStop}%
	\bibitem [{\citenamefont {Bauswein}\ \emph {et~al.}(2019)\citenamefont
		{Bauswein}, \citenamefont {Friedrich~Bastian}, \citenamefont {Blaschke},
		\citenamefont {Chatziioannou}, \citenamefont {Clark}, \citenamefont
		{Fischer}, \citenamefont {Janka}, \citenamefont {Just}, \citenamefont
		{Oertel},\ and\ \citenamefont {Stergioulas}}]{Bauswein2019}%
	\BibitemOpen
	\bibfield  {author} {\bibinfo {author} {\bibfnamefont {A.}~\bibnamefont
			{Bauswein}}, \bibinfo {author} {\bibfnamefont {N.-U.}\ \bibnamefont
			{Friedrich~Bastian}}, \bibinfo {author} {\bibfnamefont {D.}~\bibnamefont
			{Blaschke}}, \bibinfo {author} {\bibfnamefont {K.}~\bibnamefont
			{Chatziioannou}}, \bibinfo {author} {\bibfnamefont {J.~A.}\ \bibnamefont
			{Clark}}, \bibinfo {author} {\bibfnamefont {T.}~\bibnamefont {Fischer}},
		\bibinfo {author} {\bibfnamefont {H.-T.}\ \bibnamefont {Janka}}, \bibinfo
		{author} {\bibfnamefont {O.}~\bibnamefont {Just}}, \bibinfo {author}
		{\bibfnamefont {M.}~\bibnamefont {Oertel}}, \ and\ \bibinfo {author}
		{\bibfnamefont {N.}~\bibnamefont {Stergioulas}},\ }\href {\doibase
		10.1063/1.5117803} {\bibfield  {journal} {\bibinfo  {journal} {AIP Conf.
				Proc.}\ }\textbf {\bibinfo {volume} {2127}},\ \bibinfo {pages} {020013}
		(\bibinfo {year} {2019})},\ \Eprint {http://arxiv.org/abs/1904.01306}
	{arXiv:1904.01306 [astro-ph.HE]} \BibitemShut {NoStop}%
	\bibitem [{\citenamefont {Coughlin}\ \emph {et~al.}(2018)\citenamefont
		{Coughlin} \emph {et~al.}}]{Coughlin2018}%
	\BibitemOpen
	\bibfield  {author} {\bibinfo {author} {\bibfnamefont {M.~W.}\ \bibnamefont
			{Coughlin}} \emph {et~al.},\ }\href {\doibase 10.1093/mnras/sty2174}
	{\bibfield  {journal} {\bibinfo  {journal} {Mon. Not. Roy. Astron. Soc.}\
		}\textbf {\bibinfo {volume} {480}},\ \bibinfo {pages} {3871} (\bibinfo {year}
		{2018})},\ \Eprint {http://arxiv.org/abs/1805.09371} {arXiv:1805.09371
		[astro-ph.HE]} \BibitemShut {NoStop}%
	\bibitem [{\citenamefont {Wang}\ \emph {et~al.}(2019)\citenamefont {Wang},
		\citenamefont {Shao}, \citenamefont {Jiang}, \citenamefont {Tang},
		\citenamefont {Ren}, \citenamefont {Zhang}, \citenamefont {Jin},
		\citenamefont {Fan},\ and\ \citenamefont {Wei}}]{Wang2018}%
	\BibitemOpen
	\bibfield  {author} {\bibinfo {author} {\bibfnamefont {Y.-Z.}\ \bibnamefont
			{Wang}}, \bibinfo {author} {\bibfnamefont {D.-S.}\ \bibnamefont {Shao}},
		\bibinfo {author} {\bibfnamefont {J.-L.}\ \bibnamefont {Jiang}}, \bibinfo
		{author} {\bibfnamefont {S.-P.}\ \bibnamefont {Tang}}, \bibinfo {author}
		{\bibfnamefont {X.-X.}\ \bibnamefont {Ren}}, \bibinfo {author} {\bibfnamefont
			{F.-W.}\ \bibnamefont {Zhang}}, \bibinfo {author} {\bibfnamefont {Z.-P.}\
			\bibnamefont {Jin}}, \bibinfo {author} {\bibfnamefont {Y.-Z.}\ \bibnamefont
			{Fan}}, \ and\ \bibinfo {author} {\bibfnamefont {D.-M.}\ \bibnamefont
			{Wei}},\ }\href {\doibase 10.3847/1538-4357/ab1914} {\bibfield  {journal}
		{\bibinfo  {journal} {Astrophys. J.}\ }\textbf {\bibinfo {volume} {877}},\
		\bibinfo {pages} {2} (\bibinfo {year} {2019})},\ \Eprint
	{http://arxiv.org/abs/1811.02558} {arXiv:1811.02558 [astro-ph.HE]}
	\BibitemShut {NoStop}%
	\bibitem [{\citenamefont {Margueron}\ \emph
		{et~al.}(2018{\natexlab{a}})\citenamefont {Margueron}, \citenamefont
		{Hoffmann~Casali},\ and\ \citenamefont {Gulminelli}}]{Margueron2018a}%
	\BibitemOpen
	\bibfield  {author} {\bibinfo {author} {\bibfnamefont {J.}~\bibnamefont
			{Margueron}}, \bibinfo {author} {\bibfnamefont {R.}~\bibnamefont
			{Hoffmann~Casali}}, \ and\ \bibinfo {author} {\bibfnamefont {F.}~\bibnamefont
			{Gulminelli}},\ }\href {\doibase 10.1103/PhysRevC.97.025805} {\bibfield
		{journal} {\bibinfo  {journal} {Phys. Rev.}\ }\textbf {\bibinfo {volume}
			{C97}},\ \bibinfo {pages} {025805} (\bibinfo {year} {2018}{\natexlab{a}})},\
	\Eprint {http://arxiv.org/abs/1708.06894} {arXiv:1708.06894 [nucl-th]}
	\BibitemShut {NoStop}%
	\bibitem [{\citenamefont {Margueron}\ \emph
		{et~al.}(2018{\natexlab{b}})\citenamefont {Margueron}, \citenamefont
		{Hoffmann~Casali},\ and\ \citenamefont {Gulminelli}}]{Margueron2018b}%
	\BibitemOpen
	\bibfield  {author} {\bibinfo {author} {\bibfnamefont {J.}~\bibnamefont
			{Margueron}}, \bibinfo {author} {\bibfnamefont {R.}~\bibnamefont
			{Hoffmann~Casali}}, \ and\ \bibinfo {author} {\bibfnamefont {F.}~\bibnamefont
			{Gulminelli}},\ }\href {\doibase 10.1103/PhysRevC.97.025806} {\bibfield
		{journal} {\bibinfo  {journal} {Phys. Rev.}\ }\textbf {\bibinfo {volume}
			{C97}},\ \bibinfo {pages} {025806} (\bibinfo {year} {2018}{\natexlab{b}})},\
	\Eprint {http://arxiv.org/abs/1708.06895} {arXiv:1708.06895 [nucl-th]}
	\BibitemShut {NoStop}%
	\bibitem [{\citenamefont {Zhang}\ \emph {et~al.}(2018)\citenamefont {Zhang},
		\citenamefont {Li},\ and\ \citenamefont {Xu}}]{Zhang2018}%
	\BibitemOpen
	\bibfield  {author} {\bibinfo {author} {\bibfnamefont {N.-B.}\ \bibnamefont
			{Zhang}}, \bibinfo {author} {\bibfnamefont {B.-A.}\ \bibnamefont {Li}}, \
		and\ \bibinfo {author} {\bibfnamefont {J.}~\bibnamefont {Xu}},\ }\href
	{\doibase 10.3847/1538-4357/aac027} {\bibfield  {journal} {\bibinfo
			{journal} {Astrophys. J.}\ }\textbf {\bibinfo {volume} {859}},\ \bibinfo
		{pages} {90} (\bibinfo {year} {2018})},\ \Eprint
	{http://arxiv.org/abs/1801.06855} {arXiv:1801.06855 [nucl-th]} \BibitemShut
	{NoStop}%
	\bibitem [{\citenamefont {Margueron}\ and\ \citenamefont
		{Gulminelli}(2019)}]{Margueron2019}%
	\BibitemOpen
	\bibfield  {author} {\bibinfo {author} {\bibfnamefont {J.}~\bibnamefont
			{Margueron}}\ and\ \bibinfo {author} {\bibfnamefont {F.}~\bibnamefont
			{Gulminelli}},\ }\href {\doibase 10.1103/PhysRevC.99.025806} {\bibfield
		{journal} {\bibinfo  {journal} {Phys. Rev.}\ }\textbf {\bibinfo {volume}
			{C99}},\ \bibinfo {pages} {025806} (\bibinfo {year} {2019})},\ \Eprint
	{http://arxiv.org/abs/1807.01729} {arXiv:1807.01729 [nucl-th]} \BibitemShut
	{NoStop}%
	\bibitem [{\citenamefont {Bombaci}\ and\ \citenamefont
		{Lombardo}(1991)}]{PhysRevC.44.1892}%
	\BibitemOpen
	\bibfield  {author} {\bibinfo {author} {\bibfnamefont {I.}~\bibnamefont
			{Bombaci}}\ and\ \bibinfo {author} {\bibfnamefont {U.}~\bibnamefont
			{Lombardo}},\ }\href {\doibase 10.1103/PhysRevC.44.1892} {\bibfield
		{journal} {\bibinfo  {journal} {Phys. Rev. C}\ }\textbf {\bibinfo {volume}
			{44}},\ \bibinfo {pages} {1892} (\bibinfo {year} {1991})}\BibitemShut
	{NoStop}%
	\bibitem [{\citenamefont {Read}\ \emph
		{et~al.}(2009{\natexlab{a}})\citenamefont {Read}, \citenamefont {Lackey},
		\citenamefont {Owen},\ and\ \citenamefont {Friedman}}]{PhysRevD.79.124032}%
	\BibitemOpen
	\bibfield  {author} {\bibinfo {author} {\bibfnamefont {J.~S.}\ \bibnamefont
			{Read}}, \bibinfo {author} {\bibfnamefont {B.~D.}\ \bibnamefont {Lackey}},
		\bibinfo {author} {\bibfnamefont {B.~J.}\ \bibnamefont {Owen}}, \ and\
		\bibinfo {author} {\bibfnamefont {J.~L.}\ \bibnamefont {Friedman}},\ }\href
	{\doibase 10.1103/PhysRevD.79.124032} {\bibfield  {journal} {\bibinfo
			{journal} {Phys. Rev. D}\ }\textbf {\bibinfo {volume} {79}},\ \bibinfo
		{pages} {124032} (\bibinfo {year} {2009}{\natexlab{a}})}\BibitemShut
	{NoStop}%
	\bibitem [{\citenamefont {\"Ozel}\ and\ \citenamefont
		{Psaltis}(2009)}]{PhysRevD.80.103003}%
	\BibitemOpen
	\bibfield  {author} {\bibinfo {author} {\bibfnamefont {F.}~\bibnamefont
			{\"Ozel}}\ and\ \bibinfo {author} {\bibfnamefont {D.}~\bibnamefont
			{Psaltis}},\ }\href {\doibase 10.1103/PhysRevD.80.103003} {\bibfield
		{journal} {\bibinfo  {journal} {Phys. Rev. D}\ }\textbf {\bibinfo {volume}
			{80}},\ \bibinfo {pages} {103003} (\bibinfo {year} {2009})}\BibitemShut
	{NoStop}%
	\bibitem [{\citenamefont {Steiner}\ \emph {et~al.}(2010)\citenamefont
		{Steiner}, \citenamefont {Lattimer},\ and\ \citenamefont
		{Brown}}]{Steiner:2010fz}%
	\BibitemOpen
	\bibfield  {author} {\bibinfo {author} {\bibfnamefont {A.~W.}\ \bibnamefont
			{Steiner}}, \bibinfo {author} {\bibfnamefont {J.~M.}\ \bibnamefont
			{Lattimer}}, \ and\ \bibinfo {author} {\bibfnamefont {E.~F.}\ \bibnamefont
			{Brown}},\ }\href {\doibase 10.1088/0004-637X/722/1/33} {\bibfield  {journal}
		{\bibinfo  {journal} {Astrophys. J.}\ }\textbf {\bibinfo {volume} {722}},\
		\bibinfo {pages} {33} (\bibinfo {year} {2010})},\ \Eprint
	{http://arxiv.org/abs/1005.0811} {arXiv:1005.0811 [astro-ph.HE]} \BibitemShut
	{NoStop}%
	\bibitem [{\citenamefont {Raithel}\ \emph {et~al.}(2016)\citenamefont
		{Raithel}, \citenamefont {Ozel},\ and\ \citenamefont
		{Psaltis}}]{Raithel:2016bux}%
	\BibitemOpen
	\bibfield  {author} {\bibinfo {author} {\bibfnamefont {C.~A.}\ \bibnamefont
			{Raithel}}, \bibinfo {author} {\bibfnamefont {F.}~\bibnamefont {Ozel}}, \
		and\ \bibinfo {author} {\bibfnamefont {D.}~\bibnamefont {Psaltis}},\ }\href
	{\doibase 10.3847/0004-637X/831/1/44} {\bibfield  {journal} {\bibinfo
			{journal} {Astrophys. J.}\ }\textbf {\bibinfo {volume} {831}},\ \bibinfo
		{pages} {44} (\bibinfo {year} {2016})},\ \Eprint
	{http://arxiv.org/abs/1605.03591} {arXiv:1605.03591 [astro-ph.HE]}
	\BibitemShut {NoStop}%
	\bibitem [{\citenamefont {Lindblom}(2010)}]{Lindblom:2010bb}%
	\BibitemOpen
	\bibfield  {author} {\bibinfo {author} {\bibfnamefont {L.}~\bibnamefont
			{Lindblom}},\ }\href {\doibase 10.1103/PhysRevD.82.103011} {\bibfield
		{journal} {\bibinfo  {journal} {Phys. Rev.}\ }\textbf {\bibinfo {volume}
			{D82}},\ \bibinfo {pages} {103011} (\bibinfo {year} {2010})},\ \Eprint
	{http://arxiv.org/abs/1009.0738} {arXiv:1009.0738 [astro-ph.HE]} \BibitemShut
	{NoStop}%
	\bibitem [{\citenamefont {Tews}\ \emph {et~al.}(2018)\citenamefont {Tews},
		\citenamefont {Margueron},\ and\ \citenamefont {Reddy}}]{Tews:2018chv}%
	\BibitemOpen
	\bibfield  {author} {\bibinfo {author} {\bibfnamefont {I.}~\bibnamefont
			{Tews}}, \bibinfo {author} {\bibfnamefont {J.}~\bibnamefont {Margueron}}, \
		and\ \bibinfo {author} {\bibfnamefont {S.}~\bibnamefont {Reddy}},\ }\href
	{\doibase 10.1103/PhysRevC.98.045804} {\bibfield  {journal} {\bibinfo
			{journal} {Phys. Rev.}\ }\textbf {\bibinfo {volume} {C98}},\ \bibinfo {pages}
		{045804} (\bibinfo {year} {2018})},\ \Eprint
	{http://arxiv.org/abs/1804.02783} {arXiv:1804.02783 [nucl-th]} \BibitemShut
	{NoStop}%
	\bibitem [{\citenamefont {Annala}\ \emph {et~al.}(2019)\citenamefont {Annala},
		\citenamefont {Gorda}, \citenamefont {Kurkela}, \citenamefont {Nättilä},\
		and\ \citenamefont {Vuorinen}}]{Annala:2019puf}%
	\BibitemOpen
	\bibfield  {author} {\bibinfo {author} {\bibfnamefont {E.}~\bibnamefont
			{Annala}}, \bibinfo {author} {\bibfnamefont {T.}~\bibnamefont {Gorda}},
		\bibinfo {author} {\bibfnamefont {A.}~\bibnamefont {Kurkela}}, \bibinfo
		{author} {\bibfnamefont {J.}~\bibnamefont {Nättilä}}, \ and\ \bibinfo
		{author} {\bibfnamefont {A.}~\bibnamefont {Vuorinen}},\ }\href@noop {} {\
		(\bibinfo {year} {2019})},\ \Eprint {http://arxiv.org/abs/1903.09121}
	{arXiv:1903.09121 [astro-ph.HE]} \BibitemShut {NoStop}%
	\bibitem [{\citenamefont {Read}\ \emph
		{et~al.}(2009{\natexlab{b}})\citenamefont {Read}, \citenamefont {Lackey},
		\citenamefont {Owen},\ and\ \citenamefont {Friedman}}]{Read2008}%
	\BibitemOpen
	\bibfield  {author} {\bibinfo {author} {\bibfnamefont {J.~S.}\ \bibnamefont
			{Read}}, \bibinfo {author} {\bibfnamefont {B.~D.}\ \bibnamefont {Lackey}},
		\bibinfo {author} {\bibfnamefont {B.~J.}\ \bibnamefont {Owen}}, \ and\
		\bibinfo {author} {\bibfnamefont {J.~L.}\ \bibnamefont {Friedman}},\ }\href
	{\doibase 10.1103/PhysRevD.79.124032} {\bibfield  {journal} {\bibinfo
			{journal} {Phys. Rev.}\ }\textbf {\bibinfo {volume} {D79}},\ \bibinfo {pages}
		{124032} (\bibinfo {year} {2009}{\natexlab{b}})},\ \Eprint
	{http://arxiv.org/abs/0812.2163} {arXiv:0812.2163 [astro-ph]} \BibitemShut
	{NoStop}%
	\bibitem [{\citenamefont {Carson}\ \emph
		{et~al.}(2019{\natexlab{a}})\citenamefont {Carson}, \citenamefont {Steiner},\
		and\ \citenamefont {Yagi}}]{Carson:2018xri}%
	\BibitemOpen
	\bibfield  {author} {\bibinfo {author} {\bibfnamefont {Z.}~\bibnamefont
			{Carson}}, \bibinfo {author} {\bibfnamefont {A.~W.}\ \bibnamefont {Steiner}},
		\ and\ \bibinfo {author} {\bibfnamefont {K.}~\bibnamefont {Yagi}},\ }\href
	{\doibase 10.1103/PhysRevD.99.043010} {\bibfield  {journal} {\bibinfo
			{journal} {Phys. Rev.}\ }\textbf {\bibinfo {volume} {D99}},\ \bibinfo {pages}
		{043010} (\bibinfo {year} {2019}{\natexlab{a}})},\ \Eprint
	{http://arxiv.org/abs/1812.08910} {arXiv:1812.08910 [gr-qc]} \BibitemShut
	{NoStop}%
	\bibitem [{\citenamefont {Carson}\ \emph
		{et~al.}(2019{\natexlab{b}})\citenamefont {Carson}, \citenamefont {Steiner},\
		and\ \citenamefont {Yagi}}]{Carson:2019xxz}%
	\BibitemOpen
	\bibfield  {author} {\bibinfo {author} {\bibfnamefont {Z.}~\bibnamefont
			{Carson}}, \bibinfo {author} {\bibfnamefont {A.~W.}\ \bibnamefont {Steiner}},
		\ and\ \bibinfo {author} {\bibfnamefont {K.}~\bibnamefont {Yagi}},\ }\href
	{\doibase 10.1103/PhysRevD.100.023012} {\bibfield  {journal} {\bibinfo
			{journal} {Phys. Rev.}\ }\textbf {\bibinfo {volume} {D100}},\ \bibinfo
		{pages} {023012} (\bibinfo {year} {2019}{\natexlab{b}})},\ \Eprint
	{http://arxiv.org/abs/1906.05978} {arXiv:1906.05978 [gr-qc]} \BibitemShut
	{NoStop}%
	\bibitem [{\citenamefont {Fortin}\ \emph {et~al.}(2016)\citenamefont {Fortin},
		\citenamefont {Providencia}, \citenamefont {Raduta}, \citenamefont
		{Gulminelli}, \citenamefont {Zdunik}, \citenamefont {Haensel},\ and\
		\citenamefont {Bejger}}]{Fortin16}%
	\BibitemOpen
	\bibfield  {author} {\bibinfo {author} {\bibfnamefont {M.}~\bibnamefont
			{Fortin}}, \bibinfo {author} {\bibfnamefont {C.}~\bibnamefont {Providencia}},
		\bibinfo {author} {\bibfnamefont {A.~R.}\ \bibnamefont {Raduta}}, \bibinfo
		{author} {\bibfnamefont {F.}~\bibnamefont {Gulminelli}}, \bibinfo {author}
		{\bibfnamefont {J.~L.}\ \bibnamefont {Zdunik}}, \bibinfo {author}
		{\bibfnamefont {P.}~\bibnamefont {Haensel}}, \ and\ \bibinfo {author}
		{\bibfnamefont {M.}~\bibnamefont {Bejger}},\ }\href {\doibase
		10.1103/PhysRevC.94.035804} {\bibfield  {journal} {\bibinfo  {journal} {Phys.
				Rev.}\ }\textbf {\bibinfo {volume} {C94}},\ \bibinfo {pages} {035804}
		(\bibinfo {year} {2016})},\ \Eprint {http://arxiv.org/abs/1604.01944}
	{arXiv:1604.01944 [astro-ph.SR]} \BibitemShut {NoStop}%
	\bibitem [{\citenamefont {Ferreira}\ \emph {et~al.}(2020)\citenamefont
		{Ferreira}, \citenamefont {Fortin}, \citenamefont {Malik}, \citenamefont
		{Agrawal},\ and\ \citenamefont {Providência}}]{Ferreira:2019bgy}%
	\BibitemOpen
	\bibfield  {author} {\bibinfo {author} {\bibfnamefont {M.}~\bibnamefont
			{Ferreira}}, \bibinfo {author} {\bibfnamefont {M.}~\bibnamefont {Fortin}},
		\bibinfo {author} {\bibfnamefont {T.}~\bibnamefont {Malik}}, \bibinfo
		{author} {\bibfnamefont {B.}~\bibnamefont {Agrawal}}, \ and\ \bibinfo
		{author} {\bibfnamefont {C.}~\bibnamefont {Providência}},\ }\href {\doibase
		10.1103/PhysRevD.101.043021} {\bibfield  {journal} {\bibinfo  {journal}
			{Phys. Rev. D}\ }\textbf {\bibinfo {volume} {101}},\ \bibinfo {pages}
		{043021} (\bibinfo {year} {2020})},\ \Eprint
	{http://arxiv.org/abs/1912.11131} {arXiv:1912.11131 [nucl-th]} \BibitemShut
	{NoStop}%
	\bibitem [{\citenamefont {Youngblood}\ \emph {et~al.}(1999)\citenamefont
		{Youngblood}, \citenamefont {Clark},\ and\ \citenamefont
		{Lui}}]{Youngblood1999}%
	\BibitemOpen
	\bibfield  {author} {\bibinfo {author} {\bibfnamefont {D.~H.}\ \bibnamefont
			{Youngblood}}, \bibinfo {author} {\bibfnamefont {H.~L.}\ \bibnamefont
			{Clark}}, \ and\ \bibinfo {author} {\bibfnamefont {Y.~W.}\ \bibnamefont
			{Lui}},\ }\href {\doibase 10.1103/PhysRevLett.82.691} {\bibfield  {journal}
		{\bibinfo  {journal} {Phys. Rev. Lett.}\ }\textbf {\bibinfo {volume} {82}},\
		\bibinfo {pages} {691} (\bibinfo {year} {1999})}\BibitemShut {NoStop}%
	\bibitem [{\citenamefont {Margueron}\ and\ \citenamefont
		{Khan}(2012)}]{Margueron2012}%
	\BibitemOpen
	\bibfield  {author} {\bibinfo {author} {\bibfnamefont {J.}~\bibnamefont
			{Margueron}}\ and\ \bibinfo {author} {\bibfnamefont {E.}~\bibnamefont
			{Khan}},\ }\href {\doibase 10.1103/PhysRevC.86.065801} {\bibfield  {journal}
		{\bibinfo  {journal} {Phys. Rev.}\ }\textbf {\bibinfo {volume} {C86}},\
		\bibinfo {pages} {065801} (\bibinfo {year} {2012})},\ \Eprint
	{http://arxiv.org/abs/1203.2134} {arXiv:1203.2134 [nucl-th]} \BibitemShut
	{NoStop}%
	\bibitem [{\citenamefont {Li}\ and\ \citenamefont {Han}(2013)}]{Li2013}%
	\BibitemOpen
	\bibfield  {author} {\bibinfo {author} {\bibfnamefont {B.-A.}\ \bibnamefont
			{Li}}\ and\ \bibinfo {author} {\bibfnamefont {X.}~\bibnamefont {Han}},\
	}\href {\doibase 10.1016/j.physletb.2013.10.006} {\bibfield  {journal}
		{\bibinfo  {journal} {Phys. Lett.}\ }\textbf {\bibinfo {volume} {B727}},\
		\bibinfo {pages} {276} (\bibinfo {year} {2013})},\ \Eprint
	{http://arxiv.org/abs/1304.3368} {arXiv:1304.3368 [nucl-th]} \BibitemShut
	{NoStop}%
	\bibitem [{\citenamefont {Lattimer}\ and\ \citenamefont
		{Lim}(2013)}]{Lattimer2013}%
	\BibitemOpen
	\bibfield  {author} {\bibinfo {author} {\bibfnamefont {J.~M.}\ \bibnamefont
			{Lattimer}}\ and\ \bibinfo {author} {\bibfnamefont {Y.}~\bibnamefont {Lim}},\
	}\href {\doibase 10.1088/0004-637X/771/1/51} {\bibfield  {journal} {\bibinfo
			{journal} {Astrophys. J.}\ }\textbf {\bibinfo {volume} {771}},\ \bibinfo
		{pages} {51} (\bibinfo {year} {2013})}\BibitemShut {NoStop}%
	\bibitem [{\citenamefont {Stone}\ \emph {et~al.}(2014)\citenamefont {Stone},
		\citenamefont {Stone},\ and\ \citenamefont {Moszkowski}}]{Stone2014}%
	\BibitemOpen
	\bibfield  {author} {\bibinfo {author} {\bibfnamefont {J.~R.}\ \bibnamefont
			{Stone}}, \bibinfo {author} {\bibfnamefont {N.~J.}\ \bibnamefont {Stone}}, \
		and\ \bibinfo {author} {\bibfnamefont {S.~A.}\ \bibnamefont {Moszkowski}},\
	}\href {\doibase 10.1103/PhysRevC.89.044316} {\bibfield  {journal} {\bibinfo
			{journal} {Phys. Rev.}\ }\textbf {\bibinfo {volume} {C89}},\ \bibinfo {pages}
		{044316} (\bibinfo {year} {2014})},\ \Eprint {http://arxiv.org/abs/1404.0744}
	{arXiv:1404.0744 [nucl-th]} \BibitemShut {NoStop}%
	\bibitem [{\citenamefont {Oertel}\ \emph {et~al.}(2017)\citenamefont {Oertel},
		\citenamefont {Hempel}, \citenamefont {Klähn},\ and\ \citenamefont
		{Typel}}]{OertelRMP16}%
	\BibitemOpen
	\bibfield  {author} {\bibinfo {author} {\bibfnamefont {M.}~\bibnamefont
			{Oertel}}, \bibinfo {author} {\bibfnamefont {M.}~\bibnamefont {Hempel}},
		\bibinfo {author} {\bibfnamefont {T.}~\bibnamefont {Klähn}}, \ and\ \bibinfo
		{author} {\bibfnamefont {S.}~\bibnamefont {Typel}},\ }\href {\doibase
		10.1103/RevModPhys.89.015007} {\bibfield  {journal} {\bibinfo  {journal}
			{Rev. Mod. Phys.}\ }\textbf {\bibinfo {volume} {89}},\ \bibinfo {pages}
		{015007} (\bibinfo {year} {2017})}\BibitemShut {NoStop}%
	\bibitem [{\citenamefont {Farine}\ \emph {et~al.}(1997)\citenamefont {Farine},
		\citenamefont {Pearson},\ and\ \citenamefont {Tondeur}}]{Farine1997}%
	\BibitemOpen
	\bibfield  {author} {\bibinfo {author} {\bibfnamefont {M.}~\bibnamefont
			{Farine}}, \bibinfo {author} {\bibfnamefont {J.~M.}\ \bibnamefont {Pearson}},
		\ and\ \bibinfo {author} {\bibfnamefont {F.}~\bibnamefont {Tondeur}},\ }\href
	{\doibase 10.1016/S0375-9474(96)00453-8} {\bibfield  {journal} {\bibinfo
			{journal} {Nucl. Phys.}\ }\textbf {\bibinfo {volume} {A615}},\ \bibinfo
		{pages} {135} (\bibinfo {year} {1997})}\BibitemShut {NoStop}%
	\bibitem [{\citenamefont {De}\ \emph {et~al.}(2015)\citenamefont {De},
		\citenamefont {Samaddar},\ and\ \citenamefont {Agrawal}}]{De2015}%
	\BibitemOpen
	\bibfield  {author} {\bibinfo {author} {\bibfnamefont {J.~N.}\ \bibnamefont
			{De}}, \bibinfo {author} {\bibfnamefont {S.~K.}\ \bibnamefont {Samaddar}}, \
		and\ \bibinfo {author} {\bibfnamefont {B.~K.}\ \bibnamefont {Agrawal}},\
	}\href {\doibase 10.1103/PhysRevC.92.014304} {\bibfield  {journal} {\bibinfo
			{journal} {Phys. Rev.}\ }\textbf {\bibinfo {volume} {C92}},\ \bibinfo {pages}
		{014304} (\bibinfo {year} {2015})},\ \Eprint
	{http://arxiv.org/abs/1506.06461} {arXiv:1506.06461 [nucl-th]} \BibitemShut
	{NoStop}%
	\bibitem [{\citenamefont {Mondal}\ \emph {et~al.}(2016)\citenamefont {Mondal},
		\citenamefont {Agrawal}, \citenamefont {De},\ and\ \citenamefont
		{Samaddar}}]{Mondal2016}%
	\BibitemOpen
	\bibfield  {author} {\bibinfo {author} {\bibfnamefont {C.}~\bibnamefont
			{Mondal}}, \bibinfo {author} {\bibfnamefont {B.~K.}\ \bibnamefont {Agrawal}},
		\bibinfo {author} {\bibfnamefont {J.~N.}\ \bibnamefont {De}}, \ and\ \bibinfo
		{author} {\bibfnamefont {S.~K.}\ \bibnamefont {Samaddar}},\ }\href@noop {}
	{\bibfield  {journal} {\bibinfo  {journal} {Phys. Rev.}\ }\textbf {\bibinfo
			{volume} {C93}},\ \bibinfo {pages} {044328} (\bibinfo {year}
		{2016})}\BibitemShut {NoStop}%
	\bibitem [{\citenamefont {Malik}\ \emph
		{et~al.}(2018{\natexlab{a}})\citenamefont {Malik}, \citenamefont {Alam},
		\citenamefont {Fortin}, \citenamefont {Providência}, \citenamefont
		{Agrawal}, \citenamefont {Jha}, \citenamefont {Kumar},\ and\ \citenamefont
		{Patra}}]{Malik2018}%
	\BibitemOpen
	\bibfield  {author} {\bibinfo {author} {\bibfnamefont {T.}~\bibnamefont
			{Malik}}, \bibinfo {author} {\bibfnamefont {N.}~\bibnamefont {Alam}},
		\bibinfo {author} {\bibfnamefont {M.}~\bibnamefont {Fortin}}, \bibinfo
		{author} {\bibfnamefont {C.}~\bibnamefont {Providência}}, \bibinfo {author}
		{\bibfnamefont {B.~K.}\ \bibnamefont {Agrawal}}, \bibinfo {author}
		{\bibfnamefont {T.~K.}\ \bibnamefont {Jha}}, \bibinfo {author} {\bibfnamefont
			{B.}~\bibnamefont {Kumar}}, \ and\ \bibinfo {author} {\bibfnamefont {S.~K.}\
			\bibnamefont {Patra}},\ }\href {\doibase 10.1103/PhysRevC.98.035804}
	{\bibfield  {journal} {\bibinfo  {journal} {Phys. Rev.}\ }\textbf {\bibinfo
			{volume} {C98}},\ \bibinfo {pages} {035804} (\bibinfo {year}
		{2018}{\natexlab{a}})},\ \Eprint {http://arxiv.org/abs/1805.11963}
	{arXiv:1805.11963 [nucl-th]} \BibitemShut {NoStop}%
	\bibitem [{\citenamefont {Li}\ \emph {et~al.}(2019)\citenamefont {Li},
		\citenamefont {Krastev}, \citenamefont {Wen}, \citenamefont {Xie},\ and\
		\citenamefont {Zhang}}]{Li2019}%
	\BibitemOpen
	\bibfield  {author} {\bibinfo {author} {\bibfnamefont {B.-A.}\ \bibnamefont
			{Li}}, \bibinfo {author} {\bibfnamefont {P.~G.}\ \bibnamefont {Krastev}},
		\bibinfo {author} {\bibfnamefont {D.-H.}\ \bibnamefont {Wen}}, \bibinfo
		{author} {\bibfnamefont {W.-J.}\ \bibnamefont {Xie}}, \ and\ \bibinfo
		{author} {\bibfnamefont {N.-B.}\ \bibnamefont {Zhang}},\ }\href {\doibase
		10.1063/1.5117808} {\bibfield  {journal} {\bibinfo  {journal} {AIP Conf.
				Proc.}\ }\textbf {\bibinfo {volume} {2127}},\ \bibinfo {pages} {020018}
		(\bibinfo {year} {2019})}\BibitemShut {NoStop}%
	\bibitem [{\citenamefont {Douchin}\ and\ \citenamefont
		{Haensel}(2001)}]{Douchin2001}%
	\BibitemOpen
	\bibfield  {author} {\bibinfo {author} {\bibfnamefont {F.}~\bibnamefont
			{Douchin}}\ and\ \bibinfo {author} {\bibfnamefont {P.}~\bibnamefont
			{Haensel}},\ }\href {\doibase 10.1051/0004-6361:20011402} {\bibfield
		{journal} {\bibinfo  {journal} {Astron. Astrophys.}\ }\textbf {\bibinfo
			{volume} {380}},\ \bibinfo {pages} {151} (\bibinfo {year} {2001})},\ \Eprint
	{http://arxiv.org/abs/astro-ph/0111092} {arXiv:astro-ph/0111092 [astro-ph]}
	\BibitemShut {NoStop}%
	\bibitem [{\citenamefont {Avancini}\ \emph {et~al.}(2008)\citenamefont
		{Avancini}, \citenamefont {Menezes}, \citenamefont {Alloy}, \citenamefont
		{Marinelli}, \citenamefont {Moraes},\ and\ \citenamefont
		{Provid\^encia}}]{Avancini08}%
	\BibitemOpen
	\bibfield  {author} {\bibinfo {author} {\bibfnamefont {S.~S.}\ \bibnamefont
			{Avancini}}, \bibinfo {author} {\bibfnamefont {D.~P.}\ \bibnamefont
			{Menezes}}, \bibinfo {author} {\bibfnamefont {M.~D.}\ \bibnamefont {Alloy}},
		\bibinfo {author} {\bibfnamefont {J.~R.}\ \bibnamefont {Marinelli}}, \bibinfo
		{author} {\bibfnamefont {M.~M.~W.}\ \bibnamefont {Moraes}}, \ and\ \bibinfo
		{author} {\bibfnamefont {C.}~\bibnamefont {Provid\^encia}},\ }\href {\doibase
		10.1103/PhysRevC.78.015802} {\bibfield  {journal} {\bibinfo  {journal} {Phys.
				Rev. C}\ }\textbf {\bibinfo {volume} {78}},\ \bibinfo {pages} {015802}
		(\bibinfo {year} {2008})}\BibitemShut {NoStop}%
	\bibitem [{\citenamefont {Grill}\ \emph {et~al.}(2012)\citenamefont {Grill},
		\citenamefont {Providencia},\ and\ \citenamefont {Avancini}}]{Grill12}%
	\BibitemOpen
	\bibfield  {author} {\bibinfo {author} {\bibfnamefont {F.}~\bibnamefont
			{Grill}}, \bibinfo {author} {\bibfnamefont {C.}~\bibnamefont {Providencia}},
		\ and\ \bibinfo {author} {\bibfnamefont {S.~S.}\ \bibnamefont {Avancini}},\
	}\href {\doibase 10.1103/PhysRevC.85.055808} {\bibfield  {journal} {\bibinfo
			{journal} {Phys. Rev.}\ }\textbf {\bibinfo {volume} {C85}},\ \bibinfo {pages}
		{055808} (\bibinfo {year} {2012})},\ \Eprint {http://arxiv.org/abs/1203.4166}
	{arXiv:1203.4166 [nucl-th]} \BibitemShut {NoStop}%
	\bibitem [{\citenamefont {Grill}\ \emph {et~al.}(2014)\citenamefont {Grill},
		\citenamefont {Pais}, \citenamefont {Providência}, \citenamefont {Vidaña},\
		and\ \citenamefont {Avancini}}]{Grill14}%
	\BibitemOpen
	\bibfield  {author} {\bibinfo {author} {\bibfnamefont {F.}~\bibnamefont
			{Grill}}, \bibinfo {author} {\bibfnamefont {H.}~\bibnamefont {Pais}},
		\bibinfo {author} {\bibfnamefont {C.}~\bibnamefont {Providência}}, \bibinfo
		{author} {\bibfnamefont {I.}~\bibnamefont {Vidaña}}, \ and\ \bibinfo
		{author} {\bibfnamefont {S.~S.}\ \bibnamefont {Avancini}},\ }\href {\doibase
		10.1103/PhysRevC.90.045803} {\bibfield  {journal} {\bibinfo  {journal} {Phys.
				Rev.}\ }\textbf {\bibinfo {volume} {C90}},\ \bibinfo {pages} {045803}
		(\bibinfo {year} {2014})},\ \Eprint {http://arxiv.org/abs/1404.2753}
	{arXiv:1404.2753 [nucl-th]} \BibitemShut {NoStop}%
	\bibitem [{\citenamefont {Lalazissis}\ \emph {et~al.}(1997)\citenamefont
		{Lalazissis}, \citenamefont {Konig},\ and\ \citenamefont {Ring}}]{nl3}%
	\BibitemOpen
	\bibfield  {author} {\bibinfo {author} {\bibfnamefont {G.~A.}\ \bibnamefont
			{Lalazissis}}, \bibinfo {author} {\bibfnamefont {J.}~\bibnamefont {Konig}}, \
		and\ \bibinfo {author} {\bibfnamefont {P.}~\bibnamefont {Ring}},\ }\href
	{\doibase 10.1103/PhysRevC.55.540} {\bibfield  {journal} {\bibinfo  {journal}
			{Phys. Rev.}\ }\textbf {\bibinfo {volume} {C55}},\ \bibinfo {pages} {540}
		(\bibinfo {year} {1997})},\ \Eprint {http://arxiv.org/abs/nucl-th/9607039}
	{arXiv:nucl-th/9607039 [nucl-th]} \BibitemShut {NoStop}%
	\bibitem [{\citenamefont {Gaitanos}\ \emph {et~al.}(2004)\citenamefont
		{Gaitanos}, \citenamefont {Di~Toro}, \citenamefont {Typel}, \citenamefont
		{Baran}, \citenamefont {Fuchs}, \citenamefont {Greco},\ and\ \citenamefont
		{Wolter}}]{ddhd}%
	\BibitemOpen
	\bibfield  {author} {\bibinfo {author} {\bibfnamefont {T.}~\bibnamefont
			{Gaitanos}}, \bibinfo {author} {\bibfnamefont {M.}~\bibnamefont {Di~Toro}},
		\bibinfo {author} {\bibfnamefont {S.}~\bibnamefont {Typel}}, \bibinfo
		{author} {\bibfnamefont {V.}~\bibnamefont {Baran}}, \bibinfo {author}
		{\bibfnamefont {C.}~\bibnamefont {Fuchs}}, \bibinfo {author} {\bibfnamefont
			{V.}~\bibnamefont {Greco}}, \ and\ \bibinfo {author} {\bibfnamefont {H.~H.}\
			\bibnamefont {Wolter}},\ }\href@noop {} {\bibfield  {journal} {\bibinfo
			{journal} {Nucl. Phys.}\ }\textbf {\bibinfo {volume} {A732}},\ \bibinfo
		{pages} {24} (\bibinfo {year} {2004})}\BibitemShut {NoStop}%
	\bibitem [{\citenamefont {Pais}\ and\ \citenamefont
		{Providência}(2016)}]{Pais16}%
	\BibitemOpen
	\bibfield  {author} {\bibinfo {author} {\bibfnamefont {H.}~\bibnamefont
			{Pais}}\ and\ \bibinfo {author} {\bibfnamefont {C.}~\bibnamefont
			{Providência}},\ }\href {\doibase 10.1103/PhysRevC.94.015808} {\bibfield
		{journal} {\bibinfo  {journal} {Phys. Rev.}\ }\textbf {\bibinfo {volume}
			{C94}},\ \bibinfo {pages} {015808} (\bibinfo {year} {2016})},\ \Eprint
	{http://arxiv.org/abs/1607.05899} {arXiv:1607.05899 [nucl-th]} \BibitemShut
	{NoStop}%
	\bibitem [{\citenamefont {Raduta}\ and\ \citenamefont
		{Gulminelli}(2010)}]{Raduta:2010ym}%
	\BibitemOpen
	\bibfield  {author} {\bibinfo {author} {\bibfnamefont {A.}~\bibnamefont
			{Raduta}}\ and\ \bibinfo {author} {\bibfnamefont {F.}~\bibnamefont
			{Gulminelli}},\ }\href {\doibase 10.1103/PhysRevC.82.065801} {\bibfield
		{journal} {\bibinfo  {journal} {Phys. Rev. C}\ }\textbf {\bibinfo {volume}
			{82}},\ \bibinfo {pages} {065801} (\bibinfo {year} {2010})},\ \Eprint
	{http://arxiv.org/abs/1009.2226} {arXiv:1009.2226 [nucl-th]} \BibitemShut
	{NoStop}%
	\bibitem [{\citenamefont {Pais}\ \emph {et~al.}(2014)\citenamefont {Pais},
		\citenamefont {Newton},\ and\ \citenamefont {Stone}}]{Pais2014}%
	\BibitemOpen
	\bibfield  {author} {\bibinfo {author} {\bibfnamefont {H.}~\bibnamefont
			{Pais}}, \bibinfo {author} {\bibfnamefont {W.~G.}\ \bibnamefont {Newton}}, \
		and\ \bibinfo {author} {\bibfnamefont {J.~R.}\ \bibnamefont {Stone}},\ }\href
	{\doibase 10.1103/PhysRevC.90.065802} {\bibfield  {journal} {\bibinfo
			{journal} {Phys. Rev. C}\ }\textbf {\bibinfo {volume} {90}},\ \bibinfo
		{pages} {065802} (\bibinfo {year} {2014})},\ \Eprint
	{http://arxiv.org/abs/1411.1885} {arXiv:1411.1885 [nucl-th]} \BibitemShut
	{NoStop}%
	\bibitem [{\citenamefont {Ducoin}\ \emph {et~al.}(2011)\citenamefont {Ducoin},
		\citenamefont {Margueron}, \citenamefont {Providencia},\ and\ \citenamefont
		{Vidana}}]{Ducoin2011}%
	\BibitemOpen
	\bibfield  {author} {\bibinfo {author} {\bibfnamefont {C.}~\bibnamefont
			{Ducoin}}, \bibinfo {author} {\bibfnamefont {J.}~\bibnamefont {Margueron}},
		\bibinfo {author} {\bibfnamefont {C.}~\bibnamefont {Providencia}}, \ and\
		\bibinfo {author} {\bibfnamefont {I.}~\bibnamefont {Vidana}},\ }\href
	{\doibase 10.1103/PhysRevC.83.045810} {\bibfield  {journal} {\bibinfo
			{journal} {Phys. Rev.}\ }\textbf {\bibinfo {volume} {C83}},\ \bibinfo {pages}
		{045810} (\bibinfo {year} {2011})},\ \Eprint {http://arxiv.org/abs/1102.1283}
	{arXiv:1102.1283 [nucl-th]} \BibitemShut {NoStop}%
	\bibitem [{\citenamefont {Tolman}(1939)}]{TOV1}%
	\BibitemOpen
	\bibfield  {author} {\bibinfo {author} {\bibfnamefont {R.~C.}\ \bibnamefont
			{Tolman}},\ }\href {\doibase 10.1103/PhysRev.55.364} {\bibfield  {journal}
		{\bibinfo  {journal} {Phys. Rev.}\ }\textbf {\bibinfo {volume} {55}},\
		\bibinfo {pages} {364} (\bibinfo {year} {1939})}\BibitemShut {NoStop}%
	\bibitem [{\citenamefont {Oppenheimer}\ and\ \citenamefont
		{Volkoff}(1939)}]{TOV2}%
	\BibitemOpen
	\bibfield  {author} {\bibinfo {author} {\bibfnamefont {J.~R.}\ \bibnamefont
			{Oppenheimer}}\ and\ \bibinfo {author} {\bibfnamefont {G.~M.}\ \bibnamefont
			{Volkoff}},\ }\href {\doibase 10.1103/PhysRev.55.374} {\bibfield  {journal}
		{\bibinfo  {journal} {Phys. Rev.}\ }\textbf {\bibinfo {volume} {55}},\
		\bibinfo {pages} {374} (\bibinfo {year} {1939})}\BibitemShut {NoStop}%
	\bibitem [{\citenamefont {Hinderer}(2008)}]{Hinderer2008}%
	\BibitemOpen
	\bibfield  {author} {\bibinfo {author} {\bibfnamefont {T.}~\bibnamefont
			{Hinderer}},\ }\href {\doibase 10.1086/533487} {\bibfield  {journal}
		{\bibinfo  {journal} {Astrophys. J.}\ }\textbf {\bibinfo {volume} {677}},\
		\bibinfo {pages} {1216} (\bibinfo {year} {2008})},\ \Eprint
	{http://arxiv.org/abs/0711.2420} {arXiv:0711.2420 [astro-ph]} \BibitemShut
	{NoStop}%
	\bibitem [{\citenamefont {Radice}\ and\ \citenamefont
		{Dai}(2019)}]{Radice2019}%
	\BibitemOpen
	\bibfield  {author} {\bibinfo {author} {\bibfnamefont {D.}~\bibnamefont
			{Radice}}\ and\ \bibinfo {author} {\bibfnamefont {L.}~\bibnamefont {Dai}},\
	}\href {\doibase 10.1140/epja/i2019-12716-4} {\bibfield  {journal} {\bibinfo
			{journal} {The European Physical Journal A}\ }\textbf {\bibinfo {volume}
			{55}} (\bibinfo {year} {2019}),\ 10.1140/epja/i2019-12716-4}\BibitemShut
	{NoStop}%
	\bibitem [{\citenamefont {Maselli}\ \emph {et~al.}(2013)\citenamefont
		{Maselli}, \citenamefont {Cardoso}, \citenamefont {Ferrari}, \citenamefont
		{Gualtieri},\ and\ \citenamefont {Pani}}]{Maselli2013}%
	\BibitemOpen
	\bibfield  {author} {\bibinfo {author} {\bibfnamefont {A.}~\bibnamefont
			{Maselli}}, \bibinfo {author} {\bibfnamefont {V.}~\bibnamefont {Cardoso}},
		\bibinfo {author} {\bibfnamefont {V.}~\bibnamefont {Ferrari}}, \bibinfo
		{author} {\bibfnamefont {L.}~\bibnamefont {Gualtieri}}, \ and\ \bibinfo
		{author} {\bibfnamefont {P.}~\bibnamefont {Pani}},\ }\href {\doibase
		10.1103/PhysRevD.88.023007} {\bibfield  {journal} {\bibinfo  {journal} {Phys.
				Rev.}\ }\textbf {\bibinfo {volume} {D88}},\ \bibinfo {pages} {023007}
		(\bibinfo {year} {2013})},\ \Eprint {http://arxiv.org/abs/1304.2052}
	{arXiv:1304.2052 [gr-qc]} \BibitemShut {NoStop}%
	\bibitem [{\citenamefont {Yagi}\ and\ \citenamefont {Yunes}(2017)}]{Yagi2016}%
	\BibitemOpen
	\bibfield  {author} {\bibinfo {author} {\bibfnamefont {K.}~\bibnamefont
			{Yagi}}\ and\ \bibinfo {author} {\bibfnamefont {N.}~\bibnamefont {Yunes}},\
	}\href {\doibase 10.1016/j.physrep.2017.03.002} {\bibfield  {journal}
		{\bibinfo  {journal} {Phys. Rept.}\ }\textbf {\bibinfo {volume} {681}},\
		\bibinfo {pages} {1} (\bibinfo {year} {2017})},\ \Eprint
	{http://arxiv.org/abs/1608.02582} {arXiv:1608.02582 [gr-qc]} \BibitemShut
	{NoStop}%
	\bibitem [{\citenamefont {Yagi}\ and\ \citenamefont {Yunes}(2013)}]{Yagi2013}%
	\BibitemOpen
	\bibfield  {author} {\bibinfo {author} {\bibfnamefont {K.}~\bibnamefont
			{Yagi}}\ and\ \bibinfo {author} {\bibfnamefont {N.}~\bibnamefont {Yunes}},\
	}\href {\doibase 10.1126/science.1236462} {\bibfield  {journal} {\bibinfo
			{journal} {Science}\ }\textbf {\bibinfo {volume} {341}},\ \bibinfo {pages}
		{365} (\bibinfo {year} {2013})},\ \Eprint {http://arxiv.org/abs/1302.4499}
	{arXiv:1302.4499 [gr-qc]} \BibitemShut {NoStop}%
	\bibitem [{\citenamefont {Chan}\ \emph {et~al.}(2016)\citenamefont {Chan},
		\citenamefont {Chan},\ and\ \citenamefont {Leung}}]{Chan2016}%
	\BibitemOpen
	\bibfield  {author} {\bibinfo {author} {\bibfnamefont {T.~K.}\ \bibnamefont
			{Chan}}, \bibinfo {author} {\bibfnamefont {A.~P.~O.}\ \bibnamefont {Chan}}, \
		and\ \bibinfo {author} {\bibfnamefont {P.~T.}\ \bibnamefont {Leung}},\ }\href
	{\doibase 10.1103/PhysRevD.93.024033} {\bibfield  {journal} {\bibinfo
			{journal} {Phys. Rev. D}\ }\textbf {\bibinfo {volume} {93}},\ \bibinfo
		{pages} {024033} (\bibinfo {year} {2016})}\BibitemShut {NoStop}%
	\bibitem [{\citenamefont {De}\ \emph {et~al.}(2018)\citenamefont {De},
		\citenamefont {Finstad}, \citenamefont {Lattimer}, \citenamefont {Brown},
		\citenamefont {Berger},\ and\ \citenamefont {Biwer}}]{De:2018uhw}%
	\BibitemOpen
	\bibfield  {author} {\bibinfo {author} {\bibfnamefont {S.}~\bibnamefont
			{De}}, \bibinfo {author} {\bibfnamefont {D.}~\bibnamefont {Finstad}},
		\bibinfo {author} {\bibfnamefont {J.~M.}\ \bibnamefont {Lattimer}}, \bibinfo
		{author} {\bibfnamefont {D.~A.}\ \bibnamefont {Brown}}, \bibinfo {author}
		{\bibfnamefont {E.}~\bibnamefont {Berger}}, \ and\ \bibinfo {author}
		{\bibfnamefont {C.~M.}\ \bibnamefont {Biwer}},\ }\href {\doibase
		10.1103/PhysRevLett.121.259902, 10.1103/PhysRevLett.121.091102} {\bibfield
		{journal} {\bibinfo  {journal} {Phys. Rev. Lett.}\ }\textbf {\bibinfo
			{volume} {121}},\ \bibinfo {pages} {091102} (\bibinfo {year} {2018})},\
	\bibinfo {note} {[Erratum: Phys. Rev. Lett.121,no.25,259902(2018)]},\ \Eprint
	{http://arxiv.org/abs/1804.08583} {arXiv:1804.08583 [astro-ph.HE]}
	\BibitemShut {NoStop}%
	\bibitem [{\citenamefont {{Baym}}\ \emph {et~al.}(1971)\citenamefont {{Baym}},
		\citenamefont {{Pethick}},\ and\ \citenamefont {{Sutherland}}}]{bps}%
	\BibitemOpen
	\bibfield  {author} {\bibinfo {author} {\bibfnamefont {G.}~\bibnamefont
			{{Baym}}}, \bibinfo {author} {\bibfnamefont {C.}~\bibnamefont {{Pethick}}}, \
		and\ \bibinfo {author} {\bibfnamefont {P.}~\bibnamefont {{Sutherland}}},\
	}\href {\doibase 10.1086/151216} {\bibfield  {journal} {\bibinfo  {journal}
			{\apj}\ }\textbf {\bibinfo {volume} {170}},\ \bibinfo {pages} {299} (\bibinfo
		{year} {1971})}\BibitemShut {NoStop}%
	\bibitem [{\citenamefont {Negele}\ and\ \citenamefont
		{Vautherin}(1973)}]{nveos}%
	\BibitemOpen
	\bibfield  {author} {\bibinfo {author} {\bibfnamefont {J.~W.}\ \bibnamefont
			{Negele}}\ and\ \bibinfo {author} {\bibfnamefont {D.}~\bibnamefont
			{Vautherin}},\ }\href@noop {} {\bibfield  {journal} {\bibinfo  {journal}
			{Nuclear Physics A}\ }\textbf {\bibinfo {volume} {207}},\ \bibinfo {pages}
		{298} (\bibinfo {year} {1973})}\BibitemShut {NoStop}%
	\bibitem [{\citenamefont {Annala}\ \emph {et~al.}(2018)\citenamefont {Annala},
		\citenamefont {Gorda}, \citenamefont {Kurkela},\ and\ \citenamefont
		{Vuorinen}}]{Annala18}%
	\BibitemOpen
	\bibfield  {author} {\bibinfo {author} {\bibfnamefont {E.}~\bibnamefont
			{Annala}}, \bibinfo {author} {\bibfnamefont {T.}~\bibnamefont {Gorda}},
		\bibinfo {author} {\bibfnamefont {A.}~\bibnamefont {Kurkela}}, \ and\
		\bibinfo {author} {\bibfnamefont {A.}~\bibnamefont {Vuorinen}},\ }\href
	{\doibase 10.1103/PhysRevLett.120.172703} {\bibfield  {journal} {\bibinfo
			{journal} {Phys. Rev. Lett.}\ }\textbf {\bibinfo {volume} {120}},\ \bibinfo
		{pages} {172703} (\bibinfo {year} {2018})}\BibitemShut {NoStop}%
	\bibitem [{\citenamefont {Malik}\ \emph
		{et~al.}(2018{\natexlab{b}})\citenamefont {Malik}, \citenamefont {Alam},
		\citenamefont {Fortin}, \citenamefont {Providência}, \citenamefont
		{Agrawal}, \citenamefont {Jha}, \citenamefont {Kumar},\ and\ \citenamefont
		{Patra}}]{Malik:2018zcf}%
	\BibitemOpen
	\bibfield  {author} {\bibinfo {author} {\bibfnamefont {T.}~\bibnamefont
			{Malik}}, \bibinfo {author} {\bibfnamefont {N.}~\bibnamefont {Alam}},
		\bibinfo {author} {\bibfnamefont {M.}~\bibnamefont {Fortin}}, \bibinfo
		{author} {\bibfnamefont {C.}~\bibnamefont {Providência}}, \bibinfo {author}
		{\bibfnamefont {B.~K.}\ \bibnamefont {Agrawal}}, \bibinfo {author}
		{\bibfnamefont {T.~K.}\ \bibnamefont {Jha}}, \bibinfo {author} {\bibfnamefont
			{B.}~\bibnamefont {Kumar}}, \ and\ \bibinfo {author} {\bibfnamefont {S.~K.}\
			\bibnamefont {Patra}},\ }\href {\doibase 10.1103/PhysRevC.98.035804}
	{\bibfield  {journal} {\bibinfo  {journal} {Phys. Rev.}\ }\textbf {\bibinfo
			{volume} {C98}},\ \bibinfo {pages} {035804} (\bibinfo {year}
		{2018}{\natexlab{b}})},\ \Eprint {http://arxiv.org/abs/1805.11963}
	{arXiv:1805.11963 [nucl-th]} \BibitemShut {NoStop}%
	\bibitem [{\citenamefont {Fattoyev}\ \emph {et~al.}(2018)\citenamefont
		{Fattoyev}, \citenamefont {Piekarewicz},\ and\ \citenamefont
		{Horowitz}}]{Fattoyev:2017jql}%
	\BibitemOpen
	\bibfield  {author} {\bibinfo {author} {\bibfnamefont {F.~J.}\ \bibnamefont
			{Fattoyev}}, \bibinfo {author} {\bibfnamefont {J.}~\bibnamefont
			{Piekarewicz}}, \ and\ \bibinfo {author} {\bibfnamefont {C.~J.}\ \bibnamefont
			{Horowitz}},\ }\href {\doibase 10.1103/PhysRevLett.120.172702} {\bibfield
		{journal} {\bibinfo  {journal} {Phys. Rev. Lett.}\ }\textbf {\bibinfo
			{volume} {120}},\ \bibinfo {pages} {172702} (\bibinfo {year} {2018})},\
	\Eprint {http://arxiv.org/abs/1711.06615} {arXiv:1711.06615 [nucl-th]}
	\BibitemShut {NoStop}%
	\bibitem [{\citenamefont {Raithel}\ \emph {et~al.}(2018)\citenamefont
		{Raithel}, \citenamefont {Özel},\ and\ \citenamefont
		{Psaltis}}]{Raithel2018}%
	\BibitemOpen
	\bibfield  {author} {\bibinfo {author} {\bibfnamefont {C.}~\bibnamefont
			{Raithel}}, \bibinfo {author} {\bibfnamefont {F.}~\bibnamefont {Özel}}, \
		and\ \bibinfo {author} {\bibfnamefont {D.}~\bibnamefont {Psaltis}},\ }\href
	{\doibase 10.3847/2041-8213/aabcbf} {\bibfield  {journal} {\bibinfo
			{journal} {Astrophys. J.}\ }\textbf {\bibinfo {volume} {857}},\ \bibinfo
		{pages} {L23} (\bibinfo {year} {2018})},\ \Eprint
	{http://arxiv.org/abs/1803.07687} {arXiv:1803.07687 [astro-ph.HE]}
	\BibitemShut {NoStop}%
	\bibitem [{\citenamefont {Raithel}(2019)}]{Raithel:2019uzi}%
	\BibitemOpen
	\bibfield  {author} {\bibinfo {author} {\bibfnamefont {C.~A.}\ \bibnamefont
			{Raithel}},\ }\href {\doibase 10.1140/epja/i2019-12759-5} {\bibfield
		{journal} {\bibinfo  {journal} {Eur. Phys. J.}\ }\textbf {\bibinfo {volume}
			{A55}},\ \bibinfo {pages} {80} (\bibinfo {year} {2019})},\ \Eprint
	{http://arxiv.org/abs/1904.10002} {arXiv:1904.10002 [astro-ph.HE]}
	\BibitemShut {NoStop}%
	\bibitem [{\citenamefont {Most}\ \emph {et~al.}(2018)\citenamefont {Most},
		\citenamefont {Weih}, \citenamefont {Rezzolla},\ and\ \citenamefont
		{Schaffner-Bielich}}]{Most18}%
	\BibitemOpen
	\bibfield  {author} {\bibinfo {author} {\bibfnamefont {E.~R.}\ \bibnamefont
			{Most}}, \bibinfo {author} {\bibfnamefont {L.~R.}\ \bibnamefont {Weih}},
		\bibinfo {author} {\bibfnamefont {L.}~\bibnamefont {Rezzolla}}, \ and\
		\bibinfo {author} {\bibfnamefont {J.}~\bibnamefont {Schaffner-Bielich}},\
	}\href {\doibase 10.1103/PhysRevLett.120.261103} {\bibfield  {journal}
		{\bibinfo  {journal} {Phys. Rev. Lett.}\ }\textbf {\bibinfo {volume} {120}},\
		\bibinfo {pages} {261103} (\bibinfo {year} {2018})}\BibitemShut {NoStop}%
\end{thebibliography}

%

\end{document}